\documentclass[a4paper,11pt]{article}

\usepackage[a4paper,top=1in,bottom=1in,left=1in,right=1in,marginparwidth=2cm]{geometry}

\usepackage[toc,page]{appendix}
\usepackage{cite} % Tidies up citation numbers.
\usepackage{url} % Provides better formatting of URLs.
\usepackage[utf8]{inputenc} % Allows Turkish characters.
\usepackage{booktabs} % Allows the use of \toprule, \midrule and \bottomrule in tables for horizontal lines
\usepackage{graphicx}
\usepackage{amsmath}
\usepackage{amsfonts}
\usepackage{amssymb}
\usepackage{xcolor}
\usepackage{graphicx}
\usepackage{float}
\usepackage{caption}
\usepackage{lipsum}
\usepackage{titlesec}
\usepackage{multirow}
\usepackage{lineno}
\usepackage{tikz}
\usepackage{upgreek}
\usepackage{jinstpub}
\usepackage{bm}

\graphicspath{{plots/}}
\usepackage[separate-uncertainty = true,multi-part-units=single]{siunitx}
\usepackage{listings}
\newcommand{\thicc}{\mathscr{t}}
\newcommand{\pitch}{\mathscr{p}}

\title{Calibration and simulation of ionization signal and electronics noise in the ICARUS liquid argon time projection chamber}
\author[28]{P. Abratenko}
\author[4]{N. Abrego-Martinez}
\author[7]{A. Aduszkiewicz}
\author[23]{F. Akbar}
\author[27]{L. Aliaga Soplin}
\author[17]{M. Artero Pons}
\author[27]{J. Asaadi}
\author[6]{W. F. Badgett}
\author[17]{B. Baibussinov}
\author[5]{B. Behera}
% ...
\affiliation[4]{Centro de Investigacion y de Estudios Avanzados del IPN (Cinvestav), Mexico City}
\affiliation[5]{Colorado State University, Fort Collins, CO 80523, USA}
\affiliation[6]{Fermi National Accelerator Laboratory, Batavia, IL 60510, USA}
\affiliation[7]{University of Houston, Houston, TX 77204, USA}
% ...
\affiliation[17]{INFN Sezione di Padova and University, Padova, Italy}
% ...
\affiliation[23]{University of Rochester, Rochester, NY 14627, USA}
% ...
\affiliation[27]{University of Texas at Arlington, Arlington, TX 76019, USA}
\affiliation[28]{Tufts University, Medford, MA 02155, USA}
% ...
\emailAdd{gputnam@fnal.gov}
\emailAdd{justin.mueller@colostate.edu}
\date{\today}
\abstract{The ICARUS liquid argon time projection chamber (LArTPC) neutrino detector has been taking physics data since 2022 as part of the Short-Baseline Neutrino (SBN) Program. This paper details the equalization of the response to charge in the ICARUS time projection chamber (TPC), as well as data-driven tuning of the simulation of ionization charge signals and electronics noise. The equalization procedure removes non-uniformities in the ICARUS TPC response to charge in space and time. This work leverages the copious number of cosmic ray muons available to ICARUS at the surface. The ionization signal shape simulation applies a novel procedure that tunes the simulation to match what is measured in data. The end result of the equalization procedure and simulation tuning allows for a comparison of charge measurements in ICARUS between Monte Carlo simulation and data, showing good performance with minimal residual bias between the two.}

\begin{document}
\maketitle
\flushbottom

\section{Introduction}
\label{sec:Introduction}
Liquid argon time projection chamber (LArTPC)  detectors track particle trajectories with high spatial resolution and precise calorimetry by imaging ionization electrons from charged particle tracks and showers. Ionization charge is drifted by a large electric field to multiple planes of readout wires which detect the charge as induced currents on each wire. The ICARUS LArTPC neutrino detector, after a previous run at Gran Sasso \cite{ICARUSOG} and subsequent refurbishment, has been installed at Fermilab since 2020 \cite{ICARUSOverhaul}. It has been taking physics data since 2022 as part of the Short-Baseline Neutrino (SBN) Program \cite{SBNProposal, SBNProgram}. ICARUS sits at the intersection of two neutrino beams; it is on-axis to the Booster Neutrino Beam (BNB) \cite{BNB} and is 5.7$^\circ$ off-axis to the Neutrinos at the Main Injector (NuMI) beam \cite{NuMI}. This paper addresses the calibration and simulation of electronic noise and charge signals in the ICARUS time projection chamber at Fermilab \cite{ICARUSResults}. The calibration reported here addresses the data taken in the first two ICARUS physics data collection periods: Run 1, from June 9th to July 9th 2022, and Run 2, from December 20th 2022 to July 14th 2023.

\begin{figure}[t]
    \centering
    \includegraphics[width=\textwidth]{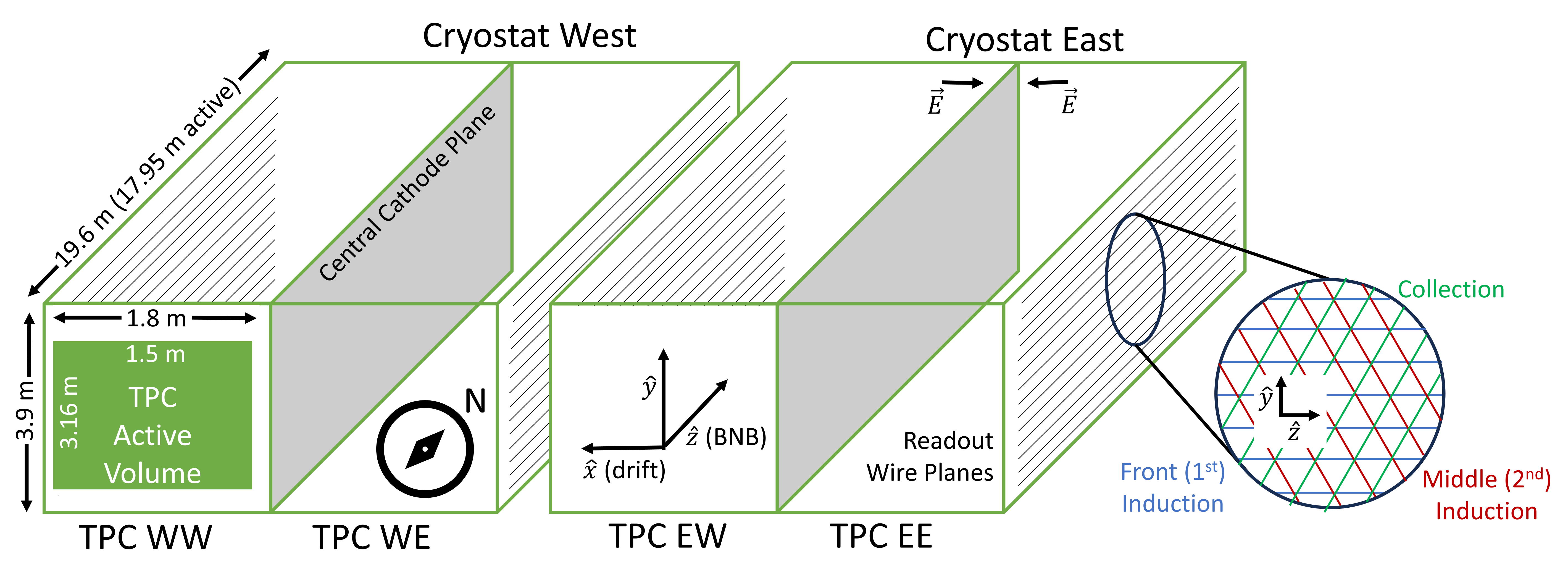}
    \caption{Diagram of the layout and enumeration of the ICARUS TPCs. Not to scale. The wire plane orientations are mirrored in opposite TPCs. The East and West cryostats have the same layout.}
    \label{fig:TPCDiagram}
\end{figure}

ICARUS is a \SI{760}{\tonne} liquid argon detector consisting of two LArTPC modules. Each module is a cryostat with dimensions 3.6$\times$3.9$\times$\SI{19.6}{\meter\cubed}. Both cryostats contain two TPCs divided by a central cathode plane. Each TPC has an active volume of 1.5$\times$3.16$\times$\SI{17.95}{\meter\cubed}. The TPCs are all operated at a drift voltage of about \SI{500}{\volt\per\centi\meter}. They all have three planes of charge sensing wires: an unshielded front induction plane, a middle induction plane, and a collection plane. The wires on the front induction plane are oriented along the horizontal (beam) direction, and the wires on the middle induction and collection plane are oriented at $\pm 60^\circ$ to the horizontal direction, depending on the TPC. The wires on each plane are each spaced \SI{3}{\milli\meter} apart and the wire planes are spaced \SI{3}{\milli\meter} from each other. The front induction wire plane is split in two at the center of the TPC by a mechanical support. In the nominal configuration, the wire bias is \SI{-250}{V} on the front induction plane, \SI{-30}{V} on the middle induction plane, and \SI{250}{V} on the collection plane. A diagram of the layout of the four ICARUS TPCs is shown in figure \ref{fig:TPCDiagram}.

Each TPC wire is instrumented to digitize the charge signals while maximizing the signal-to-noise ratio \cite{ICARUSElectronics}. Signals are run through a signal processing chain which subtracts noise that is coherent across wires in the same readout board and deconvolves the signal to provide a Gaussian shape with further reduced noise \cite{ICARUSResults}. These signals provide the input to reconstruction algorithms which group together hits into tracks (from muons, protons, or other charged hadrons) and electromagnetic showers (from electrons or photons) for use in analysis. The reconstruction applied here is supplied by the Pandora framework \cite{Pandora, DUNEPandora}, optimized for the ICARUS detector.

The detection of charge in the ICARUS detector is not perfectly uniform in space and time. Effects such as argon impurities \cite{ArImpurity2, ArImpurity} and space charge effects \cite{MikeSpaceCharge, uBSpaceCharge} can perturb the amount of charge measured across the detector. Furthermore, we have observed a non-uniform in-transparency on the middle induction plane across the ICARUS TPCs that perturbs the charge response on all three planes. We have developed a procedure to calibrate these effects so that the non-uniformity they induce can be removed from the data. This procedure relies on the copious number of cosmic ray muons available to ICARUS, which operates under only \SI{10}{m}.w.e.\ of concrete overburden. The procedures we have developed leverage many of the ideas first developed by the MicroBooNE surface LArTPC experiment \cite{CalibrationuB}, applied to the specific conditions observed in ICARUS.

The simulation of TPC signals in ICARUS is organized in the LArSoft framework \cite{LArSoft}. A Geant4 simulation \cite{GEANT} tuned for use in argon propagates the trajectories of particles in the detector and computes their ionization and scintillation depositions in the active volume. The Wire-Cell package \cite{WireCell}, with these depositions as input, drifts the ionization charge to the wire planes, including parameterized effects from attenuation due to argon impurities and ionization diffusion. Wire-Cell simulates the signal of ionization electrons on each wire plane by applying a field response computation from the GARFIELD program \cite{Garfield} with the nominal ICARUS wire plane configuration as input. The field response computation is two-dimensional (2D): it includes the dependence of the induced current on both the drift time of the ionization and the pitch of the ionization in the direction perpendicular to each wire. This accounts for long-range induced currents which are important for accurately modeling the charge signal.

We have measured the characteristics of the ICARUS TPC noise and signal for use in the simulation. Electronic noise observed in the detector includes sources intrinsic to each readout channel and sources which are common to readout channels sharing electronics. Updating the simulation with a data-driven electronic noise model is an important first step in carrying out tuning of the ionization signal response given that noise can lead to ``smearing'' in the estimated signal response shape, necessitating use of an accurate Monte Carlo simulation to account for this effect. The signal shapes we observe in the detector depart from the nominal prediction made by Wire-Cell and GARFIELD. This departure is significant, although it is not drastically different from the level of disagreement observed by prior experiments applying the same (2D) simulation \cite{SignalShapeuB}. It is critical to precisely simulate the signal shapes in LArTPCs in order to accurately characterize the performance of signal processing and its impact on physics analysis. The leading systematic uncertainty associated with detector performance in prior LArTPC experiments has been the TPC signal shape (see, e.g., Ref. \cite{uBxsecOld, uBxsecNew, uBLEE}). We have developed a novel approach to tune the underlying field responses input to Wire-Cell which match the simulated signal shapes precisely to what is measured in the detector.

This paper is organized as follows. In section \ref{sec:Equalization}, we describe the procedure used to remove non-uniformities in the ionization charge response of the ICARUS TPCs and demonstrate its impact on the charge resolution of the detector. In section \ref{sec:Noise}, we detail the measurement of electronics noise in the ICARUS TPCs. In section \ref{sec:SignalShape}, we describe the measurement of TPC signal shapes in ICARUS data, as well as the novel procedure we have developed to tune simulation to match the data. Section \ref{sec:ChargeResComp} shows the comparison of charge resolution performance between ICARUS Monte Carlo simulation and data after the calibration techniques described here are used in improvements to the simulation. Finally, section \ref{sec:Conclusion} concludes the paper.

\section{Charge Scale Equalization}
\label{sec:Equalization}
The goal of the charge scale equalization procedure is to make the TPC response to charge uniform in space and time. This is expressed in terms of the charge per length, or $dQ/dx$, of hits along particle tracks and showers. This quantity is used to compute energy loss ($dE/dx$) after correcting for electron-ion recombination \cite{ImelRecomb, ICARUSrecomb, NEUTRecomb}. As is detailed below (section \ref{sec:q=effect}), a number of effects perturb the charge response in ICARUS. To account for these effects, we have elected to equalize the charge response in three steps: an equalization in the drift direction (section \ref{sec:q=drift}), an equalization in the two wire plane directions, $\hat{y}$ and $\hat{z}$ (section \ref{sec:q=YZ}), and a final TPC equalization (section \ref{sec:q=TPC}). The performance of charge reconstruction in ICARUS after these equalization steps is shown in section \ref{sec:q=result}.

As a surface detector, ICARUS has access to a copious number of cosmic muon tracks for use in these calibrations. Most muon tracks pass through the detector as nearly minimum-ionizing particles (MIPs). We use a selection of cosmic muons to do these calibrations. The muon tracks are required to cross the cathode. For such tracks, matching the energy depositions in both TPCs on either side of the central cathode enables the identification of the arrival time ($t_0$) of the track. Knowledge of this time is needed to properly compute and apply the drift time correction.

\subsection{Effects Leading to Non-Uniformities in Charge Scale}
\label{sec:q=effect}

\subsubsection{Argon Impurities}

As the ionization cloud from a particle deposition drifts to the wire plane, impurities in the argon (primarily $\text{O}_2$ and $\text{H}_2\text{O}$ \cite{ArImpurity2, ArImpurity}) absorb electrons. The attenuation is exponential and can be described by an electron lifetime, which is the mean time an electron will survive in the argon before it is absorbed. The electron lifetime in ICARUS ranged from 3-\SI{8}{\milli\second} over the dataset considered here, which corresponds to a $\sim$5-15\% average attenuation in the charge signal across the $\sim$\SI{1}{\milli\second} ICARUS drift time.

\subsubsection{Drift Field Distortions}

The drift electric field in ICARUS is not perfectly uniform. While it is very stable across time, a few effects perturb its value spatially across the detector. The constant rate of cosmic muons ionizing the argon produces a build-up of positive argon ions, or space-charge, that affect the electric field \cite{uBSpaceCharge,MikeSpaceCharge}. In addition, the cathode plane in ICARUS is not perfectly flat. This is an effect that was previously observed during the ICARUS run at Gran Sasso \cite{ICARUSCathodeBending}. It is still present in the refurbished ICARUS installation at Fermilab at a much reduced magnitude. The biggest bending is in the East Cryostat, where the cathode is shifted by up to \SI{1.5}{cm}. This perturbs the electric field by a few percent, especially close to the cathode. Finally, there is a failure in the field cage in TPC EE that distorts the drift electric field in that TPC.

The drift field distortions can affect the charge scale in two ways. First, changes to the drift field affect the quantity of electric charge that recombines with argon ions at the point of ionization. Second, distortions to the drift field can deflect ionization tracks and therefore bias the reconstruction of the track pitch -- the $dx$ in  $dQ/dx$. 

At this time, we have not specifically calibrated the impact of drift field distortions. We have measured the broad magnitude of the distortions and found them to be small -- distortions to the drift field of at most a couple percent which deflect tracks by at most a couple centimeters. The charge scale calibrations here should be understood as folding in the (small) impact of drift field distortions.

\subsubsection{Diffusion}
\label{subsec:diffusion}

\begin{figure}[]
    \centering
    \includegraphics[width=0.5\textwidth]{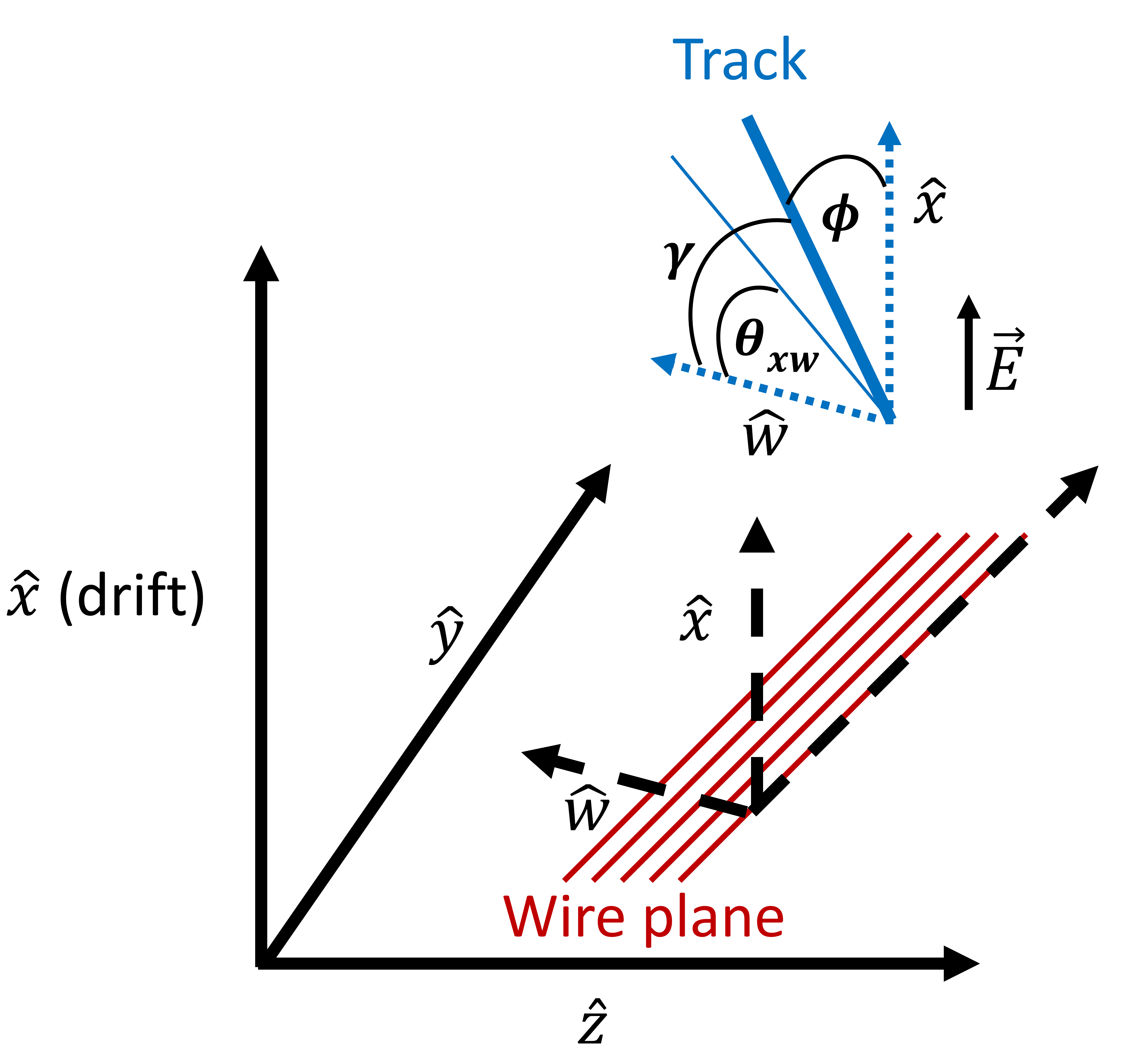}
    \caption{Diagram of the orientation of a track in a TPC and its relevant angles.}
    \label{fig:anglediagram}
\end{figure}

Diffusion in the two directions transverse to the drift field direction has been shown to impact the measured charge scale from cosmic muons \cite{GrayDiff,MichelleDiff}. This effect is due to a dependence of the Landau-Vavilov \cite{Landau,Vavilov} distribution of energy loss from muons on the magnitude of diffusion. The shape of the Landau-Vaviolv distribution depends on the length of the segment of the muon track observed by the individual readout wire. In particular, as this segment length increases, the width of the distribution narrows and the location of the peak (which is typically used as the observable of the distribution \cite{PDG}) rises, approaching the mean energy loss. Transverse diffusion smears the energy depositions observed by each wire, and thus broadens the length of the muon track segment observed by each wire. In the presence of diffusion, this length ($\thicc$) is given by \cite{GrayDiff}
\begin{equation}
    \label{eq:thicc}
    \begin{split}
        \thicc(t_\text{drift}, \gamma) &= \frac{\pitch}{\text{cos}\gamma}\  \text{exp}\left(-\int \frac{dx}{\pitch} w[\sigma_T(t_\text{drift}), x]\  \text{log}\, w[\sigma_T(t_\text{drift}), x] \right)\\ 
        \sigma_T(t_\text{drift}) &= \sqrt{2 D_T t_\text{drift}}\\
        w(\sigma_T, x) &= \int\limits_{-\pitch/2}^{\pitch/2} \frac{dx'}{\sigma_T \sqrt{2\pi}} e^{-\frac{(x-x')^2}{2\sigma_T^2}}\,,
    \end{split}
\end{equation}
where $\gamma$ is a track angle (see figure \ref{fig:anglediagram}), $t_\text{drift}$ is the drift time, $\pitch$ is the wire pitch (\SI{3}{\milli\meter} in ICARUS), $\sigma_T$ is the transverse smearing width, and $D_T$ is the transverse diffusion constant (which has been estimated to be around 5-\SI{12}{\centi\meter\squared\per\second} \cite{ICDT, LiDL, uBDL}). In the limit of no diffusion, $\thicc$ approaches the track pitch ($\pitch / \cos \gamma$). At the maximum ICARUS drift time ($\sim$\SI{1}{\milli\second}), the transverse smearing width is $\sim$1.0-\SI{1.5}{\milli\meter}, on the order of the wire pitch.

Therefore, transverse diffusion does not affect the detector response to charge (except perhaps indirectly through any impact from the broadening of the charge signal), but rather makes the ``standard-candle'' used to equalize the charge scale -- cosmic muons -- not truly standard in the drift direction. As a result, using cosmic muon depositions to equalize the charge scale produces a biased result, since such a procedure applies a non-uniform $dE/dx$ distribution.
The impact of diffusion can be mitigated by summing together adjacent hits on a cosmic muon track into a ``coarse-grained" $dQ/dx$ \cite{GrayDiff}. For example, summing together 10 hits obtains a $dQ/dx$ with an effective spacing of 10 wires, or \SI{3}{\centi\meter}, much larger than than the smearing width of transverse diffusion. 

The coarse-graining method also allows us to study the impact of diffusion in data. The drift direction profile of the ``coarse-grained" $dQ/dx$ is impacted exactly the same by detector non-uniformities (mostly argon impurities and drift field distortions) as a ``wire-by-wire" $dQ/dx$, but the two observables have different underlying $dE/dx$ distributions which are impacted differently by diffusion. 

\begin{figure}[t]
    \centering
    \includegraphics[width=0.9\textwidth]{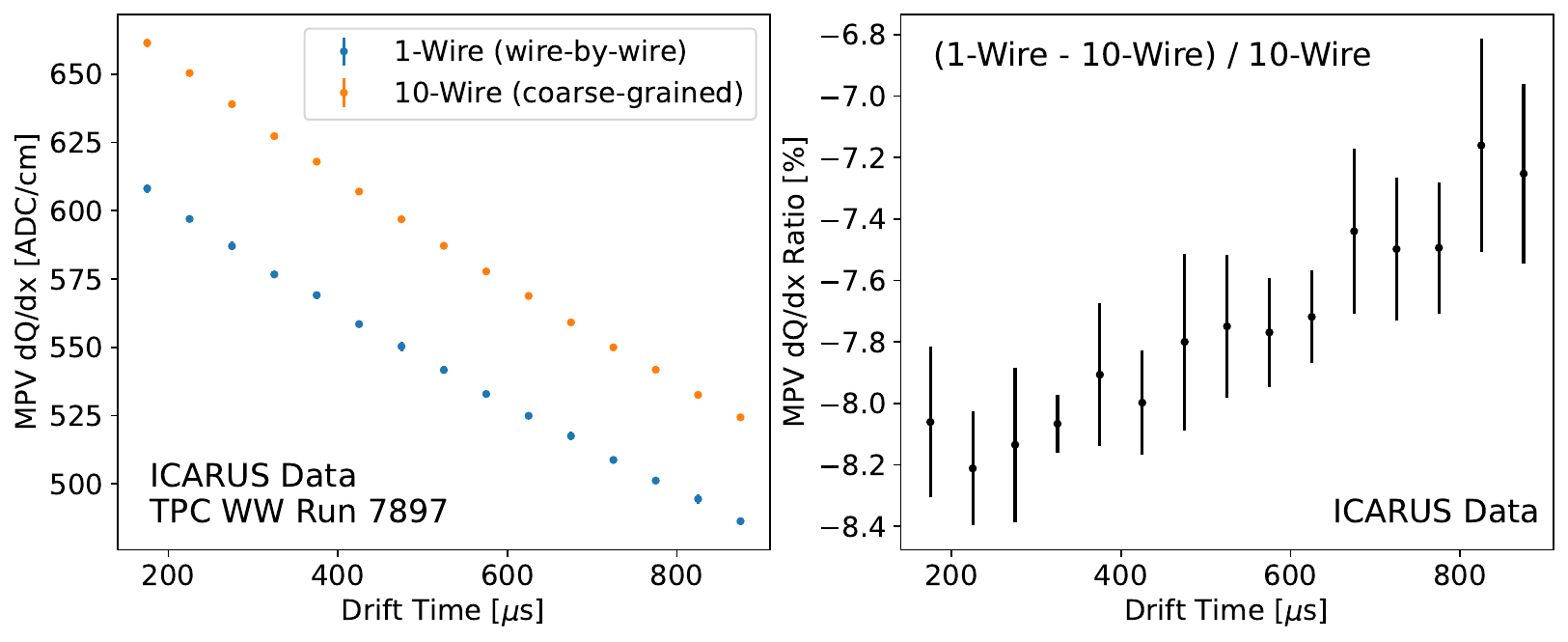}
    \caption{Comparison of the 1-Wire (wire-by-wire) and 10-Wire (coarse grained) charge measurements for one ICARUS DAQ run in TPC WW. In the left panel, the most-probable-value (MPV) of $dQ/dx$ is plotted as a function of the drift time for both measurements. In the right panel, a ratio of the two measurements is shown. The 1-Wire measurement MPV approaches the value of the 10-Wire measurement MPV due to diffusion in the direction transverse to the drift direction.}
    \label{fig:comparecoarsegrain}
\end{figure}

This is demonstrated in ICARUS data in figure \ref{fig:comparecoarsegrain}. Both the coarse-grained and wire-by-wire $dQ/dx$ attenuate across the drift direction, since the biggest effect is from argon impurities. However, the two values also get closer at larger drift times. The coarse-grained measurement has a longer track sensitive length that is constant across the detector, and thus a larger $dE/dx$ peak value. The wire-by-wire measurement has a smaller track sensitive length that increases with increasing transverse diffusion across the drift direction. Therefore, the wire-by-wire peak $dE/dx$ is smaller but approaches the coarse-grained $dE/dx$ peak at large drift times. The magnitude of the effect is hard to predict because it depends on the unknown momentum of the through-going muons used in the measurement. Its direction though is consistent with expectation. This is the first confirmation in data of the impact of transverse diffusion on the muon charge scale, which has previously only been predicted from simulation. It validates the approach we have taken in ICARUS towards drift direction charge equalization, as detailed in section \ref{sec:q=drift}.

\subsubsection{Induction Wire Plane Intransparency}

The induction wire planes in ICARUS (primarily the middle induction plane) absorb charge in a position dependent way across the detector. This effect induces significant variations ($\sim 20$\%) in the charge response on all three readout wire planes. On the collection plane, it directly reduces the amount of visible charge. On the induction planes, the unipolar collection signal from absorbing charge competes with the bipolar induction signal from non-absorbing charge.

A GARFIELD simulation \cite{Garfield} of the nominal ICARUS wire plane configuration predicts that the induction planes absorb 7\% of the charge. Thus, to explain the observed variations, the wire plane configuration of ICARUS must depart from the design specification in a position dependent way across the detector. We have checked all components of the wire bias outside of the cryostat and have found only a couple of discrete failures, which have been correlated to features in the non-uniformity but do not explain all of them. Inside the cryostat, some departure of either the wire bias or the inter-plane wire spacing from the nominal configuration must conspire to produce the spatial variations observed in ICARUS. 

We have investigated the possibility of changing the ICARUS detector configuration to mitigate the impact of this effect. The supplied wire bias cannot be turned any higher due to the rating of cables carrying the bias inside the cryostat. We tested operating at a reduced drift electric field of \SI{350}{\volt\per\centi\meter} (which increases the relative effect of the wire bias) in Summer 2022. We found that the increased effect of recombination and larger attenuation from argon impurities from the longer drift time reduced the signal-to-noise in ICARUS by too much to be feasible, especially on the induction planes. 

Although significant, the variation in the induction wire plane intransparency has been found to be very stable across time. Thus, we can calibrate out the effect using the very large sample of cosmic muons available to ICARUS.

\subsubsection{Channel Gain Variation}
\label{subsec:channelgain}

Ionization signals in ICARUS on charge-sensing wires are transmitted via cables through a set of feed-trough flanges. Each flange connects a group of 64 channels to a readout board, which amplifies and digitizes the signal on each channel. The process of signal transmittance to the front-end, amplification, and digitization may lead to channel-to-channel variations in gain and electronics response. To characterize this variation, a study was performed using the injection of test pulses at distinct points in the electronics chain. 

\paragraph{Methodology}
\label{subsec:methodology}

\begin{figure}[t]
    \centering
    \includegraphics[width=\textwidth]{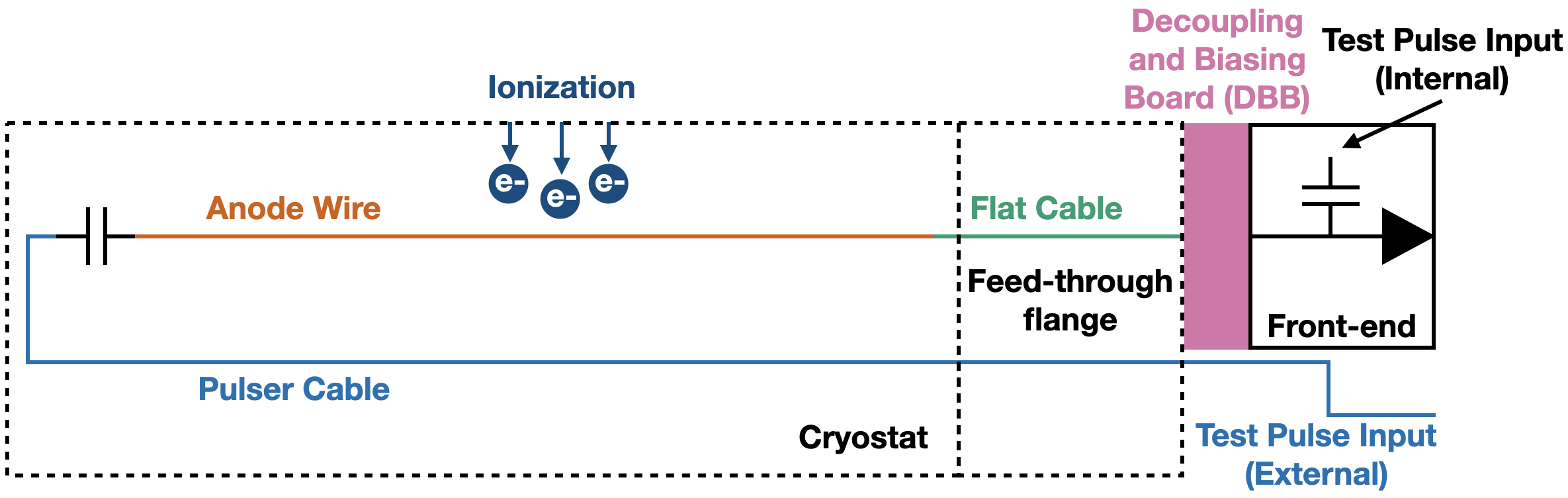}
    \caption{A simplified view of the full electronics chain at ICARUS. Signal from ionization is sensed via wires and transmitted from the chamber to the decoupling and biasing boards before being amplified and digitized in the front-end. Injection of test pulses can be done at two distinct locations: at the end of the wire through a capacitor via inputs on the external side of the flange and directly onto the pre-amplifier through a capacitor.}
    \label{fig:testpulse_readout}
\end{figure}

The ICARUS electronics chain and the test pulse injection points are shown schematically in Figure \ref{fig:testpulse_readout}. The electronics can be pulsed with either an external or an internal signal.

The external pulse (EP) method allows for arbitrary signals to be propagated to the end of the wires through a simple connection available on the external side of the feed-through flange. The signal is transmitted internally via coaxial cables that connect through a capacitor to the wires. Only wires which terminate on the bottom-edge of the detector are instrumented in this way, meaning that most of middle induction and collection wires and none of the front induction wires can be characterized with this method.

The internal pulse (IP) method allows for a \SI{2}{\kilo\hertz} square wave with configurable amplitude to be injected through a capacitor onto the pre-amplifier of either odd or even channels. This is a feature that is integrated into the readout boards and is configurable in software during data acquisition. The method allows for the simultaneous pulsing of channels across the entire detector with no hardware changes required, but allows less freedom of the test pulse signal parameters.

Neither method is individually capable of providing a complete characterization of the full electronics chain due to limitations in precision, ease of configuration, channel coverage, and the portion of the electronics chain probed. However, by comparing results obtained from both methods it is possible to make quantitative conclusions. Consequently, a data-taking campaign was performed to produce a dataset for each of these two methods. Both methods of test pulse injection were performed with a \SI{2}{\kilo\hertz} square wave using amplitudes large enough to span most of the range of the 12-bit ADC.

\begin{figure}[t]
    \centering
    \includegraphics[width=0.5\textwidth]{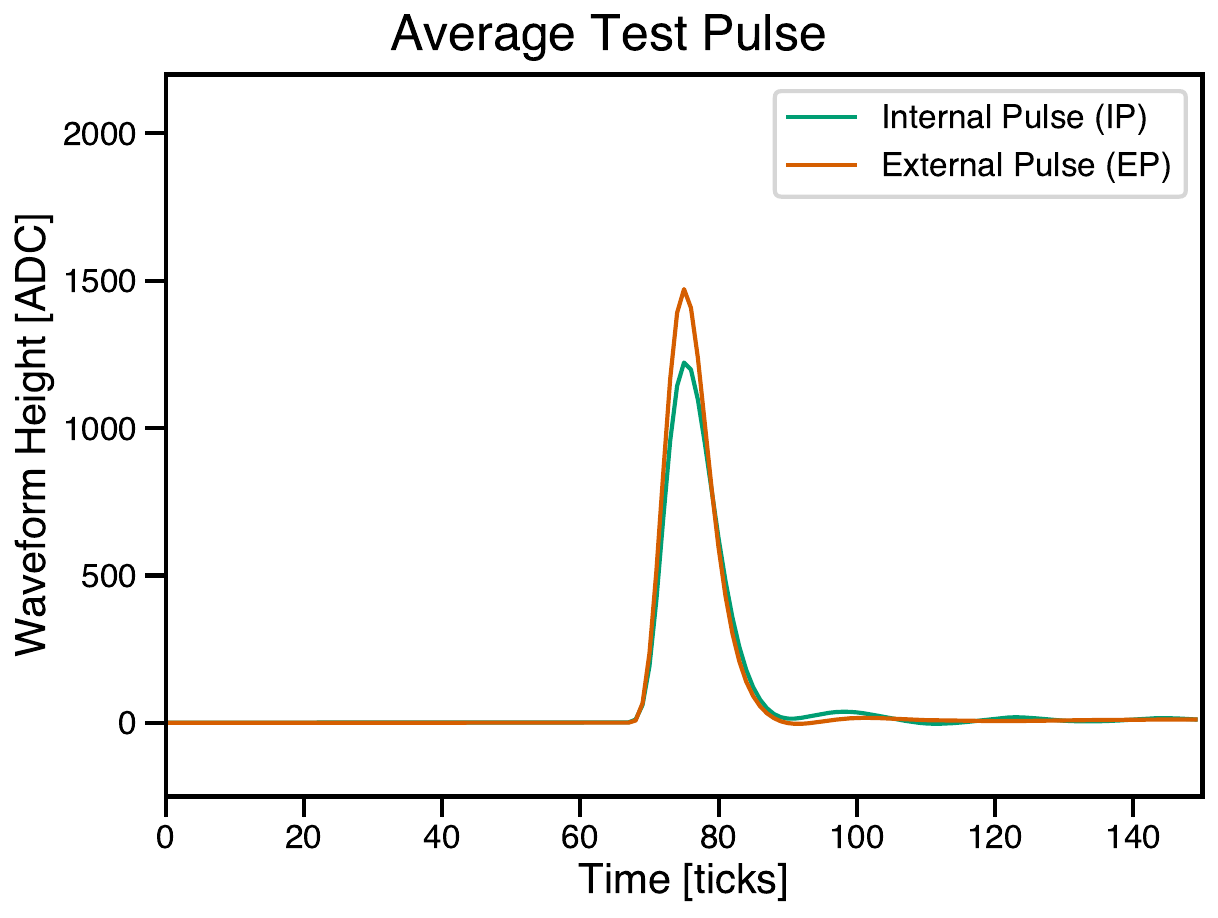}
    \caption{Average injected pulse signal for a single channel for signal injected in the front-end and at the end of the wire. Although both input signals have a similar amplitude, they are not identical and thus subsequent comparisons will be normalized.}
    \label{fig:testpulse_pulse}
\end{figure}

The waveforms collected with each method are averaged many readouts. The inherent noise in the system sums to zero on average, whereas the signal adds coherently. The result is a \SI{500}{\micro\second} average pulse waveform per channel as shown in figure \ref{fig:testpulse_pulse}. Note that although both signals have a similar amplitude, they are not identical and all subsequent comparisons are normalized to account for this. All pulses were then fit with a function resulting from a sum of several orders of Bessel functions of the first kind:

\[
f\left(x, \vec{a}\right) = \sum_{\beta=0}^{5} a_{2\beta+1} * J_{\beta}\left(x - a_{2\beta}\right),{\hskip 2em} \alpha:\ \mathrm{Order\ parameter}
\]

This functional form was chosen primarily to empirically characterize the shape of the pulse; the fit parameters themselves do not have a physical interpretation. The shape is impacted both by the intrinsic electronics response of the ICARUS electronics and by the width of the injected pulse. The fit result is used to calculate the peak height, full width at half maximum (FWHM), and the integral. The continuous nature of the fit result mitigates the bias associated with the discrete nature of the signal.

\begin{figure}[t]
    % TODO - Why does the vertical axis have one additional sigfig than the horizontal axis? Are the axes limits explicitly set the same?
    \centering
    \includegraphics[width=0.48\textwidth]{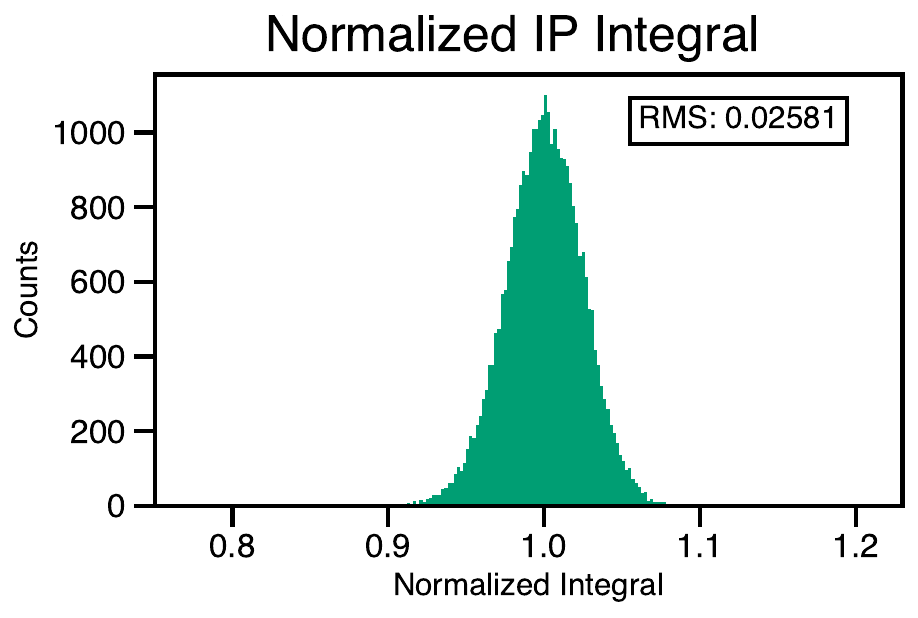}
    \includegraphics[width=0.48\textwidth]{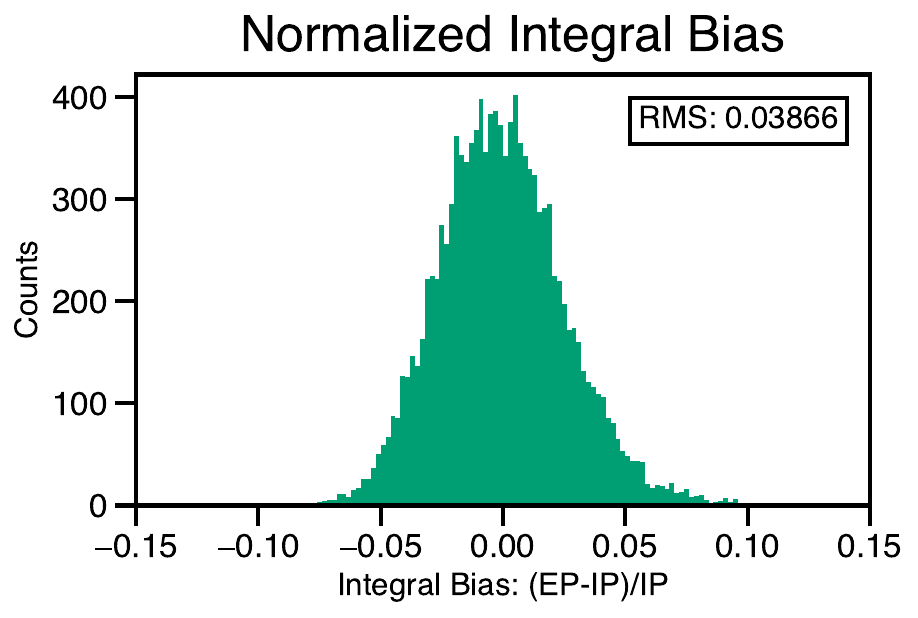}
    \caption{Distribution of the pulse integral as measured by the internal method of pulse inject (left) and the relative bias in pulse integral between the two methods (right). The relative bias is large compared to the uniformity observed in each method, as measured by the width of each distribution.}
    \label{fig:testpulse_integral_bias}
\end{figure}

\paragraph{Results}
\label{subsec:results}
Both methods of pulse injection show variations in the resulting pulse integral that are not well correlated with each other. This is demonstrated in figure \ref{fig:testpulse_integral_bias} which shows the distribution of the measured pulse integral for the internal method of pulse injection and the relative bias between the two methods. Both methods of pulse injection show a higher degree of uniformity across the detector than the relative bias between the two methods, as measured by the widths of each distribution. This motivates the decision to not apply a channel-dependent gain calibration based on the observed pulse integral. Instead, this comparison can be used to bound the amount of channel-to-channel gain variation as less than the width of the relative difference between the methods, 3.9\%, as shown in the right plot of figure \ref{fig:testpulse_integral_bias}.

\begin{figure}[t]
    % TODO - Font size relative to plot dimensions needs to be increased. Maybe reducing the figure dimensions while maintaining the same ratio will work?
    \centering
    \includegraphics[width=\textwidth]{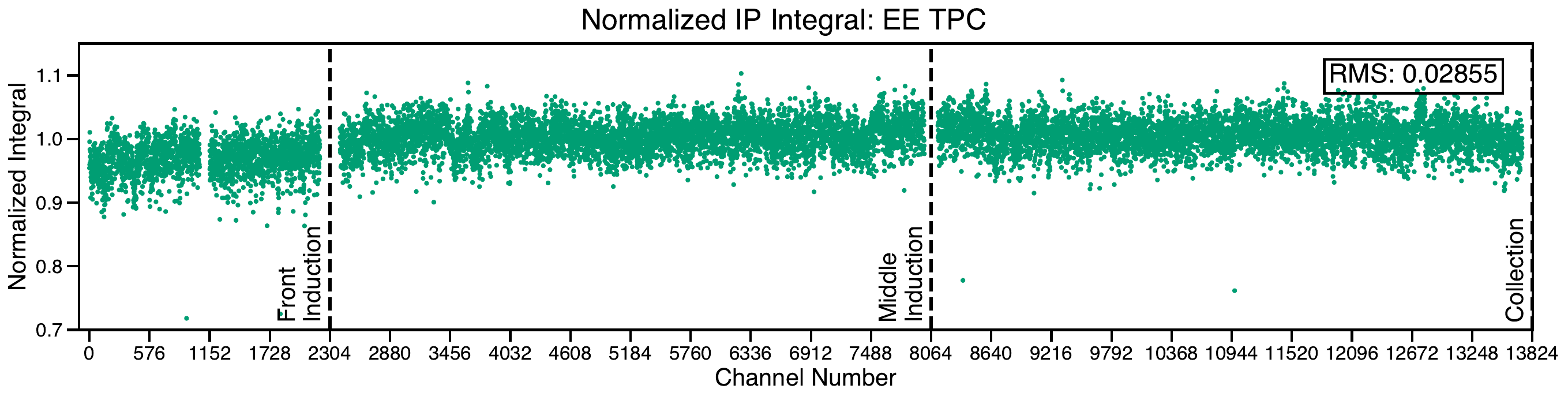}
    \caption{Variation in the normalized pulse integral across the EE TPC. The normalized pulse integral is generally uniform, but there is a systematic decrease of about 5\% associated with wires that are lower down in the detector. This is visible to the left of channel 2304 as a trend in the normalized pulse integral that is correlated with channel number.}
    \label{fig:testpulse_integral_scatter}
\end{figure}

In addition, there is a small systematic decrease of about 5\% in the pulse integral on the front induction plane as shown in Figure \ref{fig:testpulse_integral_scatter}. This correlates with the increasing channel capacitance from the longer cable length needed to reach wires lower down in the detector. Due to the imperfections in the pulsing methods, we elect to use only muon ionization signals to calibrate non-uniformities along each wire plane. This method, described in section \ref{sec:q=YZ}, is able to calibrate out all non-uniformities with 10$\times$\SI{10}{\centi\meter} bins. This spatial resolution is adequate to correct for the coarse non-uniformities in the gain observed by this study.

\begin{figure}
    \centering
    \includegraphics[width=0.48\textwidth]{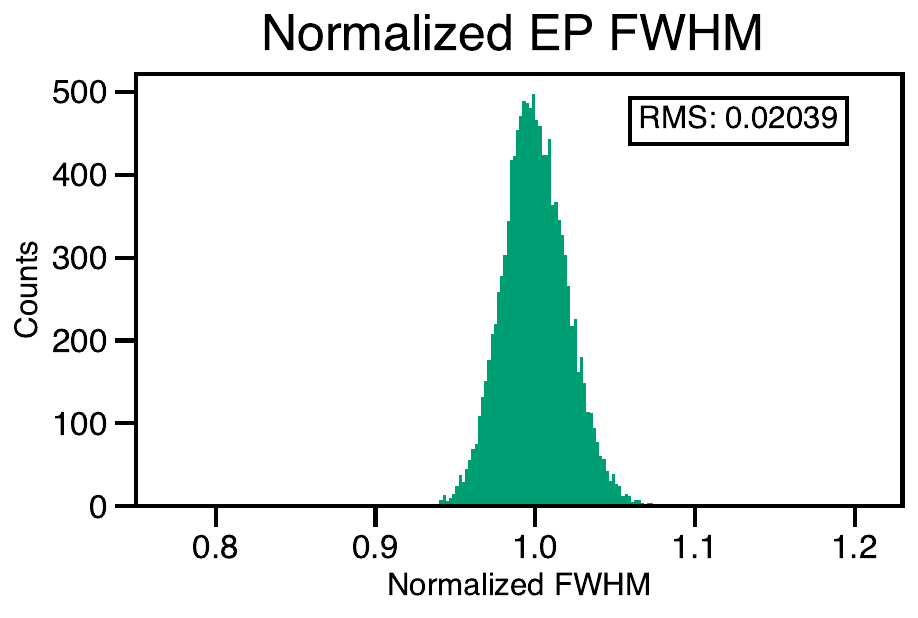}
    \includegraphics[width=0.48\textwidth]{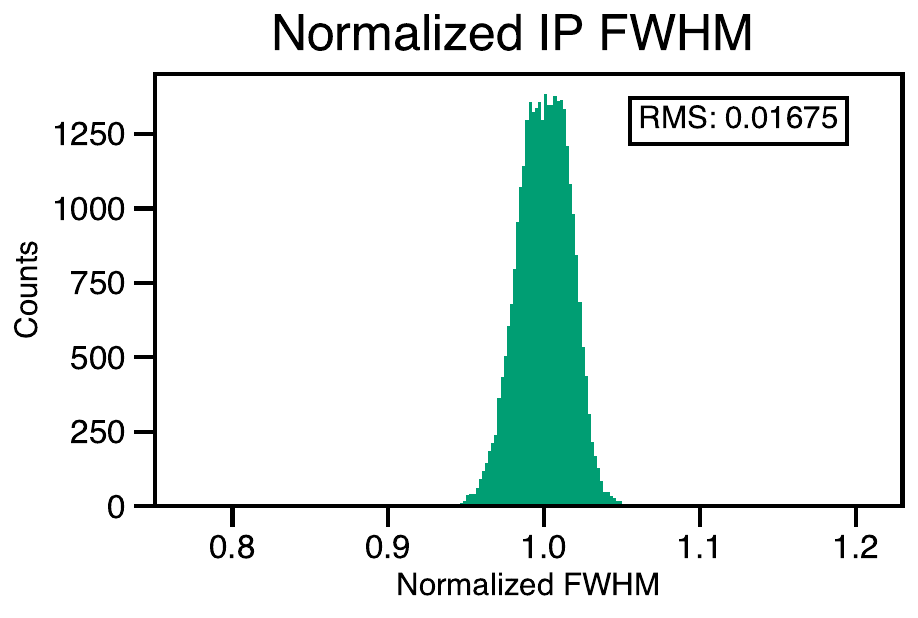}
    \caption{Distribution of the fit pulse width as observed by the external (left) and internal (right) methods of test pulse injection. A distinct pattern across each group of 32 channels is the dominant driver of a consistent variation of about 2\% or less in both measurements.}
    \label{fig:testpulse_width_variation}
\end{figure}

A variation in the pulse width of about 2\% or less is observed across all channels in the detector using both pulses injected on the ends of the wires and internally in the front-end. This is shown for both methods in Figure \ref{fig:testpulse_width_variation}. This variation appears to be driven by a trend in the pulse width as a function of the channel's position amongst the 64 channels on the readout board. The size of this variation is negligible relative to other uniformity calibrations, so no correction is applied.

\subsection{Drift Direction Equalization}
\label{sec:q=drift}

The drift direction equalization step corrects charge reconstruction for effects that vary with the ionization drift time. The primary such effect is attenuation from argon impurities. Since the electron lifetime varies across the ICARUS dataset, this equalization is done per-DAQ-run. One DAQ run in ICARUS lasts from a few hours to a few days, and the electron lifetime does not vary significantly over such a period. 

The calibration is done with depositions from anode-cathode crossing tracks. The cathode crossing identification is done by matching a pair of aligned tracks in the two TPCs on either side of the central cathode plane. A cut on the drift direction length of the track in either TPC ensures that it also crosses the anode in that TPC. As described in section \ref{subsec:diffusion}, to mitigate the impact of diffusion, the charge is summed in groups of 10 wires into a ``coarse-grained" $dQ/dx$. The coarse-grained depositions are grouped by drift time and are fit with a Landau distribution convolved with a Gaussian distribution to extract the most-probable-value (MPV) of the distribution. The MPV as a function of drift time is fit to an exponential to obtain an effective electron lifetime that parameterizes the non-uniformity. We have found that an exponential is able to model the charge non-uniformity in all runs across the ICARUS dataset. This electron lifetime should be understood to be effective because, while argon impurities are the dominant effect, the measured lifetime also includes impacts from field distortions and imperfections in signal processing. 

\begin{figure}[]
    \centering
    \includegraphics[width=1.0\textwidth]{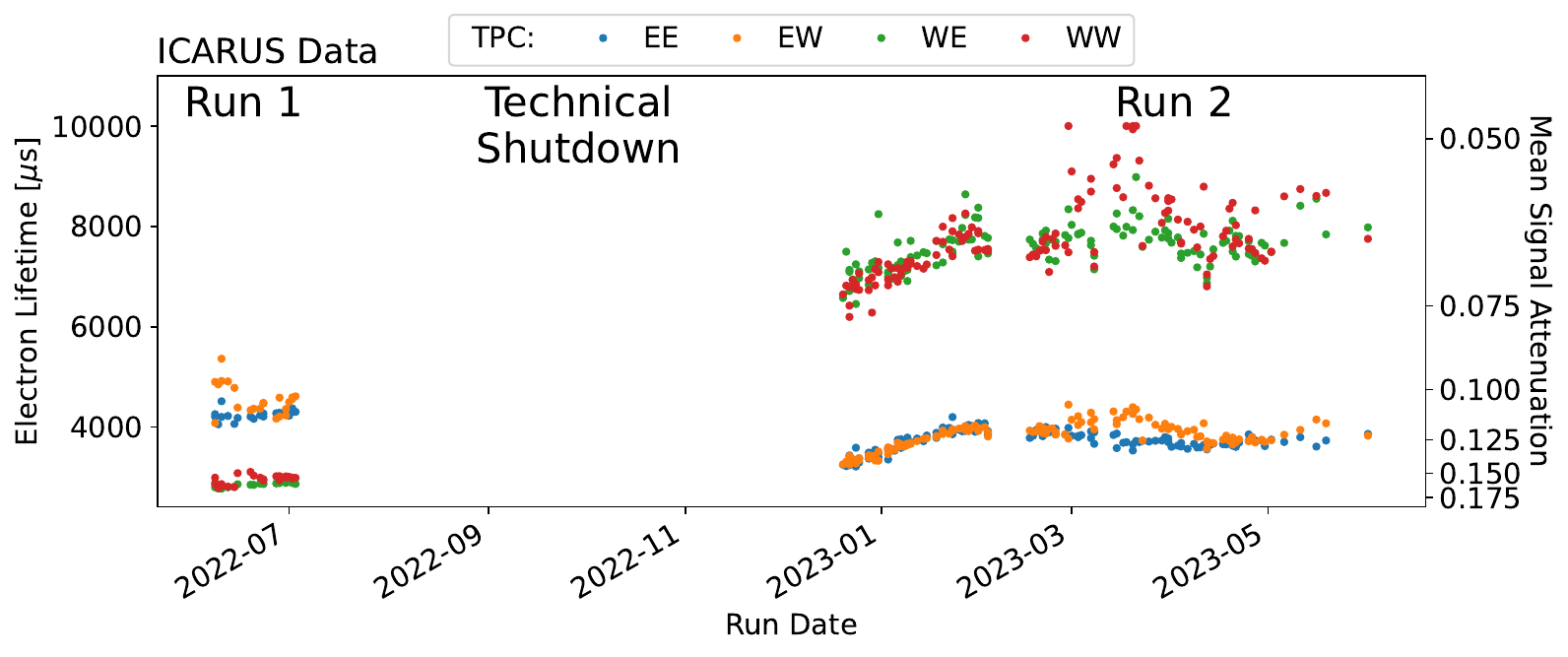}
    \caption{Measured effective electron lifetime values measured in each ICARUS TPC across the Run 1 and Run 2 datasets. The right axis shows the corresponding average signal attenuation across the $\sim$\SI{1}{\milli\second} ICARUS drift period.}
    \label{fig:t=taus}
\end{figure}

Figure \ref{fig:t=taus} shows the electron lifetime across the ICARUS dataset, as well as the corresponding mean signal attenuation in the drift direction. The electron lifetime was maintained at $\sim$\SI{3}{\milli\second} in the West cryostat and at $\sim$\SI{5}{\milli\second} in the East cryostat across the Run A and Run 1 datasets. During Run 2, the lifetime in the West cryostat reached 8-\SI{10}{\milli\second}. There are slight differences between the East and West TPCs in both Cryostats. Differences in the purity between the TPCs in each Cryostat would have to be small since the same argon circulates in both TPCs in a cryostat. This effect may also be an indication of different field distortions in the TPCs which perturb the effective electron lifetime measured here.

\subsection{Wire Plane Equalization}
\label{sec:q=YZ}

The wire plane equalization step corrects charge reconstruction for detector effects that vary across the two directions in the plane of the readout wires: $\hat{y}$, the vertical direction, and $\hat{z}$, the (BNB) beamline direction (see figure \ref{fig:TPCDiagram}). The calibration is done with coarse-grained depositions from through-going cathode-crossing tracks. Depositions are binned in terms of their $\hat{y}-\hat{z}$ location on the wire plane in $10\times$\SI{10}{\centi\meter\squared} bins. This is as small a spatial resolution as is possible given the statistics of cosmic muons ($\sim$ 3 million) available for the calibration. As in the drift direction equalization, in each spatial bin the distribution of $dQ/dx$ values is fit with a Landau distribution convolved with a Gaussian distribution to extract the MPV. The MPV in each spatial bin is converted into a scale factor computed to keep the mean MPV across the TPC fixed. The scale factors are therefore relative on each plane and are not computed with reference to an absolute gain (unlike the drift direction equalization). The scale factors are computed separately for both runs on each wire plane in each TPC.

%%%
% RUN 1 YZ MAPS
%%%
\begin{figure}[t]
    \centering
    \vspace{2.5cm}
    \includegraphics[width=\textwidth]{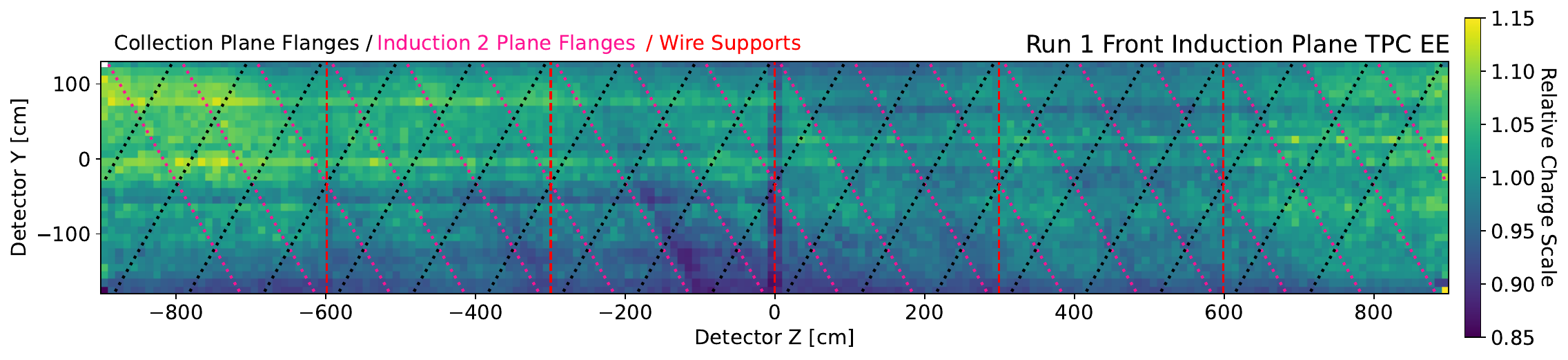}
    \includegraphics[width=\textwidth]{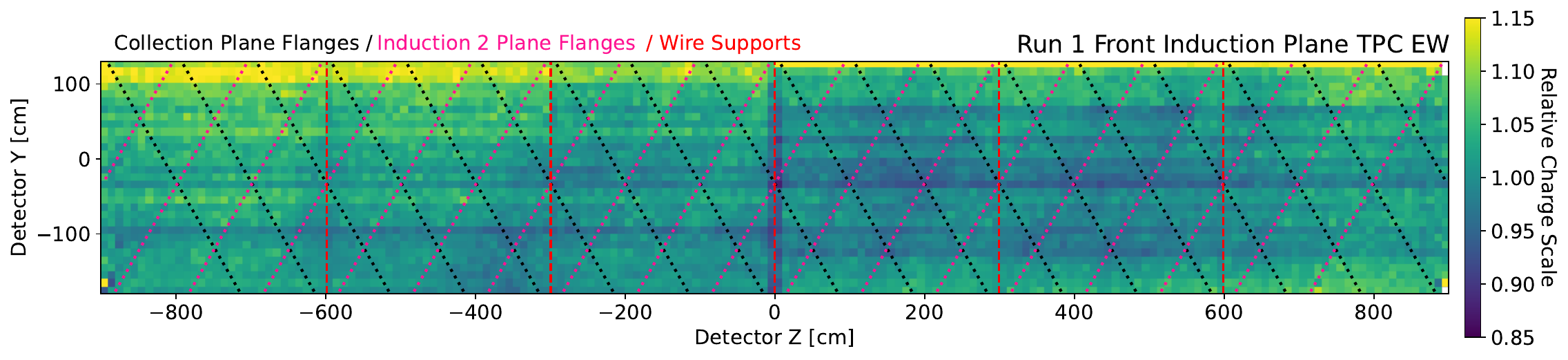}
    \includegraphics[width=\textwidth]{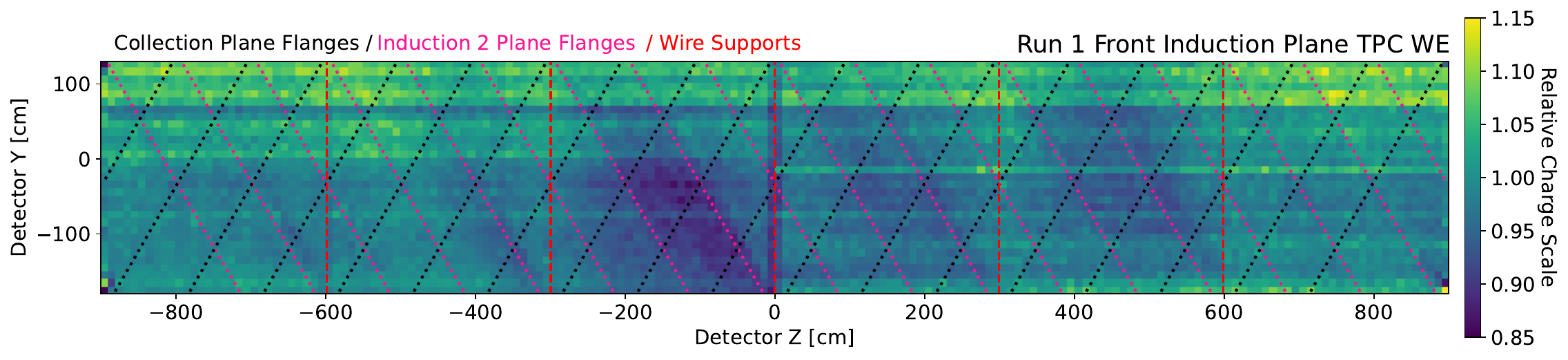}
    \includegraphics[width=\textwidth]{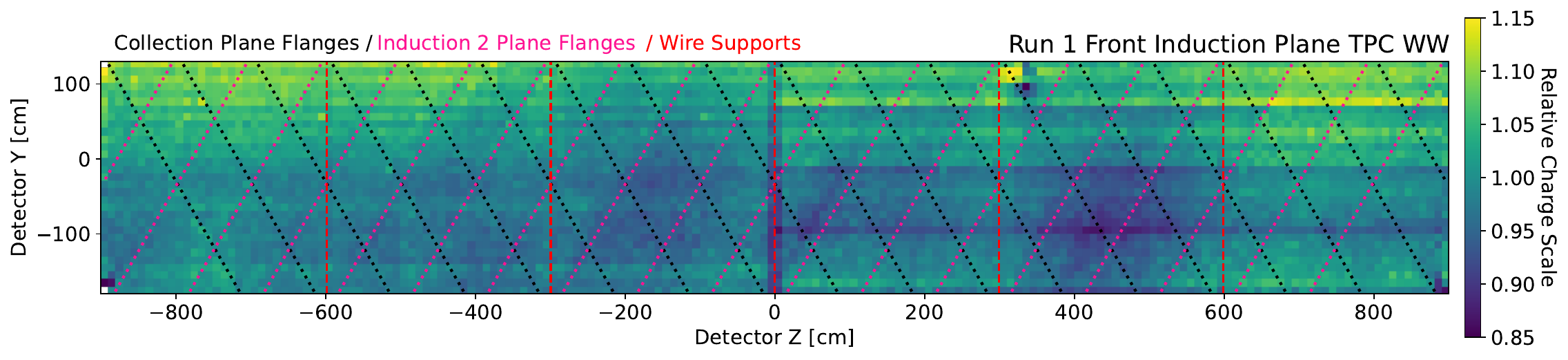}
    \caption{Coarse-grained MPV $dQ/dx$ values in $10\times$\SI{10}{\centi\meter} bins in each of the ICARUS TPCs on the front induction plane across the Run 1 dataset. The borders between different flanges that supply the wire bias to groups of wires on collection and induction are overlaid, as well as the location of mechanical support structures on the front induction wire plane.}
    \label{fig:w=run1Ind0}
    \vspace{2.5cm}
\end{figure}

\begin{figure}[t]
    \centering
    \vspace{2.5cm}
    \includegraphics[width=\textwidth]{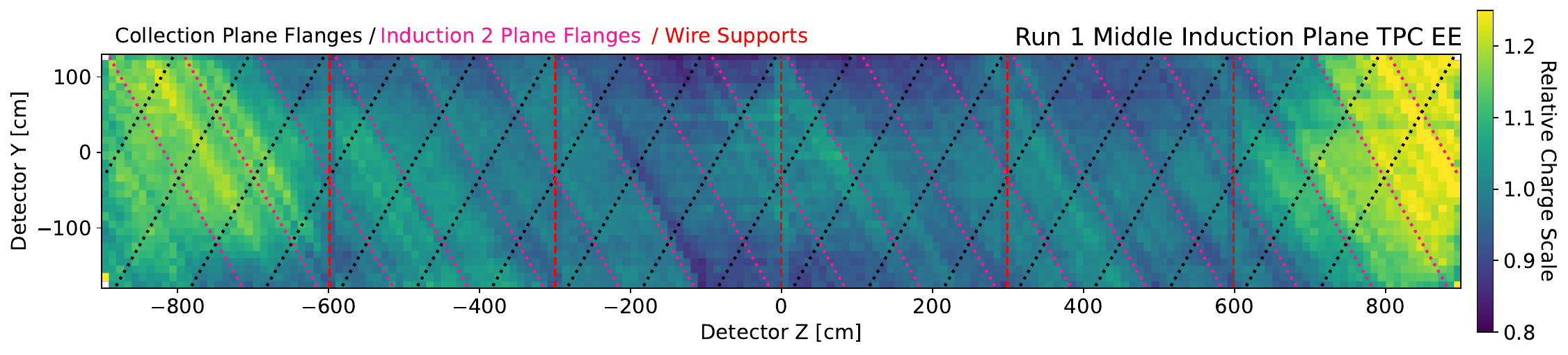}
    \includegraphics[width=\textwidth]{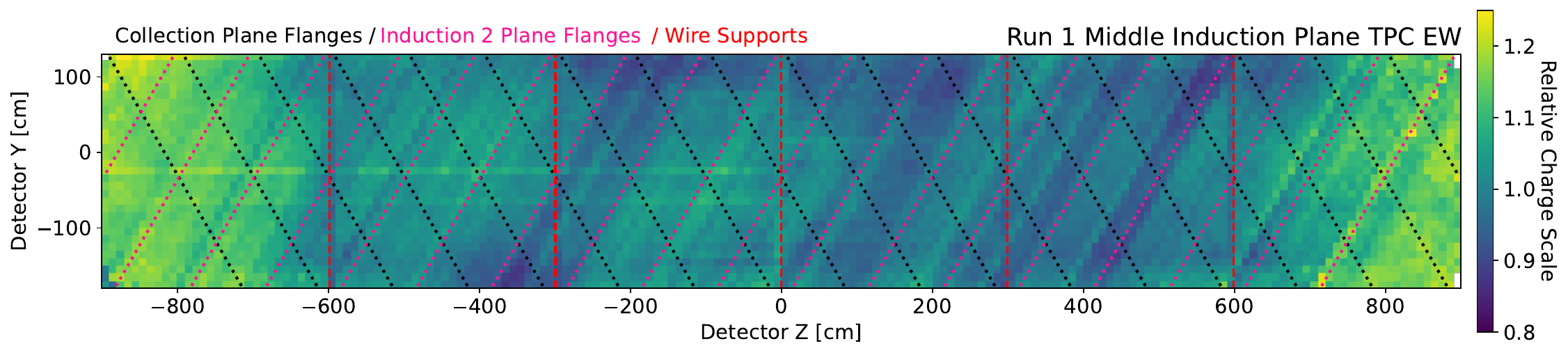}
    \includegraphics[width=\textwidth]{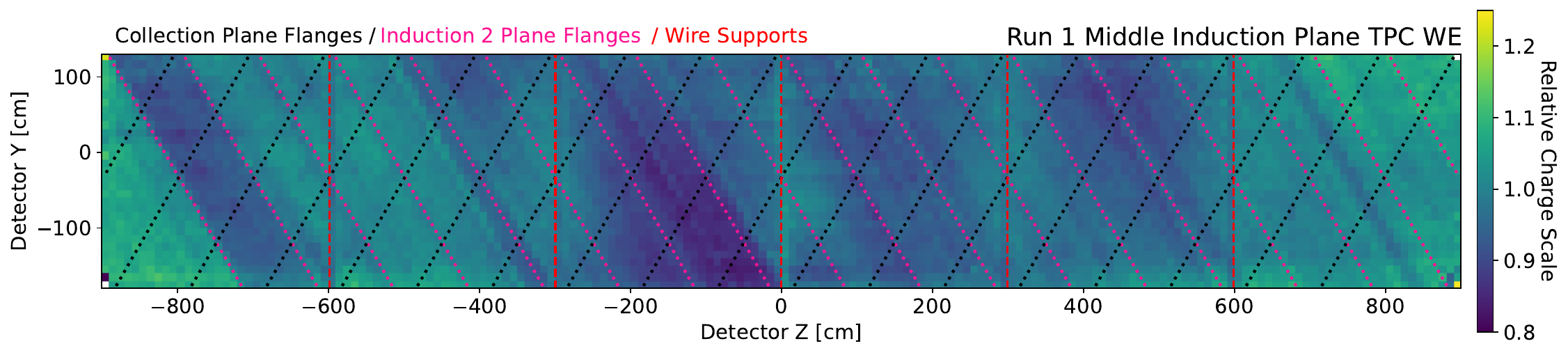}
    \includegraphics[width=\textwidth]{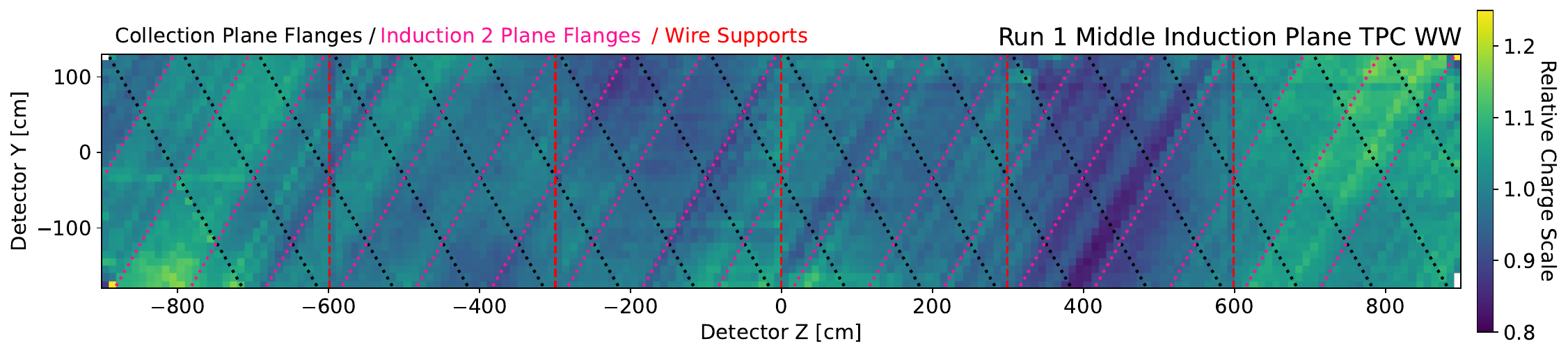}
    \caption{Coarse-grained MPV $dQ/dx$ values in $10\times$\SI{10}{\centi\meter} bins in each of the ICARUS TPCs on the middle induction plane across the Run 1 dataset. The borders between different flanges that supply the wire bias to groups of wires on collection and induction are overlaid, as well as the location of mechanical support structures on the front induction wire plane.}
    \label{fig:w=run1Ind1}
    \vspace{2.5cm}
\end{figure}

\begin{figure}[t]
    \centering
    \vspace{2.5cm}
    \includegraphics[width=\textwidth]{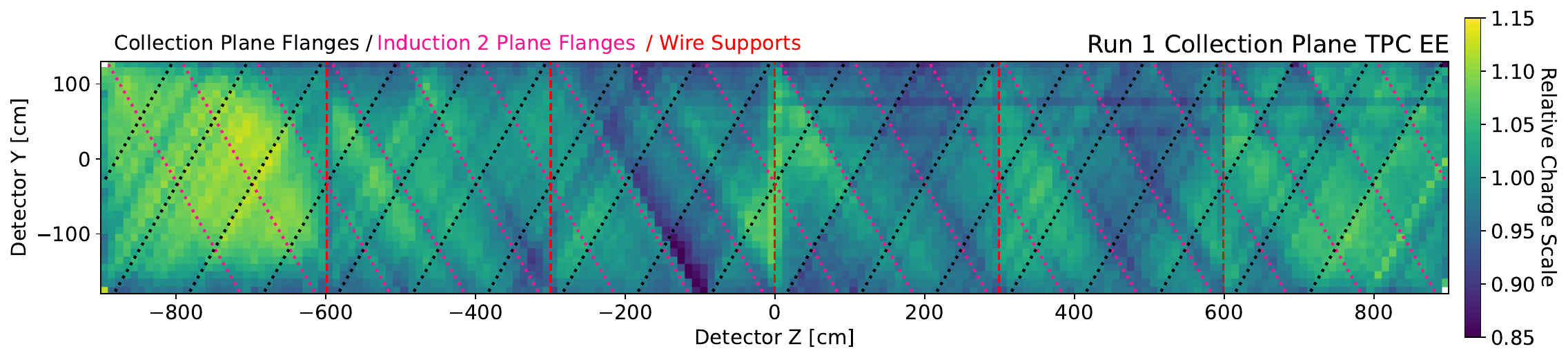}
    \includegraphics[width=\textwidth]{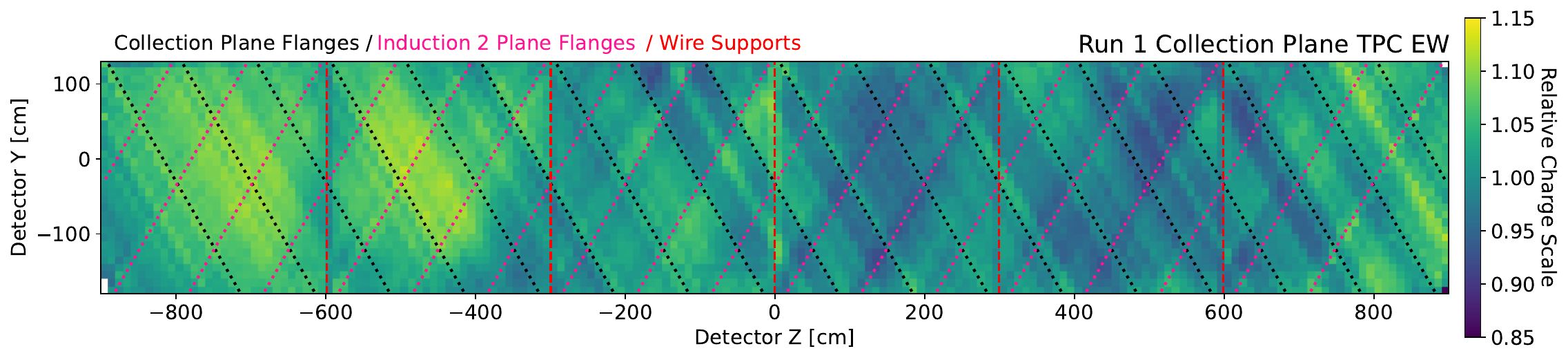}
    \includegraphics[width=\textwidth]{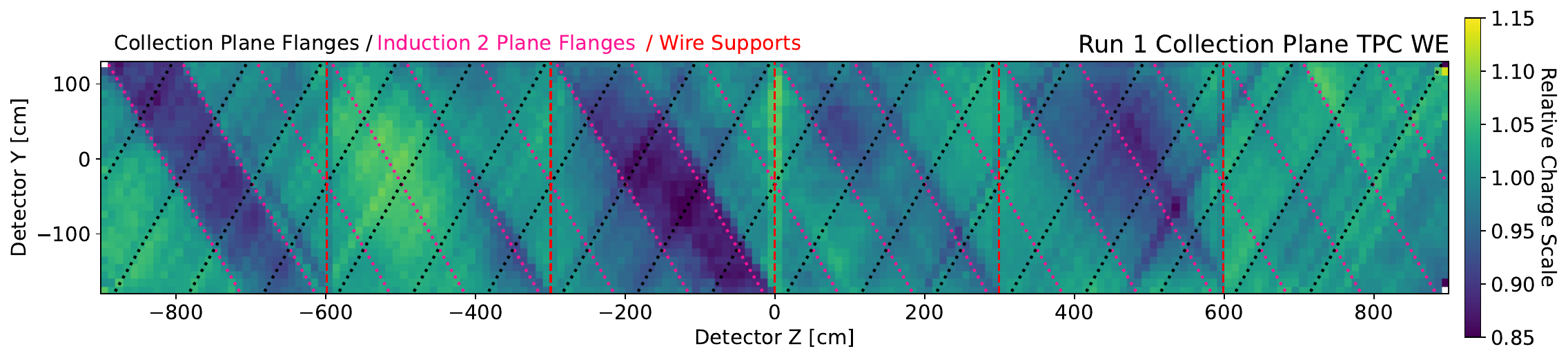}
    \includegraphics[width=\textwidth]{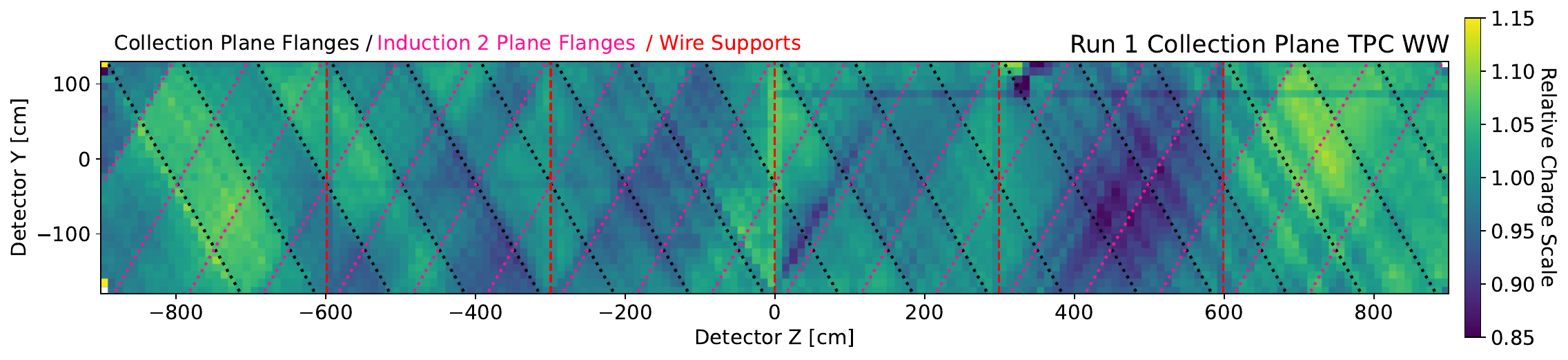}
    \caption{Coarse-grained MPV $dQ/dx$ values in $10\times$\SI{10}{\centi\meter} bins in each of the ICARUS TPCs on the collection plane across the Run 1 dataset. The borders between different flanges that supply the wire bias to groups of wires on collection and induction are overlaid, as well as the location of mechanical support structures on the front induction wire plane.}
    \label{fig:w=run1Col}
    \vspace{2.5cm}
\end{figure}

% RUN 1 VS RUN 2
\begin{figure}[t]
    \centering
    \vspace{2.75cm}
    \includegraphics[width=\textwidth]{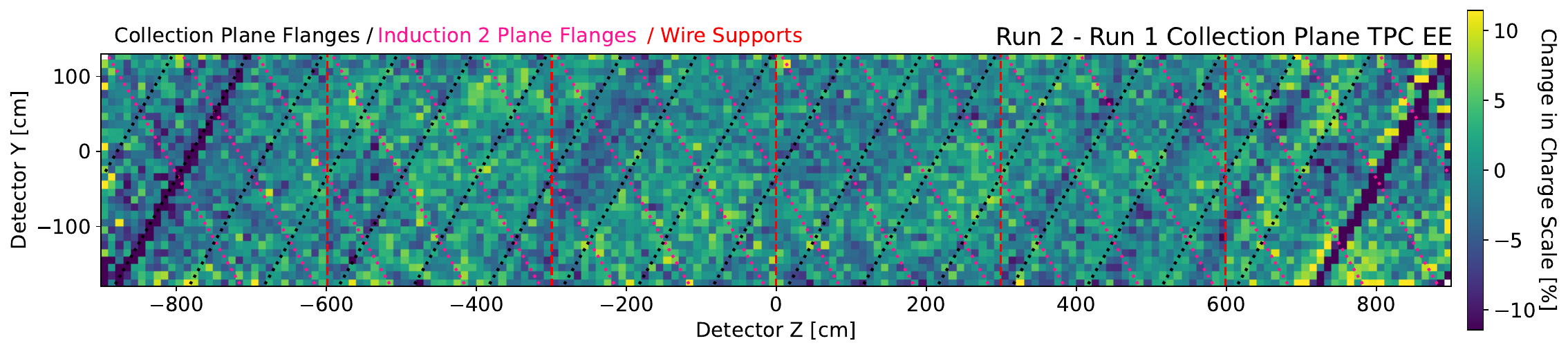}
    \includegraphics[width=\textwidth]{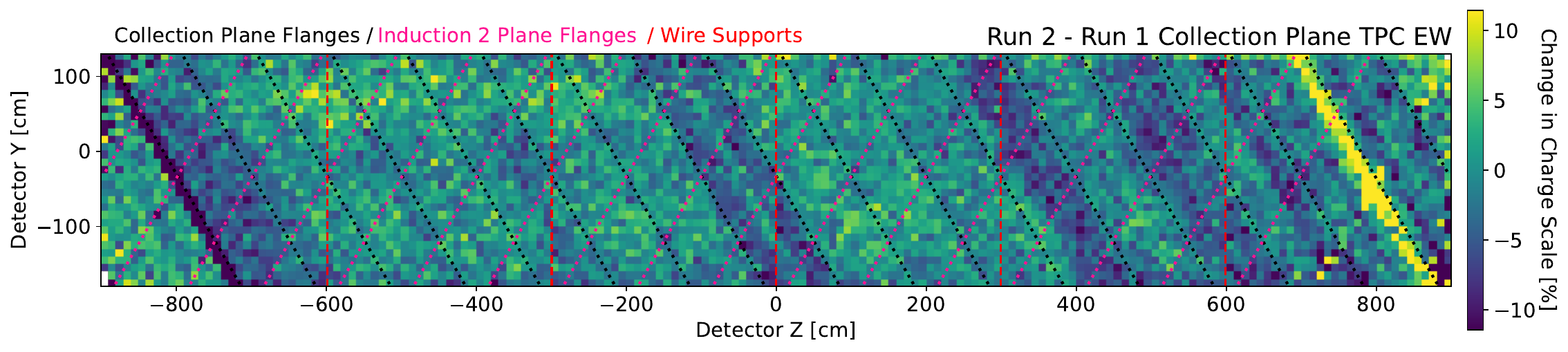}
    \includegraphics[width=\textwidth]{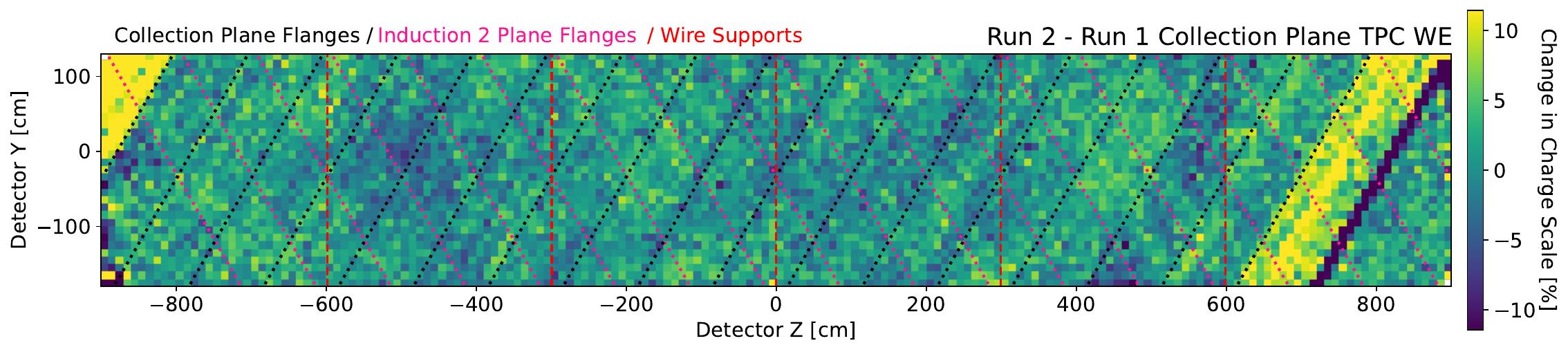}
    \includegraphics[width=\textwidth]{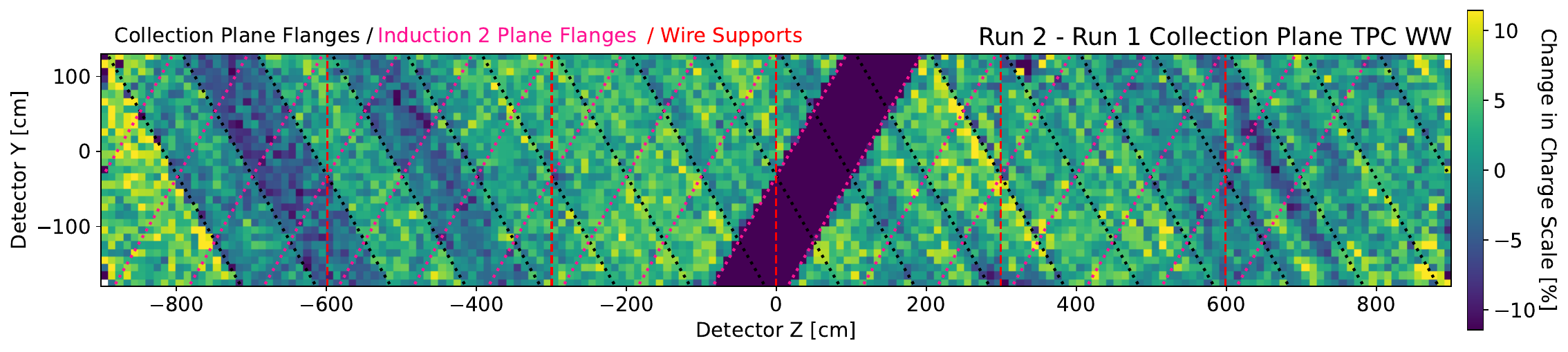}
    \caption{Differences between the Run 1 and Run 2 datasets of coarse-grained MPV $dQ/dx$ values in $10\times$\SI{10}{\centi\meter} bins in each of the ICARUS TPCs on the collection plane. The borders between different flanges that supply the wire bias to groups of wires on collection and induction are overlaid, as well as the location of mechanical support structures on the front induction wire plane.}
    \label{fig:w=run1run2diffCol}
    \vspace{2.75cm}

\end{figure}

The spatial uniformity maps are shown in the Run 1 dataset for the front induction, middle induction, and collection planes in figures \ref{fig:w=run1Ind0}, \ref{fig:w=run1Ind1}, and \ref{fig:w=run1Col}, respectively. Some, but not all, of the features in the map have been traced to known faults in the detector. For example, the band of small $dQ/dx$ around $z=0$ in each TPC is due to perturbations to the applied wire bias field by the presence of a mechanical bar supporting the front induction wires. The uniformity maps are positively correlated between the three planes: where the collection plane measures less charge, so do the induction planes. This correlation is caused by the induction plane intransparency. Where the induction plane is opaque, less charge reaches the collection plane to be observed by that plane. In addition, the intransparency reduces the measured charge on the induction planes, for two reasons. First, the unipolar collection pulse interferes partially destructively with the bipolar induction pulse. Second, the collection pulse causes the induction plane signal shape to depart from the deconvolution kernel (which is computed with no intransparency), reducing the efficacy of the deconvolution in forming a Gaussian pulse shape and therefore reducing the measured charge.

There are a couple of discrete changes in the uniformity between Runs 1 and 2. The difference in the uniformity between Run 1 and Run 2 is shown for the collection plane in figure \ref{fig:w=run1run2diffCol}. These changes have been traced to a couple changes to the detector operation during the 2022 technical shutdown: two additional failures of middle induction plane wire bias voltage supplies, and a few readout board replacements on the collection plane (which have a slightly different gain). We have not observed any time dependence of the spatial uniformity within either Run 1 or Run 2.

\subsection{TPC Equalization}
\label{sec:q=TPC}

As a final step, the gains in the four separate TPCs in ICARUS are equalized. This equalization is done separately for both ICARUS runs. This corrects for any broad differences in the gain between the different TPCs or runs. The charge scale for this equalization is computed using stopping cosmic muons, as opposed to throughgoing muons. This choice is made because stopping cosmic muons are used to measure the absolute gain in ICARUS in the ionization energy scale calibration. Equalizing the TPC gain with the same sample ensures that different TPCs are completely consistent in the gain fit. 

\begin{figure}[]
    \centering
    
    \includegraphics[width=0.45\textwidth]{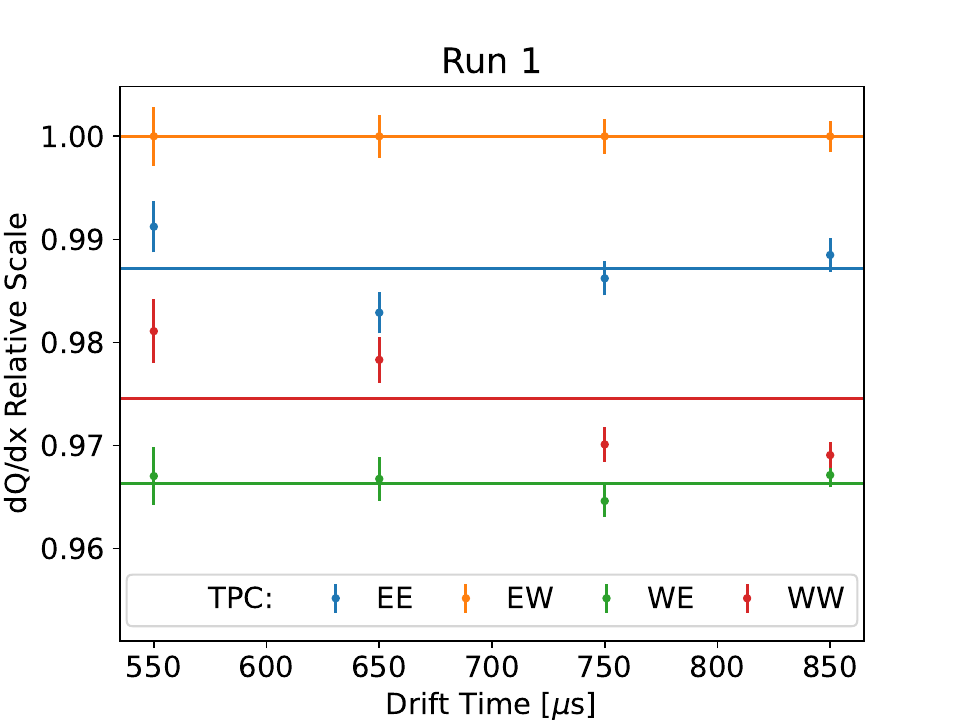}
    \includegraphics[width=0.45\textwidth]{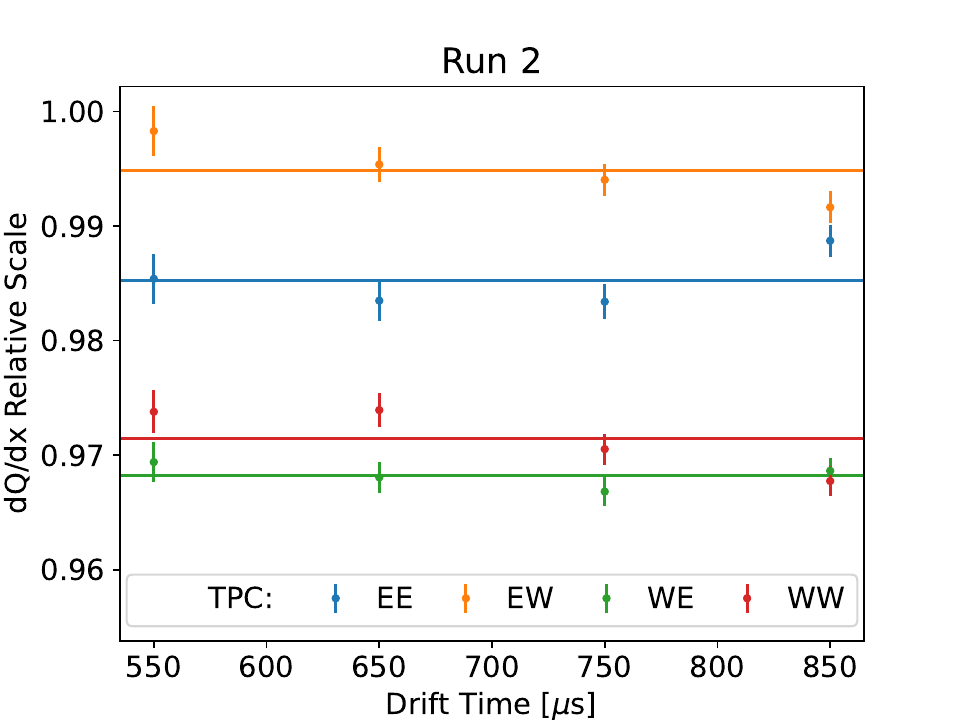}
    
    \caption{Computation of TPC equalization scale factors in each TPC in both runs. The scale factors are computed by equalizing the MPV of stopping cosmic muon depositions across four drift time bins, averaged across a residual range from 200-\SI{300}{\centi\meter}. These are shown by the data points per-TPC. The scale factors are given by the horizontal lines. They are normalized so that TPC EW in Run 1 has a scale of 1. The uncertainty on each scale factor is computed from the uncertainty on the MPV in the Landau-Gaussian fit.}
    \label{fig:tpc=scalefactors}
\end{figure}

The charge scale is computed from distributions of coarse-grained $dQ/dx$ with the drift and wire plane direction equalizations applied. The distributions are split up in terms of the stopping muon track residual range and drift time to select for a single peak $dE/dx$ in each distribution. The residual range is binned in steps of \SI{5}{\centi\meter} from 200-\SI{300}{\centi\meter}. The drift time is binned in steps of \SI{100}{\micro\second} from 500-\SI{900}{\micro\second}. Each histogram of equalized $dQ/dx$ is fit to a Landau distribution convolved with a Gaussian distribution to extract the MPV. The MPVs are averaged over residual range to obtain a single average for each drift time. Then, a scale factor is fit to the mean MPVs per each TPC per each run. This scale factor is normalized so that TPC EW in Run 1 has a scale of 1. The average MPVs and the scale factors are shown in figure \ref{fig:tpc=scalefactors}.

\subsection{Equalization Results}
\label{sec:q=result}

\begin{figure}[]
    \centering
    \includegraphics[width=0.32\textwidth]{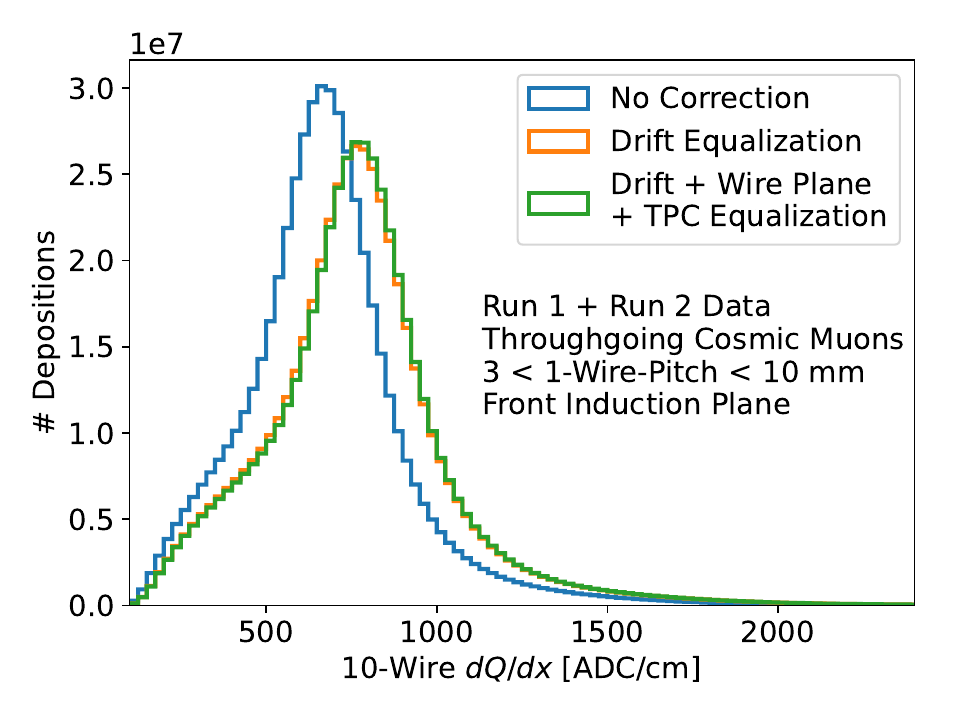}
    \includegraphics[width=0.32\textwidth]{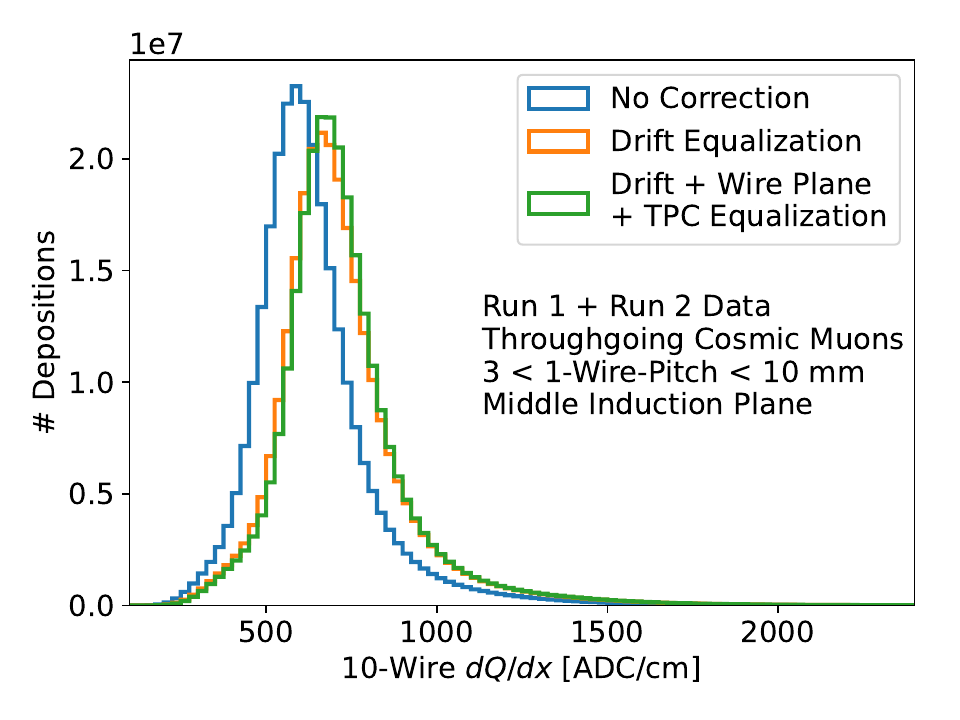}
    \includegraphics[width=0.32\textwidth]{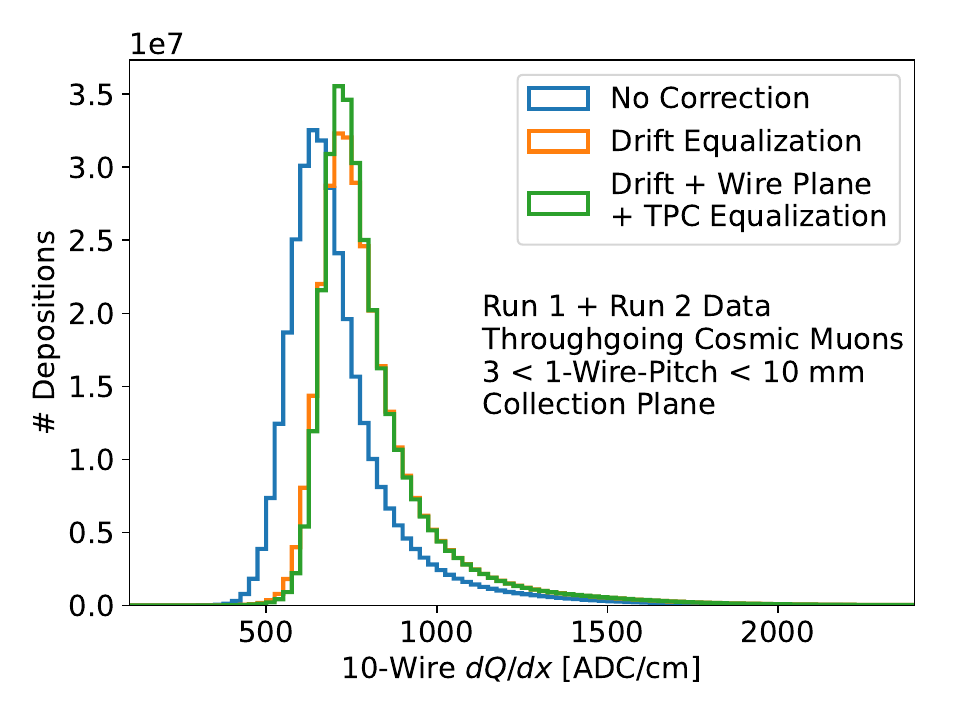}
    \caption{Distribution of coarse-grained $dQ/dx$ values for throughgoing cosmic muons in the Run 1 and Run 2 datasets uncorrected, and with the drift and wire plane equalization corrections applied. Shown for the front induction plane (left), the middle induction plane (middle), and the collection plane (right).}
    \label{fig:q=dist}
\end{figure}

Figure \ref{fig:q=dist} plots the distribution of coarse-grained $dQ/dx$ values from throughgoing cathode-crossing cosmic muons before and after the equalization procedure. After both corrections are applied, the Gaussian width divided by the MPV of the distribution decreases by 13\% on the front induction plane, 11\% on the middle induction plane, and 43\% on the collection plane. The narrowing is most significant on the collection plane because the inherent broadening from readout noise 
is the smallest on that plane.

\section{Noise Measurement and Simulation}
\label{sec:Noise}
The characterization of noise is important for understanding its impact on ionization signals and the reconstruction of particle interactions in the detector volume. The ICARUS TPC noise is characterized principally through measurements of the absolute noise scale, frequency characteristics, and channel-to-channel correlations, as is detailed below (section \ref{sec:noise_measure}). These measurements of the noise are used as input to a data-driven model (section \ref{sec:noise_model}). The performance of this noise model is presented in section \ref{sec:noise_datamc_comparisons}, highlighting areas for future improvement.

\subsection{Noise Measurement}
\label{sec:noise_measure}
The geometry of the ICARUS TPC readout is important for understanding the noise observed in the detector. Wires are connected in groups of 32 to cables, which serve to transmit the signal on the wires up to the feed-through flanges and to the readout crates. A readout crate holds nine readout boards, each having the capability to digitize signals from two cables for a total of 576 channels per readout crate. The TPC electronics are described in more detail in \cite{ICARUSElectronics} and \cite{ICARUSOverhaul}.

Data taken with the cathode voltage turned off was chosen for the measurements of noise due to the lack of signal in the waveforms from drifting ionization electrons. The channel-to-channel correlation, $\rho$, can be defined as
\[
\rho_{ij} = \frac{\vec{w_i} \cdot \vec{w_j}}{\sigma_i \sigma_j}
\]
where $\vec{w_i}$ is the waveform for channel $i$ and $\sigma_i$ is the root-mean-square (RMS) of its waveform. This can be calculated pairwise for each channel within a readout crate and averaged across many events. Channel-to-channel correlations between channels not in the same readout crate are not significant and are subsequently not shown here.

The geometry of the readout motivates two distinct classifications for readout crates: crates which connect only to front induction wires and crates which connect to a mix of middle induction and collection wires. The additional wire and cable length for channels in the front induction plane results in higher overall noise. The cables connecting the wires to the front-end are also in significantly closer proximity due to the path down to the wires and the presence of three sets of cables in a single feed-through flange.

\begin{figure}[t]
    % TODO - Reduce redundancy and combine into single plot with subplots.
    % TODO - Consider the effects of non-wired channels on correlation substructure.
    \centering
    \includegraphics[width=0.43\textwidth]{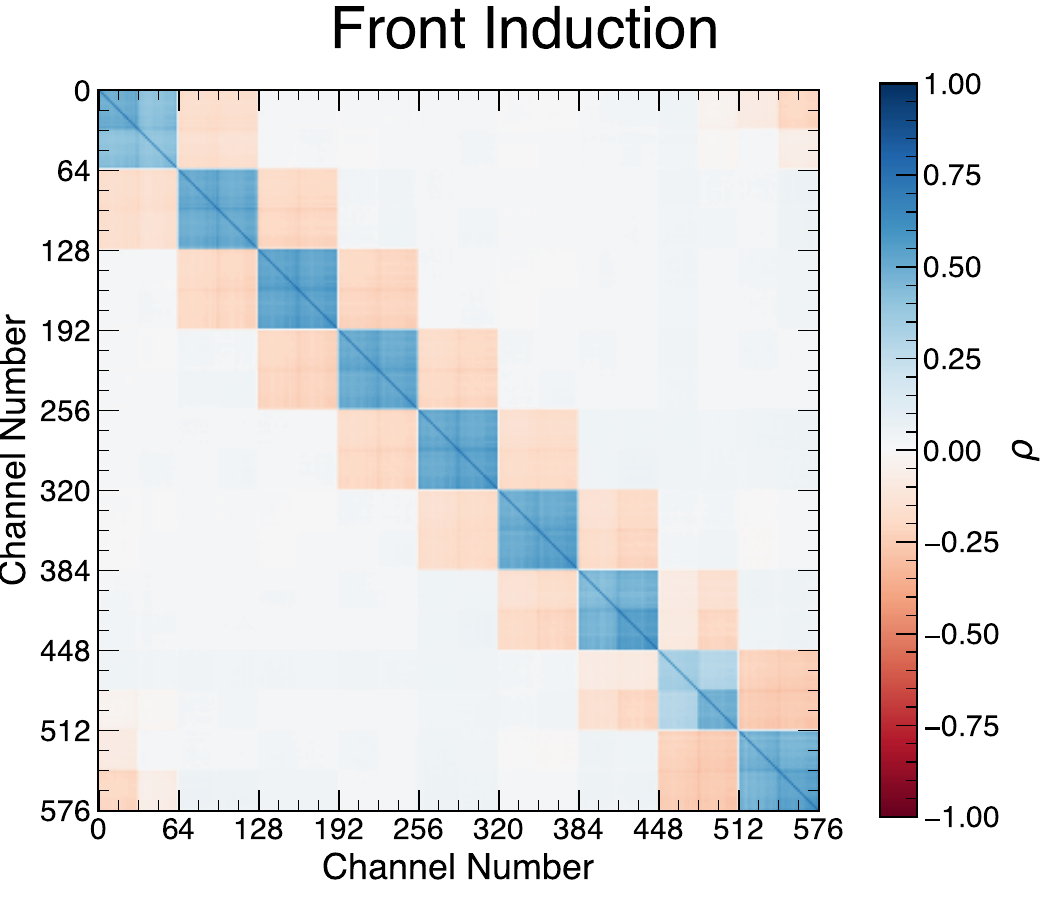}        \includegraphics[width=0.43\textwidth]{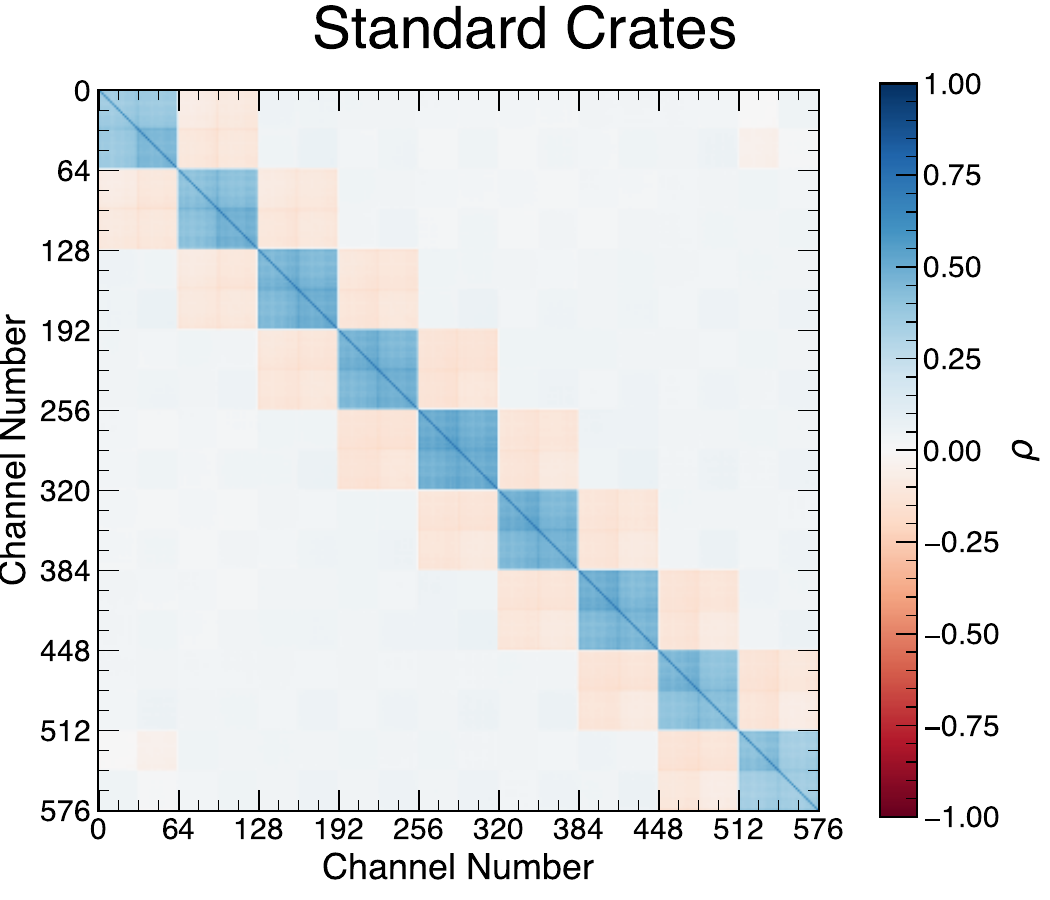}
    \caption{Correlation matrix of noise across each readout crate. Correlation coefficients are calculated pairwise for channels in the readout crate, then averaged across all selected readout crates. The left plot shows only crates which connect to front induction wires, whereas the right plot shows the crates which connect to a mix of middle induction and collection wires.}
    \label{fig:correlations_data}
\end{figure}

Figure \ref{fig:correlations_data} shows the channel-to-channel correlations for channels within the same readout crate. The left plot shows only channels belonging to readout crates serving front induction wires, whereas the right plot shows only channels belonging to readout crates which serve a mix of middle induction and collection wires. The main block diagonal structure of highly-correlated channels reflects the presence of noise that is coherent across channels of the same readout board. The off-diagonal structure of anti-correlation is believed to be due to capacitive or inductive coupling between the cables of adjacent boards. This effect is observed to be stronger in front induction, which is consistent with the closer proximity and higher path overlap of cables for these wires. Though each cable represents 32 wires, the broader correlated component for channels on the same readout board introduces anti-correlations are nearly uniform across the full group of 64 channels. 

The presence of these significant correlations necessitates some degree of noise filtering. ICARUS employs a coherent noise removal algorithm similar to the one used by MicroBooNE \cite{ubooneNoiseCharacterization}. The coherent noise removal algorithm operates on a group of channels and first defines the coherent noise component for that group using the median value of the waveforms for each time tick. This produces a waveform which is expected to represent noise fluctuations that are common to the entire group, which are visually identifiable in images of the raw waveforms. The resulting waveform is then subtracted from each waveform in the group to produce a set of corrected waveforms.

\begin{figure}
    % TODO - Add colorbar?
    % TODO - Combine into a single plot?
    \centering
    \includegraphics[width=0.85\textwidth]{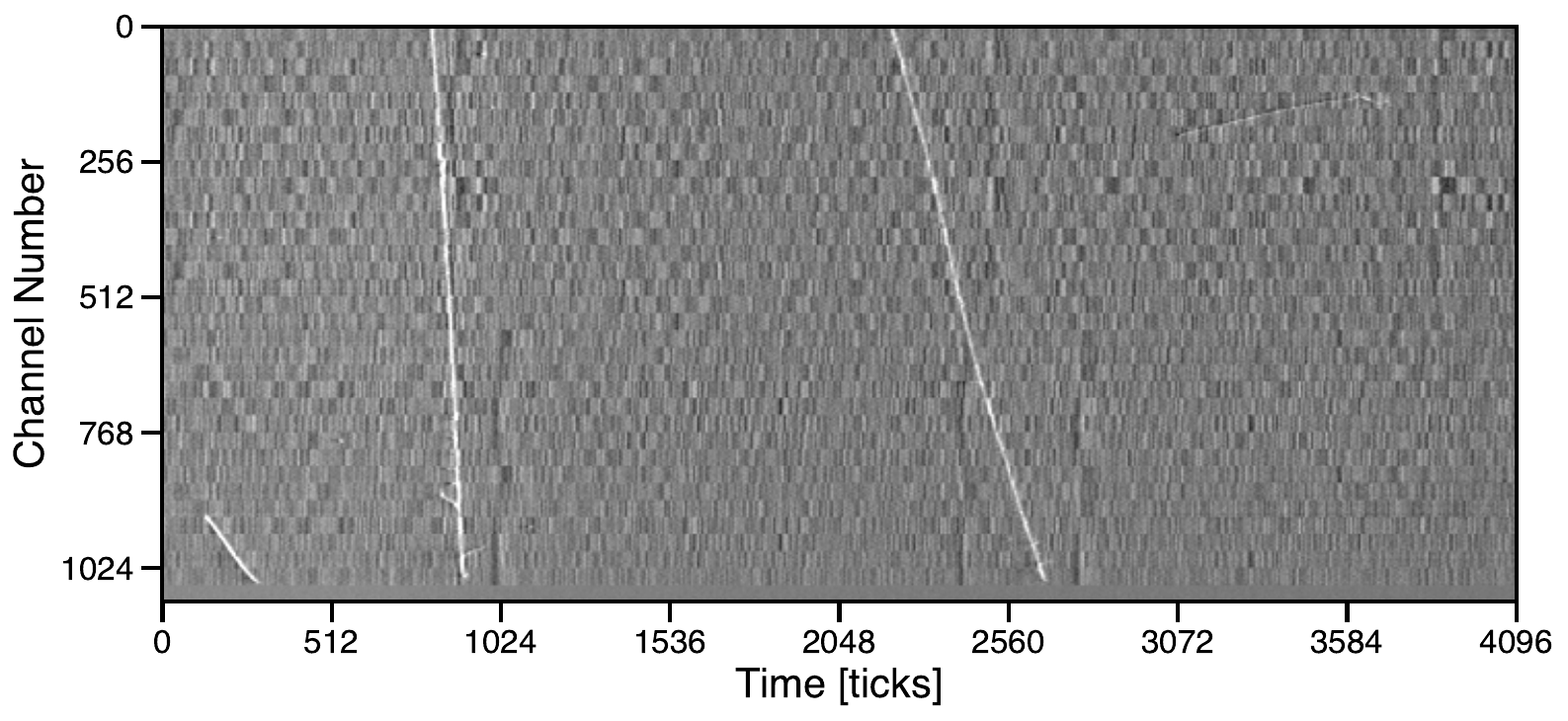}
    \includegraphics[width=0.85\textwidth]{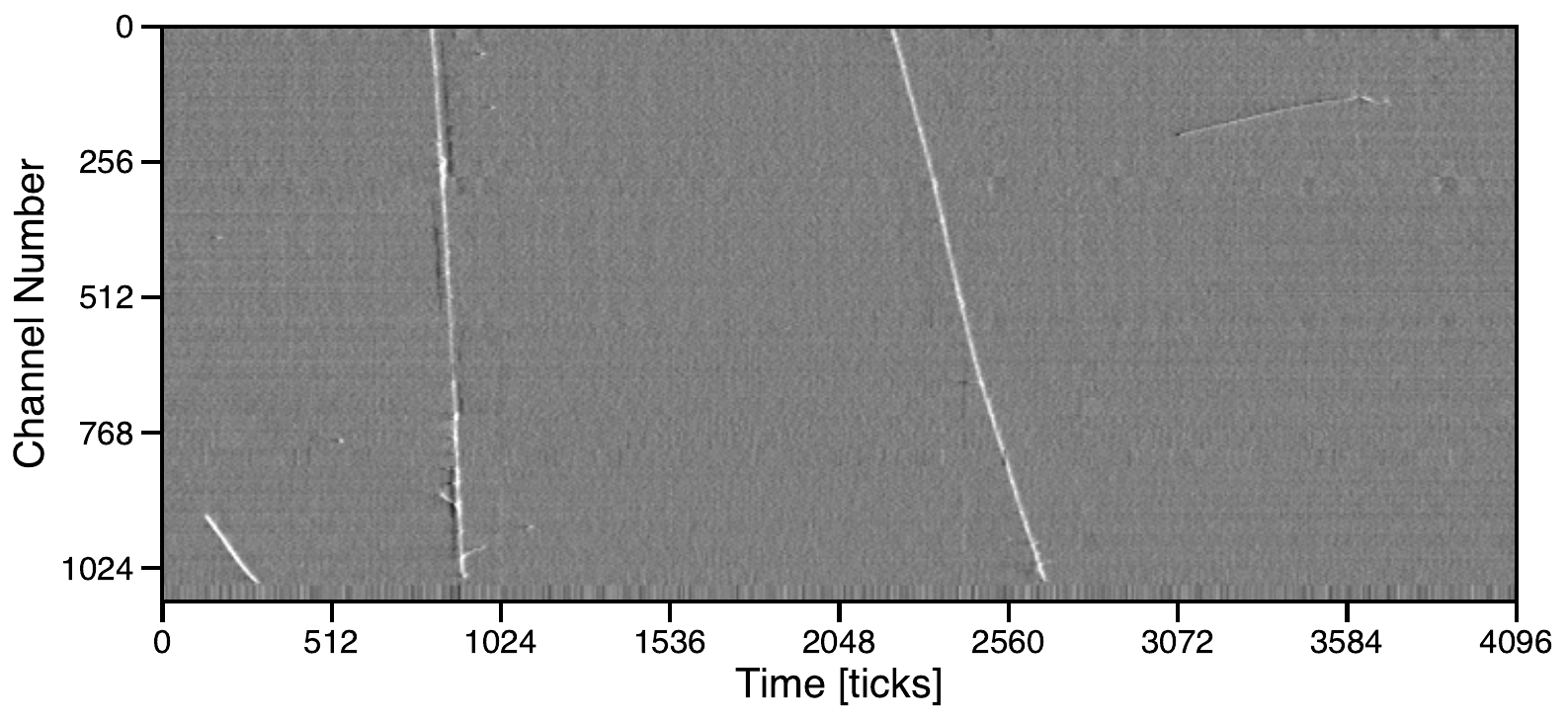}
    \caption{An example set of waveforms containing signal from cosmic muons before and after noise filtering. The image spans the South half of the front induction plane from one of the TPCs.}
    \label{fig:waveforms_ind1}
\end{figure}

In ICARUS, channels belonging to the same readout board are used to define the groups for this noise filtering as motivated by the noise correlation matrices. A downside of this algorithm is that signal from tracks which are isochronous across the group of channels may be impacted as the track itself is coherent in the same manner as the targeted noise. This is partially mitigated by the fact that no group of 64 channels contains signal from a single spatially connected region - each sub-group of 32 channels always sees a distinct region of the detector. Figure \ref{fig:waveforms_ind1} shows an event display with cosmic muon tracks before and after the coherent noise removal algorithm is applied. The coherent noise is visible as the vertical streaks running through adjacent channels.

\begin{figure}
    % TODO - Flip placement of median values to coincide with order of distributions.
    \centering
    \includegraphics[width=0.95\textwidth]{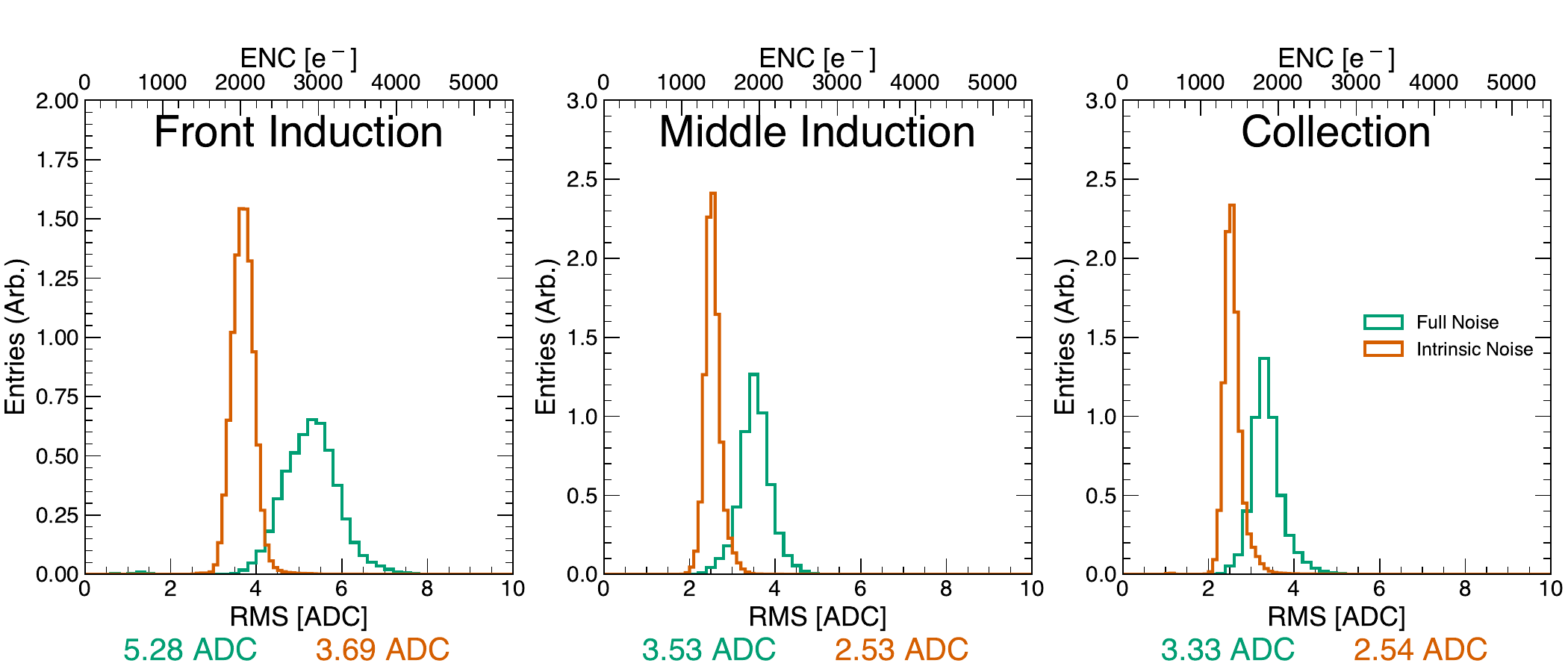}
    \caption{Noise distributions per plane as characterized by the RMS of each waveform before and after coherent noise removal. Shown below each plot are the medians associated with each distribution.}
    \label{fig:absolute_data_noise}
\end{figure}

The portion of the waveform that remains after the removal of the coherent noise component is representative of the noise naturally present on the channel due to intrinsic noise sources. The separation between coherent and intrinsic components of the noise allows for a more detailed characterization of the noise. The absolute noise levels are measured by the RMS of each waveform before and after the removal of coherent noise using data taken with the cathode voltage turned off. The distributions per plane are shown in figure \ref{fig:absolute_data_noise} in both units of ADC and Equivalent Noise Charge (ENC) using the conversion factor of 550 e$^-$/ADC found in \cite{ICARUSElectronics}. Front induction exhibits significantly higher noise due to the longer flat cables and wires. Middle induction and collection have similar noise levels owing to the fact that they have similar wire and cable lengths. The substantial variation in cable length for front induction channels drives most of the additional width of front induction noise distributions. The signal-to-noise ratio for each plane can be calculated using the mean hit amplitude from a sample of throughgoing cosmic muons and the measured value of noise after coherent noise filtering. Respectively, these are 4.7, 7.8, and 10.8 for the front induction, middle induction, and the collection planes. Over the course of Run 1 and Run 2, the detector noise has been very stable: less than a few percent on average.

\begin{figure}
    % TODO - Verify that units are correct.
    \centering
    \includegraphics[width=0.85\textwidth]{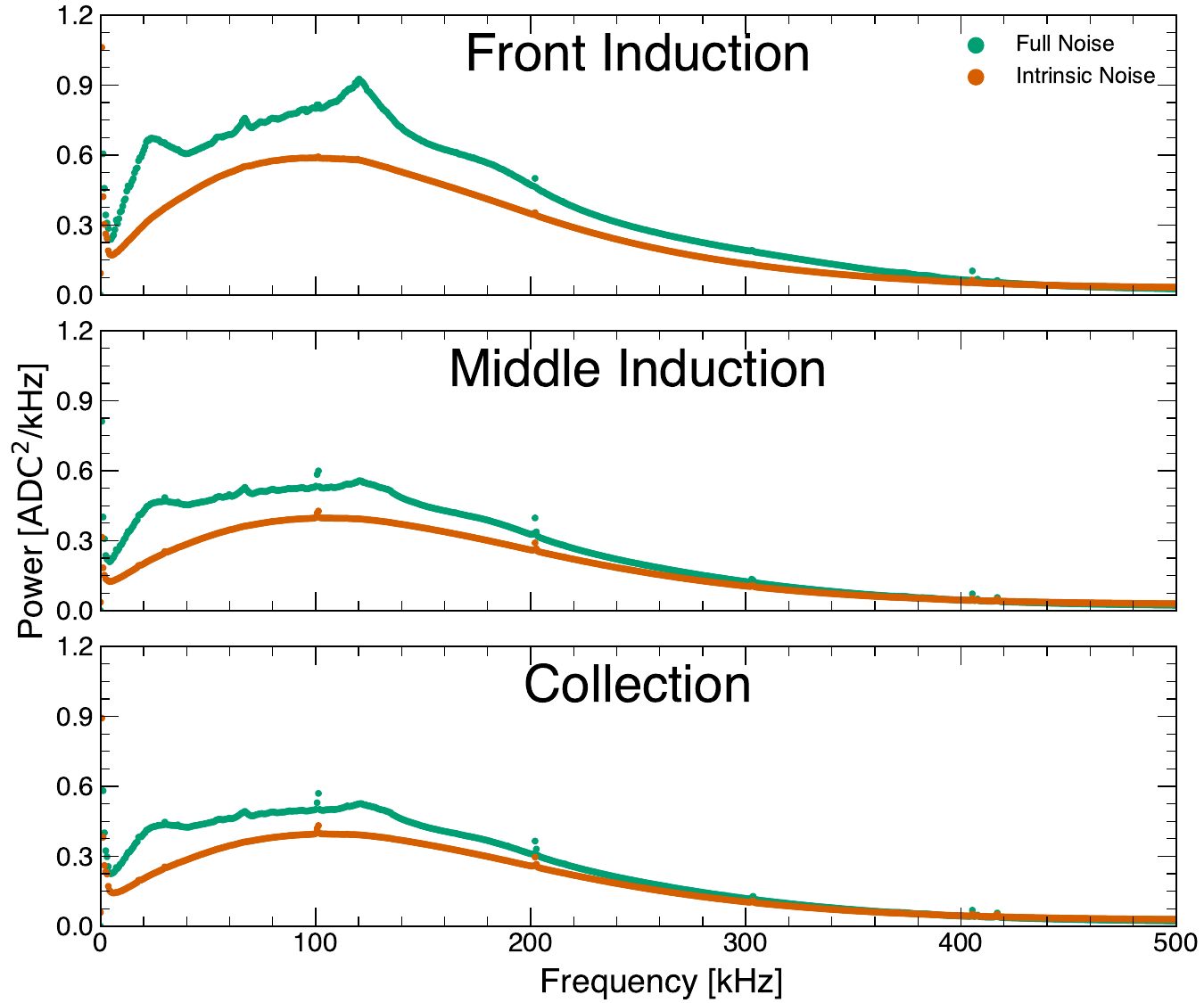}
    \caption{Frequency characteristics of the noise per plane as measured by the FFT spectra before and after coherent noise removal.}
    \label{fig:ffts_data_noise}
\end{figure}

The frequency characteristics of each component of the noise can be measured using the discrete Fast Fourier Transform (FFT). Figure \ref{fig:ffts_data_noise} shows the FFT spectra per plane before and after the coherent noise removal. The underlying intrinsic noise populates the expected Rayleigh distribution, and is similar for all three planes. The coherent noise is present as an additional, less smooth spectra on top of the intrinsic noise. The coherent noise also exhibits two broad peaks at specific frequencies that are not yet attributed to a specific source. At the lowest frequency bins, there is a sharp increase due to low-frequency oscillations in the waveforms. These oscillations are not coherent across groups of channels as evidenced by the full noise and intrinsic noise spectra exhibiting the same low-frequency trend. 

%\begin{figure}
%    \centering
%    \includegraphics[width=0.95\textwidth]{plots/noise/noise_history_intrinsic.pdf}
%    \caption{Detector noise after noise-filtering plotted as a function of the calendar year. The error bars represent the inner 68.27\% of the distribution centered at median value.}
%    \label{fig:noise_historical}
%\end{figure}

%Outside of hardware interventions, the detector noise has been historically very stable, as shown in Figure \ref{fig:noise_historical}. The biggest source of noise reduction was the disconnection of the Resistive Temperature Devices (RTDs) that read the temperature at various locations within the argon, after which the noise was both significantly reduced and considerably less variable. Due to the location of the RTDs on the inner walls of the cryostat, the collection plane was the most directly exposed plane, though the effect was highly localized and on average larger on front induction. 

\subsection{Noise Simulation}
\label{sec:noise_model}
The noise model is implemented with an algorithm provided by the Wire-Cell toolkit \cite{WireCell}. The algorithm allows for the freedom to configure a noise component by defining groups of channels, the noise spectrum associated with each group, and whether the noise component is coherent across the group or intrinsic. The noise component is simulated for each channel by drawing the amplitudes from the associated spectrum with a randomly chosen phase, then applying the inverse FFT. Coherent noise components have the additional requirement that the phases are shared for all channels within the same coherent grouping. The total waveform for each channel is calculated as the sum of the signal waveform and all noise waveforms from all components.

As discussed in the previous section, the dominant coherent noise component observed is across channels within the same readout board. After removal of this coherent component, the noise shows no significant remaining correlated components at other channel groupings. Therefore, the noise model is configured with two noise components: an intrinsic one that is uncorrelated and a coherent one that is correlated across the full group of 64 channels. In both cases, the input spectra reflect the average of the group of 64 wires.

\subsection{Data and Monte Carlo Simulation Comparisons}
\label{sec:noise_datamc_comparisons}

After configuring the noise model, it is used to generate a sample of events for a comparison with data. The data sample used for this comparison is the same as was used to generate the input spectra. Both the Monte Carlo sample and data sample were analyzed with the exact same code, so the only differences in the results are expected to be from the noise model.

\begin{figure}
    \centering
    \includegraphics[width=0.48\textwidth]{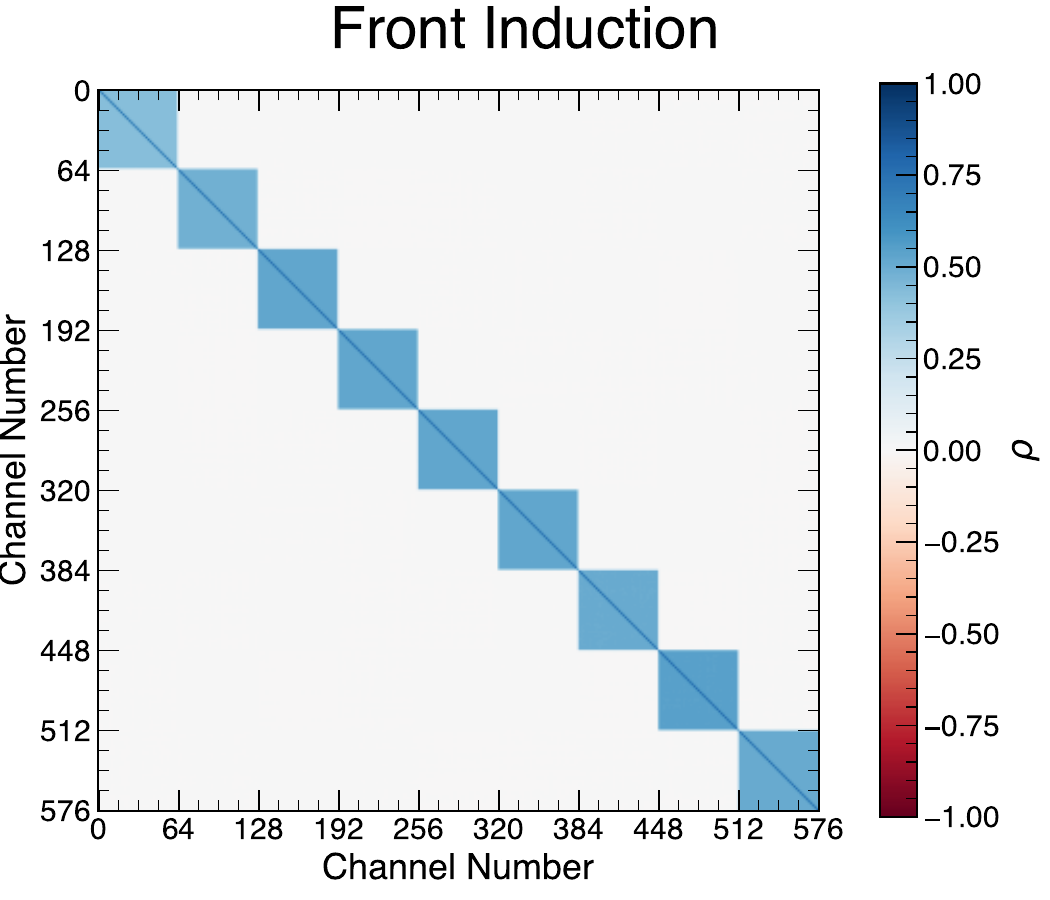}
    \includegraphics[width=0.48\textwidth]{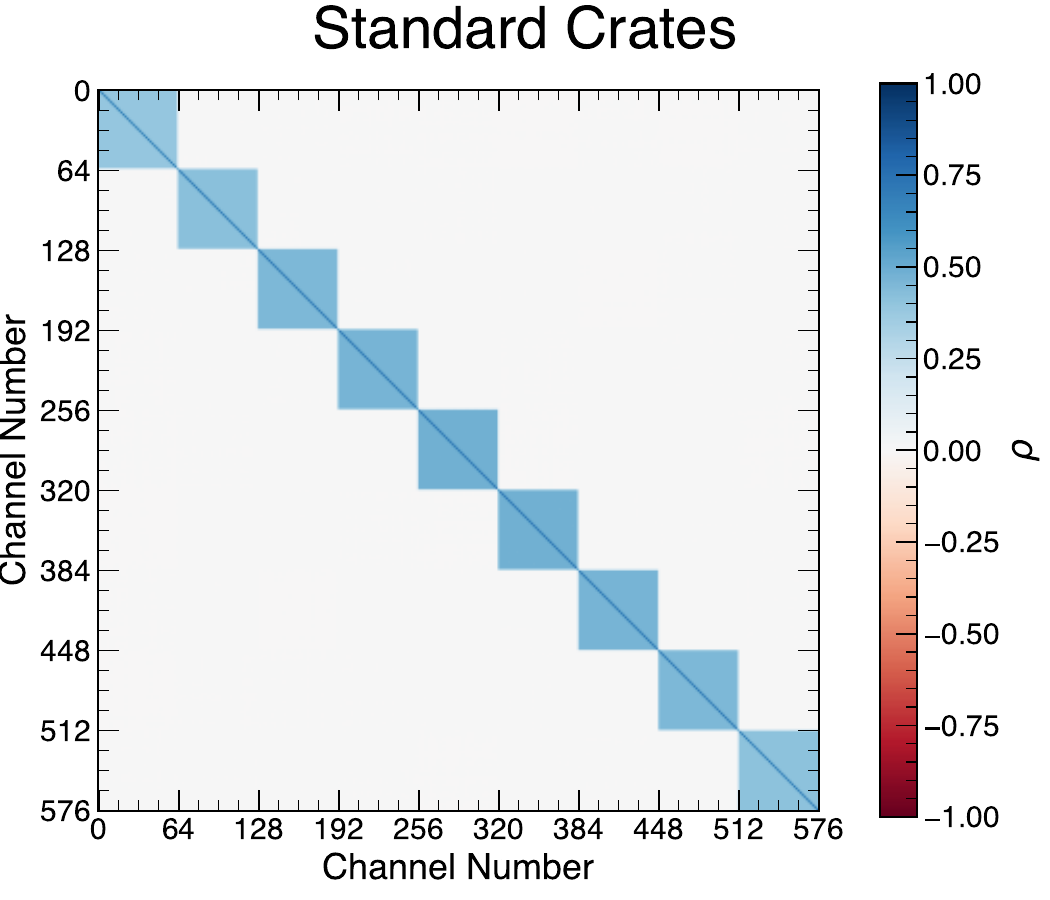}
    \caption{Channel-to-channel correlation matrix from Monte Carlo averaged across all crates serving front induction wires (left) and crates serving a mix of middle induction and collection wires (right).}
    \label{fig:correlations_montecarlo}
\end{figure}

\begin{figure}
    % TODO - Reduce redundancy in legends.
    % TODO - "Intrinsic Noise" looks attached to upper plot. Add vertical separation.
    % TODO - "[arb. units]"
    \centering
    \includegraphics[width=0.95\textwidth]{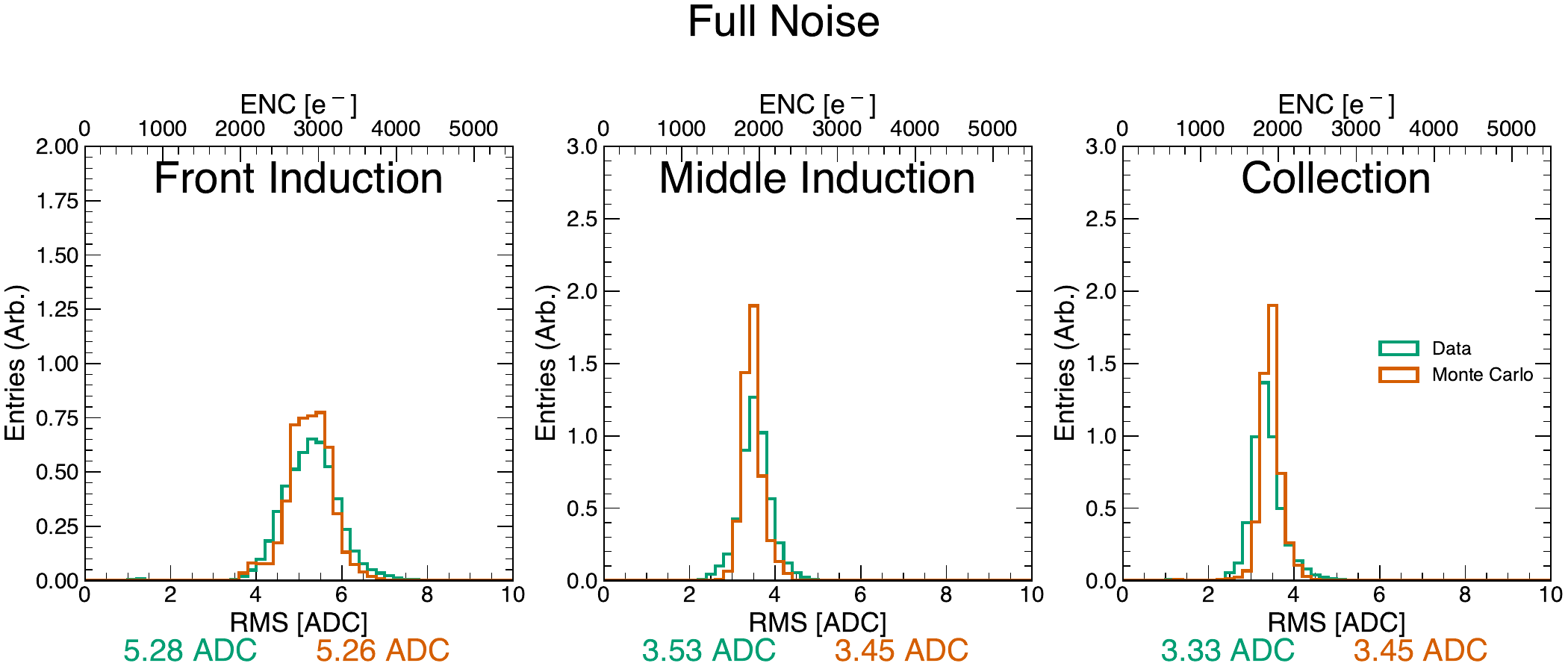}
    \includegraphics[width=0.95\textwidth]{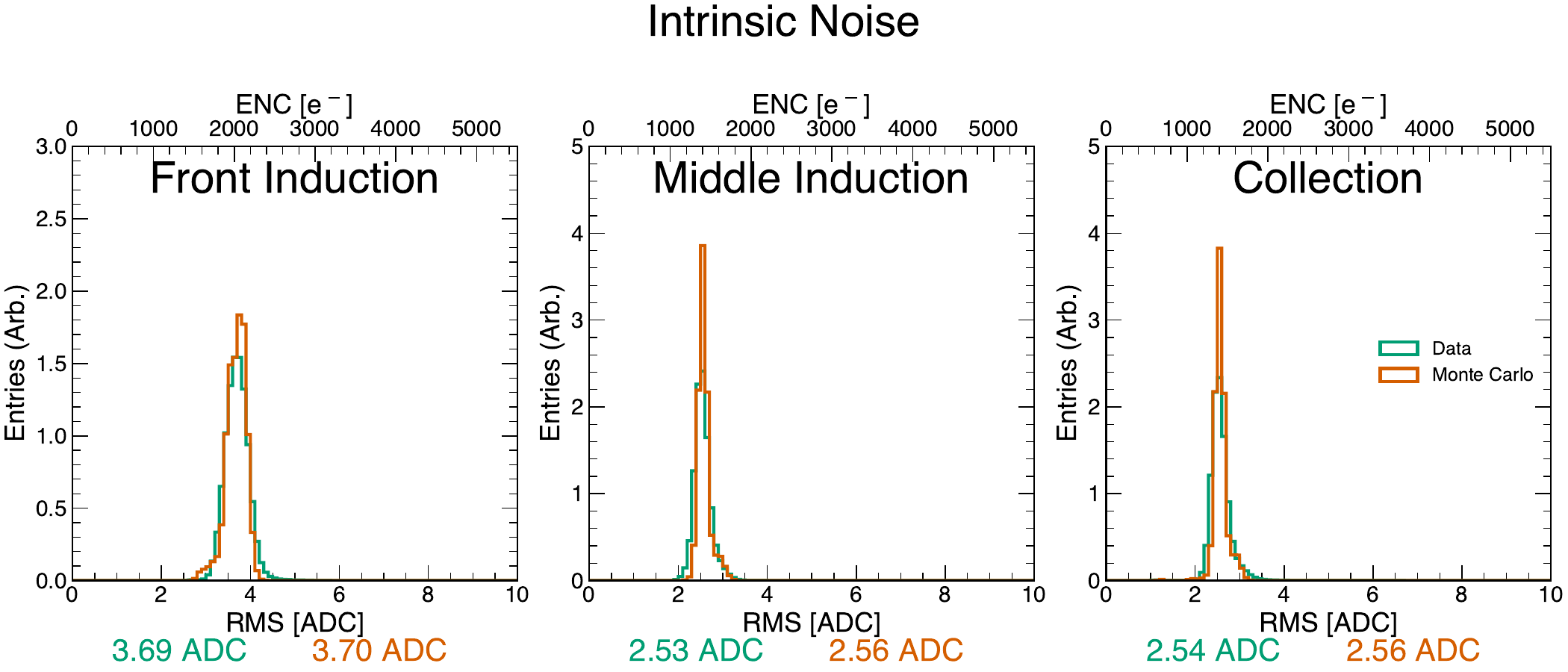}
    \caption{Noise distributions in data and Monte Carlo per plane as characterized by the RMS of each waveform. The top plot shows the full noise and the bottom plot shows the noise after coherent noise removal. The reported values are the medians associated with each distribution.}
    \label{fig:absolute_comparison}
\end{figure}

The correlation matrices for front induction crates and standard crates is shown in figure \ref{fig:correlations_montecarlo}. The noise model accurately reproduces the coherent component of the noise common to channels within the same readout board. The anti-correlated component between channels of adjacent readout boards is not modeled and therefore not reproduced. This anti-correlated component observed in data is itself coherent across the readout board and is removed by the coherent noise removal process, so this is not expected to have a noticeable impact on the overall noise levels.

The RMS calculated per channel and per event can be used to characterize the performance of the noise model in modeling the absolute noise scale. Figure \ref{fig:absolute_comparison} shows these distribution per plane before (top) and after (bottom) coherent noise removal. The full noise shows good agreement for front induction, but there is a systematic bias in middle induction and collection. Each readout board in a standard crate contains equal amounts of middle induction and collection wires, so the noise model effectively models the average of the two. This, along with the fact that middle induction exhibits slightly higher noise, results in a bias in opposite directions for both planes. After coherent noise removal, the agreement between data and Monte Carlo improves significantly, and the bias shown by middle induction and collection is reduced. It is worth noting that the waveforms after coherent noise removal are used downstream in the reconstruction, so inefficiencies in the noise model that appear in the full noise and not in the intrinsic noise are expected to have only second-order effects.

%\begin{figure}
%    \centering
%    \includegraphics[width=0.95\textwidth]{plots/noise/bias_noise.pdf}
%    \caption{The bias between data and Monte Carlo calculated as the same-channel difference in RMS normalized by the measured RMS. The left plot shows the bias in the full noise and the right plot shows the bias in the intrinsic noise.}
%    \label{fig:noise_bias}
%\end{figure}

\begin{figure}
    % TODO - Introduce better labeling?
    % TODO - Make clear which plot is which in labeling.
    \centering
    \includegraphics[width=0.48\textwidth]{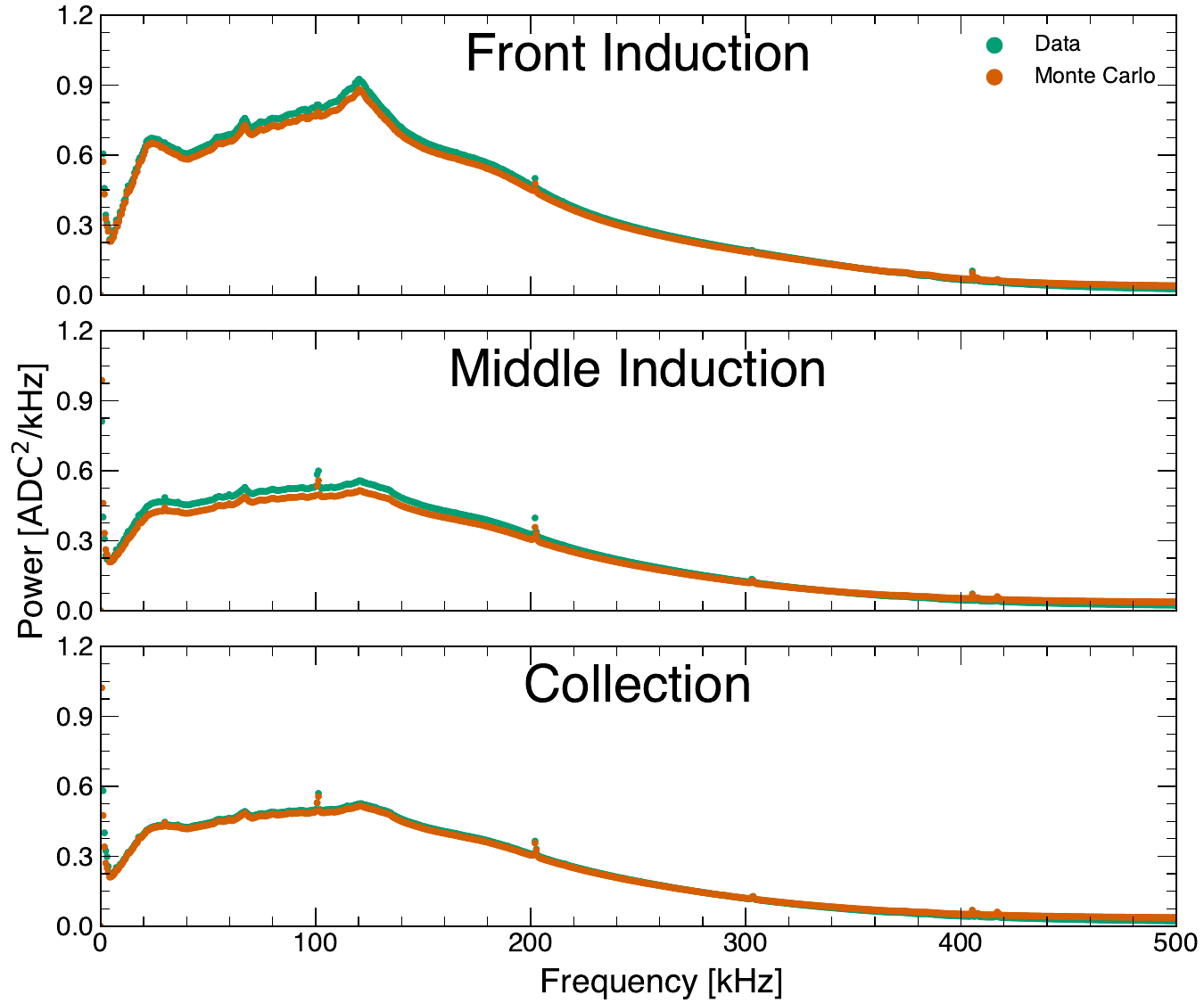}
    \includegraphics[width=0.48\textwidth]{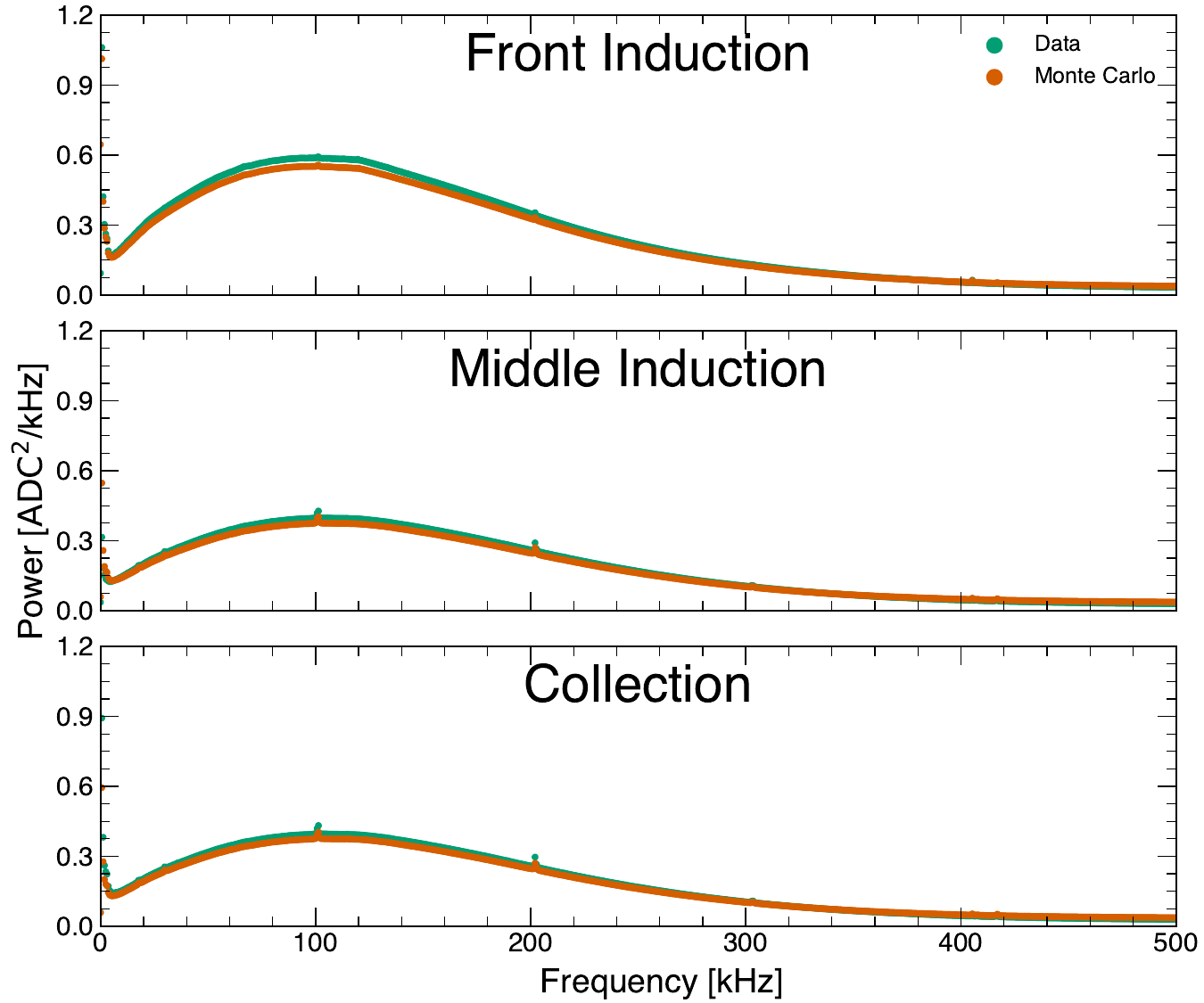}
    \caption{Frequency characteristics of the noise in data and Monte Carlo per plane as measured by the FFT spectra. The left plot shows the spectra for the full noise and the right plot shows the spectra after coherent noise removal.}
    \label{fig:ffts_comparison}
\end{figure}

The noise spectra in data and Monte Carlo can be compared to verify that the same spectra that were used as input are observed. Figure \ref{fig:ffts_comparison} shows a comparison of the average FFT spectra per plane for data and Monte Carlo. The shapes are reproduced nearly identically, but there are some minor discrepancies in the overall magnitude. Front and middle induction are consistently slightly under-predicted by Monte Carlo, whereas collection is slightly under-predicted after coherent noise removal. These observations are consistent with the results from the RMS noise comparisons shown earlier.

\begin{figure}
    \centering
    \includegraphics[width=0.95\textwidth]{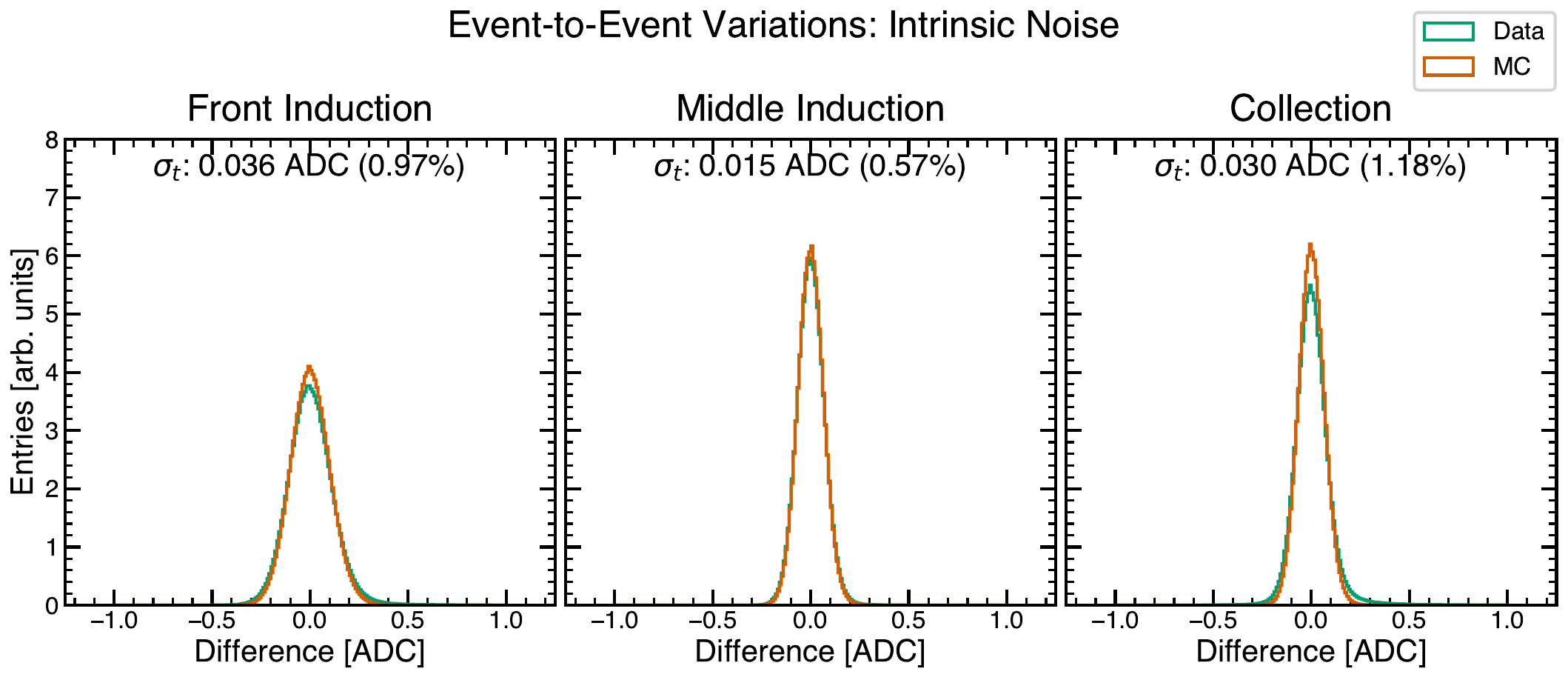}
    \caption{Noise variations in data and Monte Carlo after coherent noise filtering. Event-to-event variations are calculated per-event as the difference between the measured RMS and the median RMS for the channel. The reported values represent the Gaussian smearing necessary to match Monte Carlo to data and the corresponding percentage is the result after normalizing by the median noise on the plane.}
    \label{fig:noise_variations_e2e}
\end{figure}

Further characterization of the noise model performance can be done by examining two different metrics: the event-to-event variations and the channel-to-channel variations. The event-to-event variations are calculated as the difference of the measured RMS value from the median for the corresponding channel. The width of this distribution is driven by statistical fluctuations from the combination of frequency components of random phases and by short-scale variations in the noise itself. The overall width of the data distribution can be represented as the quadrature sum of the Gaussian widths associated with the two underlying processes. Monte Carlo, which uses a static noise distribution, is only broadened by the statistical fluctuations. By comparing data and Monte Carlo, the additional Gaussian broadening necessary to match Monte Carlo to data, parameterized by $\sigma_t$, can be extracted and used to directly characterize the variations in the noise on short time scales. Figure \ref{fig:noise_variations_e2e} shows these distribution and the associated results after coherent noise filtering for all three planes.

\begin{figure}
    \centering
    \includegraphics[width=0.95\textwidth]{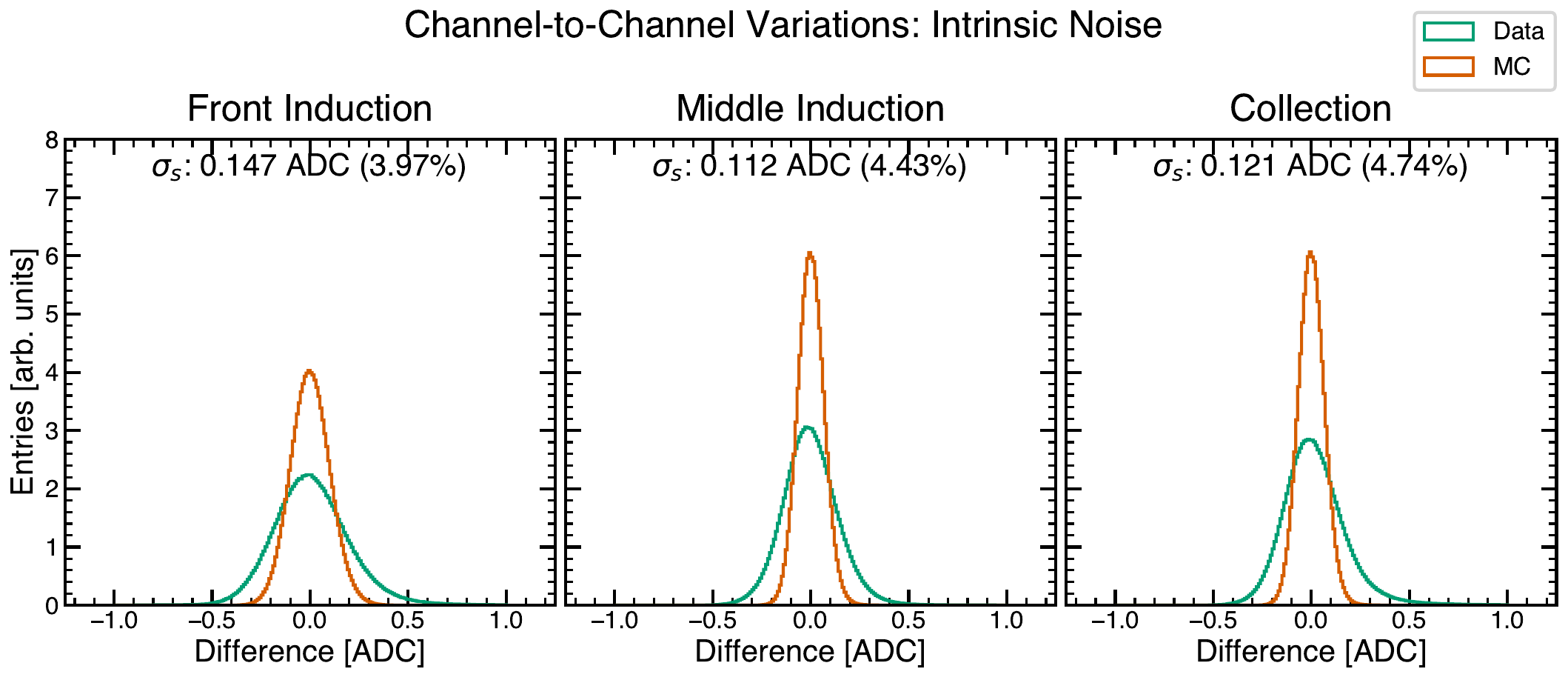}
    \caption{Noise variations in data and Monte Carlo after coherent noise filtering. Channel-to-channel variations are calculated per-event and per-channel as the difference of the measured RMS from the median RMS for the entire group of sixty four channels. The reported values represent the Gaussian smearing necessary to match Monte Carlo to data and the corresponding percentage is the result after normalizing by the median noise on the plane.}
    \label{fig:noise_variations_c2c}
\end{figure}

Channel-to-channel variations are calculated as the difference of the measured RMS value from the median for the group of 64 channels. In addition to the statistical and temporal variations discussed above, this distribution experiences broadening from the spatial variation in noise levels across channels within the same readout board. As before, the quadrature difference of the widths of the data and Monte Carlo distributions, parameterized as $\sigma_s$, characterizes the inefficiency of the noise model at modeling spatial variations smaller than the 64 channel grouping used to configure each noise component. These distributions are shown in figure \ref{fig:noise_variations_c2c}. The relative sizes of $\sigma_s$ and $\sigma_t$ suggest that spatial variations in the noise are the dominant contribution to mis-modeling of the noise. Correspondingly, the temporal variations in the noise over short time scales are negligible and do not need to be modeled. Further improvements to the noise model should target spatial variations in the noise by decreasing the group size for the intrinsic noise or by adding additional coherent components for smaller group sizes. 

\section{Signal Shape Measurement and Simulation}
\label{sec:SignalShape}
Potential disagreement of the TPC electronics response shape or ionization electron field response shape (the convolution of which determines the shape of ionization signal on the ICARUS TPC waveforms) between data and Monte Carlo simulation can lead to associated biases in charge extraction and thus $dQ/dx$ measurements. In order to minimize these biases, we carry out a data-driven tuning of the ionization signal shape used in ICARUS Monte Carlo simulation. Section \ref{sec:signalshape_measure} details the methodology used for extracting an estimate of the ionization signal shape in both data and Monte Carlo simulation, both of which are used in the tuning procedure described in section \ref{sec:signalshape_fitoverview}.

\subsection{Signal Shape Measurement}
\label{sec:signalshape_measure}

Reconstructed signal shapes are parameterized in terms of the angle $\theta_{xw}$, the angle of the track projected into the $\hat{x}-\hat{w}$ plane with respect to the $\hat{w}$ axis, where $\hat{x}$ is the drift direction of the ionization electrons in the TPC and $\hat{w}$ is the direction perpendicular to the wire orientation within the wire plane. This single angle controls the shape of the ionization signal response observed within the TPC waveform \cite{WireCell, SignalShapeuB}; it is diagrammed in figure \ref{fig:anglediagram}.

We estimate the average signal waveform associated with anode-cathode-crossing cosmic muon tracks for each angle ($\theta_{xw}$) bin in our angle range, carried out independently for each of the three wire planes. The requirement of angle-cathode-crossing tracks allows us to obtain signal shape measurements associated with ionization charge undergoing very little diffusion due to the ionization originating near the anode plane; as for the studies discussed in section~\ref{sec:Equalization}, the track crossing the cathode allows for the position of the track in the drift direction to be known. We use an angle range of $20^\circ$ to $76^\circ$ in track angle bins of $2^\circ$. This is the accessible range of angles from the anode-cathode-crossing track selection, as angles lower than $20^\circ$ will not yield many tracks as cosmic muons are likely to range out due to energy loss in the argon over the longer path length, while angles higher than $76^\circ$ are more difficult to reliably line up in time across different tracks, particularly on the two induction planes where the bipolar nature of the signal shape on the waveform leads to complications.
        
For each of the track angle bins and for each of the three TPC wire planes, we select anode-cathode-crossing track waveform data that is near the anode, chosen to be 13--16 cm away from the anode. This distance from the anode plane is chosen to be small enough to have minimal impact from diffusion that could broaden the signal shape on the waveform while being large enough to not lead to significant bias in estimation of the signal shape on the unshielded first induction plane. Coherent noise across common electronics channels is removed from the waveforms before the waveform data is included in the running average. We then pick the wire with peak waveform signal (negative for the two induction planes given the bipolar nature of the signals and positive for the unipolar collection plane signals) closest to the center of the above range (14.5 cm) using the known ionization electron drift velocity to convert the drift time to distance in the drift direction. Waveform data is saved within $\pm200$ time ticks (80 $\upmu$s) of the peak signal. The waveforms associated with individual tracks in this selection are then aligned by the minimization of a nonlinearity metric, $\eta$. For a given track, $\eta$ is defined as 
\begin{equation}
\eta = \log_{10} \left(\sum \limits_i\limits^\beta \frac{ [(t_i - t_{i, \mathrm{exp}})\cos(\theta_{xw})]^2}{N} \right),
\label{eq:non-lin}
\end{equation}
where $i$ denotes the index of individual charge measurements (in ADC) on the waveform within a given time bin, spanning both the waveform of interest and the associated waveforms from the nearest $\pm5$ wires, $\beta$ represents the condition that only charge above threshold is considered, $t_i$ is the actual drift time associated with a charge measurement, $t_{i,\mathrm{exp}}$ is the expected drift time for the ionization charge for a straight track in the corresponding track angle bin, and $N$ is the total number of charge measurements above threshold. A threshold of 8 ADC is used for the second induction plane and collection plane, while a threshold of 12 ADC is used for the first induction plane given the higher noise levels on that plane. The entire track is  shifted by up to $\pm5$ time ticks to minimize the nonlinearity metric on a track-by-track basis. This is to ensure that the tracks are lined up with each other in time as well as possible. We do this instead of lining up by the peak waveform signal on the primary waveform of interest as this is susceptible to noise fluctuations on the waveform. Tracks with a lower $\eta$ show less deviation from a hypothetical straight line associated with a given angle bin. Such deviations may be caused by delta ray production along the track which would interfere with the signal response shape measurement. Tracks with $\eta > 1.2$ are excluded from the measurement for the first induction plane and collection plane, while a tighter exclusion cut value of $\eta > 0.9$ is used for the second induction plane given that the bipolar nature of the signal waveforms leads to smaller $\eta$ values on average.

Next, we combine all selected waveform data for a given track angle bin and wire plane, lined up in time as described above. For each angle bin and plane we truncate the smallest and largest 10\% of ADC measurements in each time bin and use the mean of the remaining ADC measurements in each time bin for the estimate of the average signal waveform. This truncation removes the impact from noise and delta rays, as well as unusually low/high signal fluctuations that may skew the results in a particular time bin. Finally, because the waveform baseline may vary wire-to-wire in our data, we add an additional baseline correction for each average waveform using linear interpolation across the waveform sidebands (where no signal is present), taking the average of time ticks [-200, -160] for the left sideband and time ticks [160, 200] for the right sideband (measured with respect to the time tick associated with the peak signal on the average waveform). 

\begin{figure}[t]
    \centering
    \includegraphics[width=0.31\textwidth]{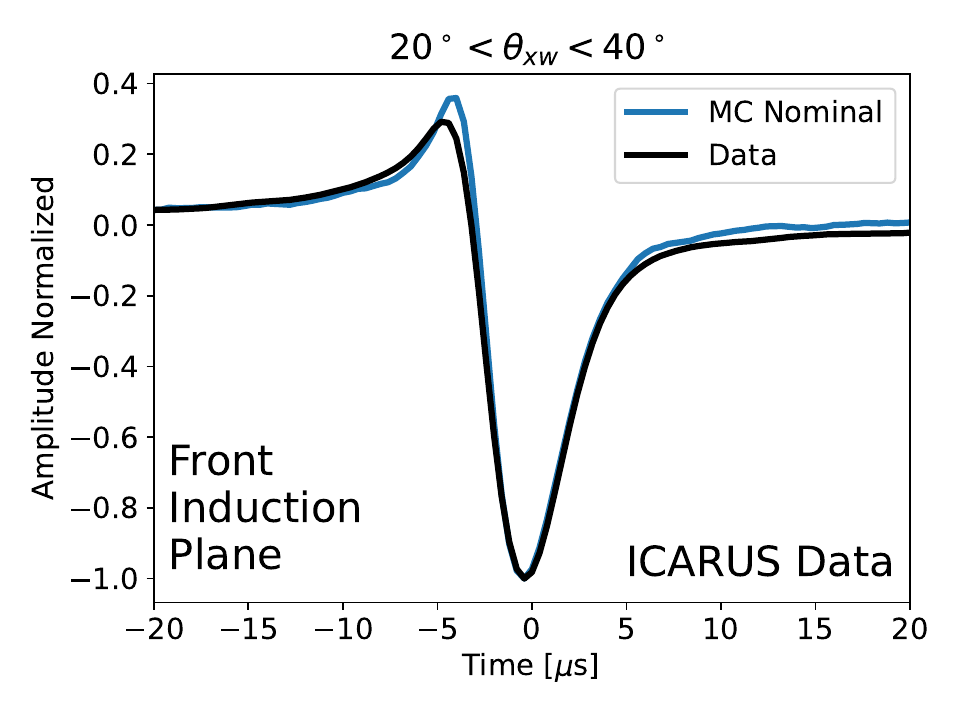}
    \includegraphics[width=0.31\textwidth]{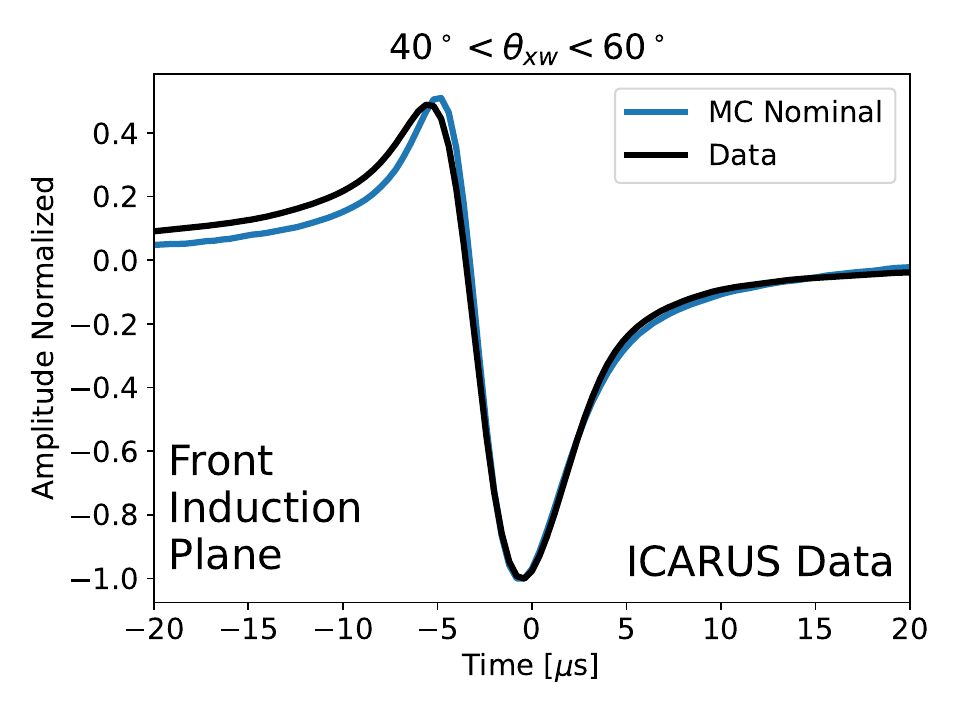}
    \includegraphics[width=0.31\textwidth]{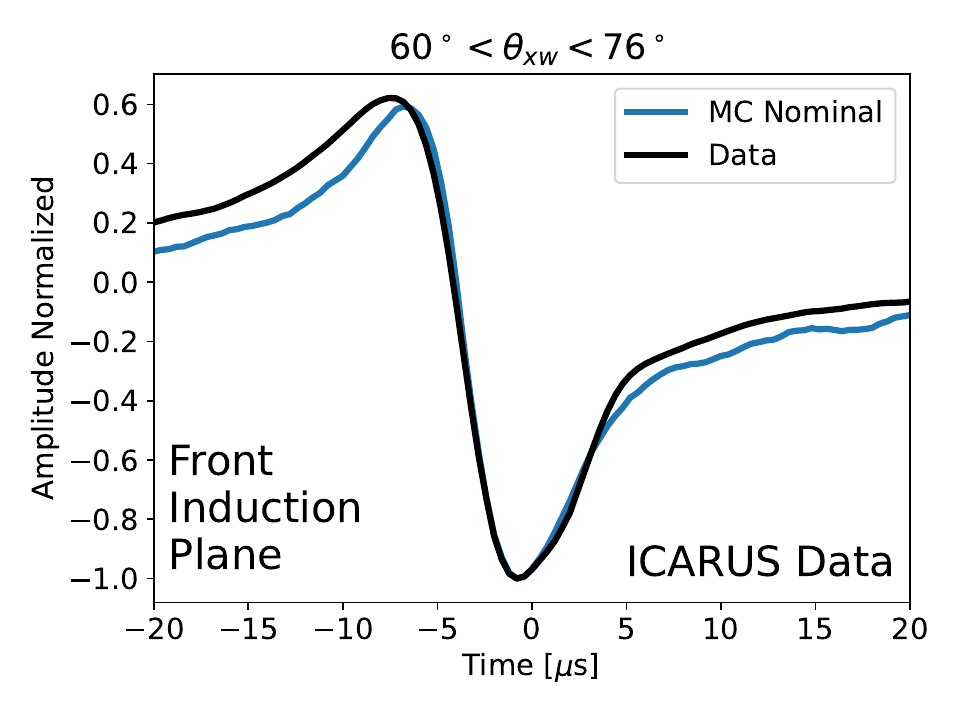}
    
    \includegraphics[width=0.31\textwidth]{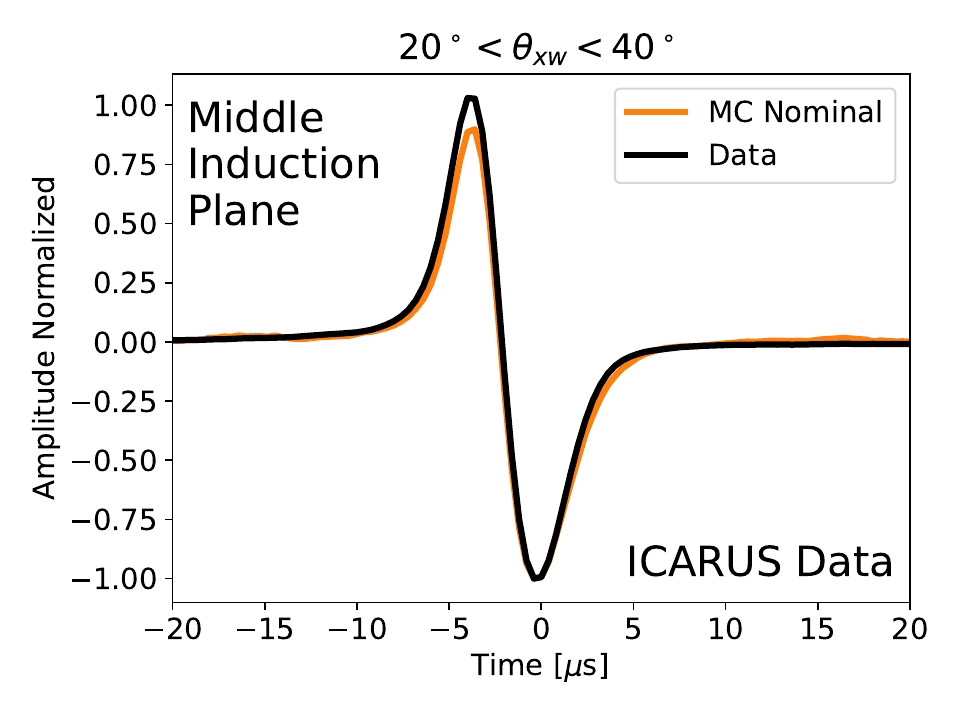}
    \includegraphics[width=0.31\textwidth]{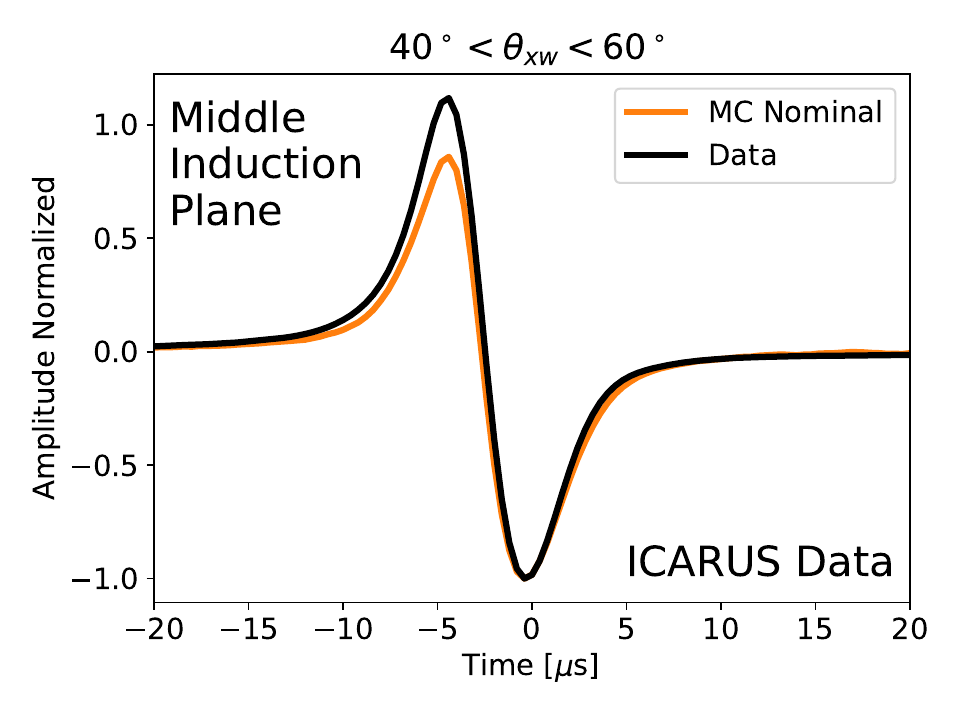}
    \includegraphics[width=0.31\textwidth]{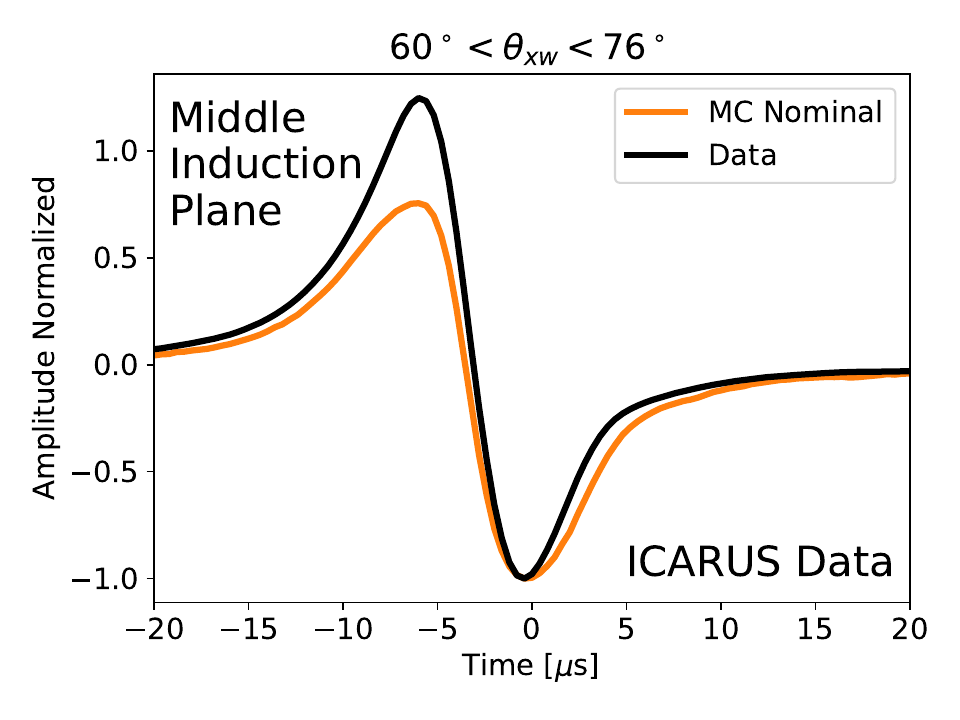}
    
    \includegraphics[width=0.31\textwidth]{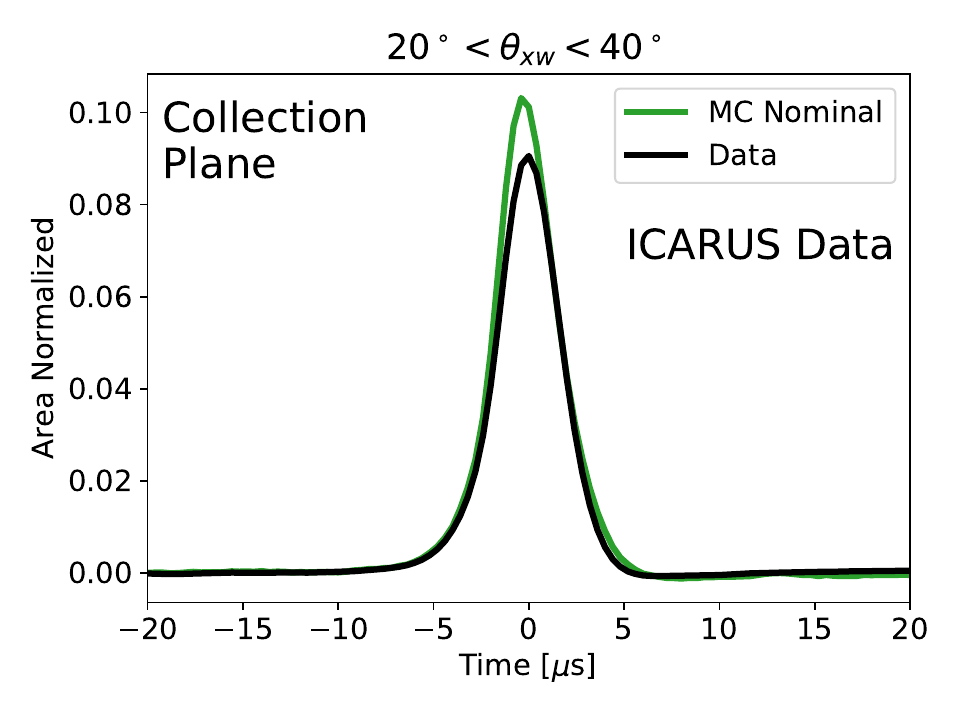}
    \includegraphics[width=0.31\textwidth]{plots/signal-shape-tuning/Validation/wvf_data_MCNominal_a20to40_P2.pdf}
    \includegraphics[width=0.31\textwidth]{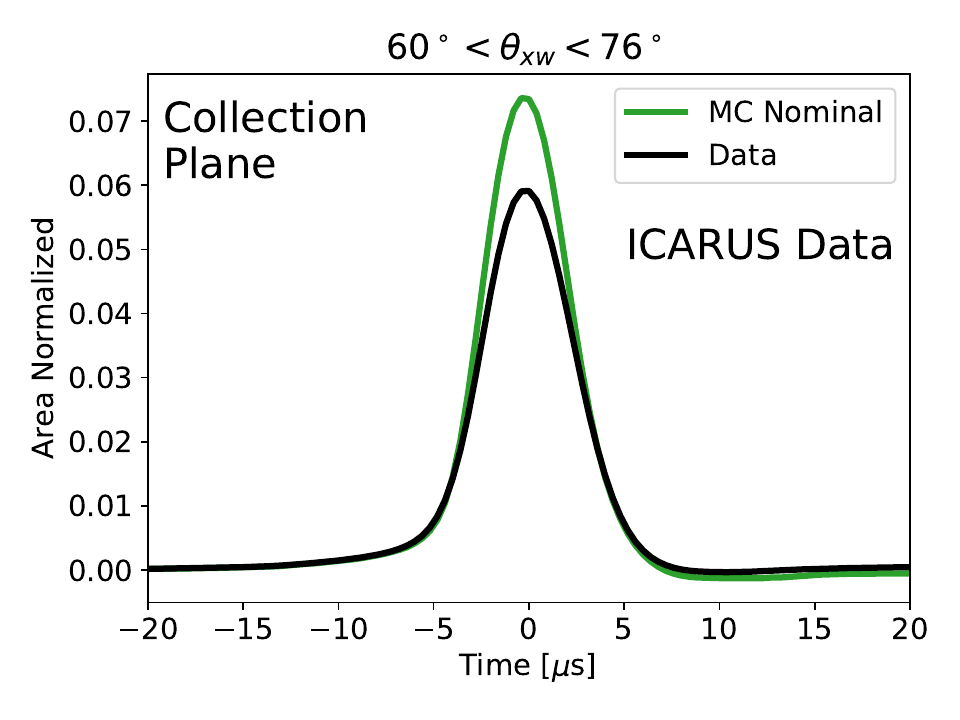}
    
    \caption{Comparison of the signal shape measurement between data and the nominal (untuned) GARFIELD-based Monte Carlo simulation (``MC Nominal") on the front induction plane (top row), middle induction plane (middle row), and collection plane (bottom row). Shown are measured signal waveforms averaged across $\theta_{xw}$ ranges of [$20^\circ$, $40^\circ$] (left column), [$40^\circ$, $60^\circ$] (middle column), and [$60^\circ$, $76^\circ$] (right column).}
    \label{fig:signalshpae_validation_nom}
\end{figure}

The measured signal shapes across three angle bins are compared between data and Monte Carlo simulation in figure  \ref{fig:signalshpae_validation_nom}. There is significant disagreement between the data and the nominal simulation on all three planes, especially at high track angle ($\theta_{xw}$). We have identified three main sources of the cause of this disagreement: a tail in the channel electronics response not in the nominal shape, distortions in the field response indicated by the intransparency of the induction planes to charge, and differences in the signal-to-noise ratio between the simulation and the data. The first two causes are directly connected to the signal shape. The last is driven by variations in the detector response to charge (such as the varying electron lifetime measured in section \ref{sec:q=drift}, and the varying induction intransparency measured in section \ref{sec:q=YZ}) that are not simulated. These variations primarily impact the front induction plane, where the noise is larger and the dependence on the exact signal-to-noise ratio is significant. In the tuning procedure we describe in the next section, we therefore elect to apply the fit only to the measurements on the middle induction and collection planes. The tuning procedure can in principle be applied to the front induction plane once the differences in the variation of signal-to-noise are addressed. 

\subsection{Fit Procedure}
\label{sec:signalshape_fitoverview}

We have developed a novel procedure to tune simulated signal shapes to match their measurement in data. This procedure tunes the input single electron field generated by GARFIELD \cite{Garfield} and applied by the Wire-Cell package, which is used in signal shape simulation for ICARUS Monte Carlo simulation. We defer to the initial paper on Wire-Cell for a detailed description of how it works \cite{WireCell}. Here we include a abbreviated discussion necessary to understand the fitting procedure. 

Wire-Cell forms signal shapes by summing the single electron field response on each wire from ionization electrons in particle energy depositions. The single electron field response depends on the location of the electron in the direction perpendicular to the wire direction (which we call $\hat{w}$, see figure \ref{fig:anglediagram}). The field response is significant even when the electron is not directly adjacent to the wire. Wire-Cell applies single electron field responses calculated every \SI{0.3}{\milli\meter} for \SI{3}{\centi\meter} (10 wire-spacings) on either side of each wire. Individual clusters of electrons in the simulation arrive in between the discrete locations where the field responses are calculated, so for a given deposition Wire-Cell linearly interpolates between the two on either side. The combined field response from all ionization electrons in a readout on a given wire is then convolved with the electronics response to create the signal shape for that readout.

The nominal ICARUS single electron field responses are computed by a GARFIELD simulation of the nominal ICARUS wire plane configuration. The nominal electronics response is a Bessel shaping function with a width of \SI{1.3}{\micro\second}. That the observed signal shapes depart from the nominal simulation indicates that the ICARUS detector departs from this nominal configuration in some way. First, we have observed a tail in the electronics response due to imperfect pole-zero cancellation that we measure for separately and include in the tuned electronics response. This step is discussed in section \ref{sec:signalshape_ertail}. The remaining differences are harder to attribute and ultimately depend on the inaccessible state of the TPCs inside both cryostats. We have thus taken the perspective that a data driven approach is an apt fix to these discrepancies. We fit the signal shapes in the simulation directly to the measurement. The objects in the fit are the position dependent single electron field responses and the electronics response\footnote{Imperfections in the means to directly pulse the ICARUS TPC readout electronics prevent a precise direct measurement of the electronics response. Instead, the fit described here produces an ``effective" electronics response adequate to describe the final signal shape.}, although we do not claim to be more accurately measuring any of these objects individually after the fit. We only attempt to model their combined impact on the signal shape. This fit is detailed in section \ref{sec:signalshape_model}. Monte Carlo simulation with the signal shapes tuned by this procedure demonstrates a much improved match in the signal shape between data and Monte Carlo simulation, as is shown in section \ref{sec:signalshape_fits}.

\subsubsection{Electronics Response Tail Measurement}
\label{sec:signalshape_ertail}

\begin{figure}[t]
    \centering
    \includegraphics[width=0.6\textwidth]{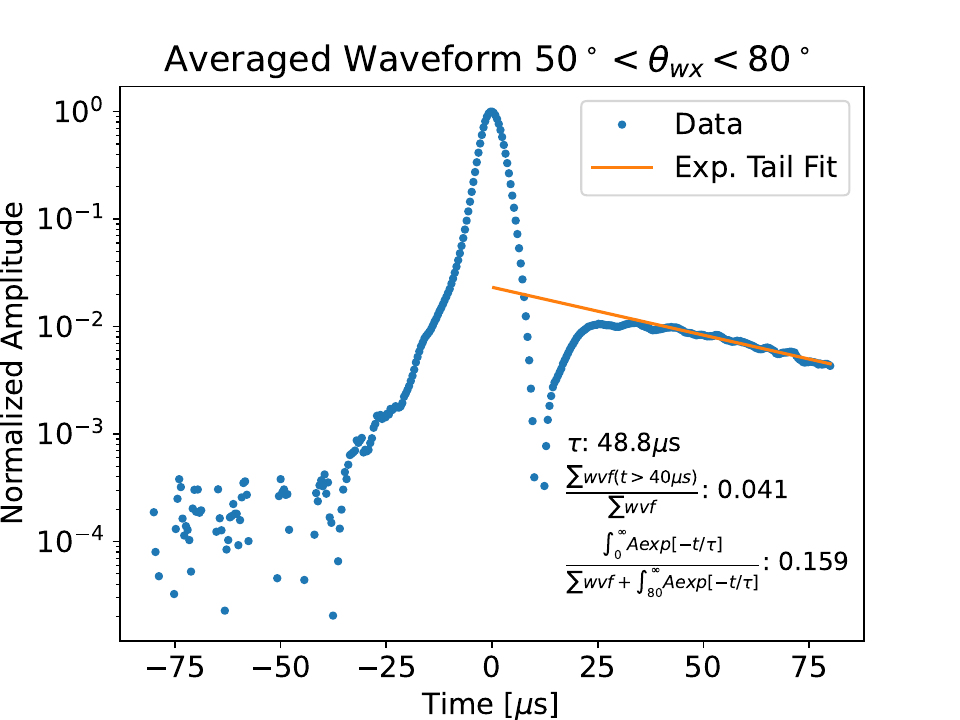}
    \caption{Averaged waveform from high angle tracks used to measure the exponential tail in the electronics response.}
    \label{fig:signalshape_tail}
\end{figure}

We have observed a long tail, with a time constant of $\sim$\SI{50}{\micro\second}, in the electronics response of ICARUS. The origin of the effect is imperfect pole-zero cancellation in the transfer function of the electronics. The tail is measured by averaging together waveforms on the collection plane from a large number of high angle muons (large $\theta_{xw}$) on the collection plane. High angle muons are used because the coherent noise subtraction can depress the effect of the tail when the track is closer to perpendicular to the drift direction. An exponential ($e^{-t/\tau}$) is fit to the averaged waveform values between 40-\SI{80}{\micro\second} (100-\SI{200}{ticks}). This time range is selected to exclude the region of the waveform where its shape is impacted by the field response, which in particular creates a visible dip in the waveform after the peak that extends out to about \SI{25}{\micro\second}. The exponential fit obtains a time constant of \SI{48.8}{\micro\second} that contains 15.9\% of the charge from the pulse. The fit exponential is convolved with the nominal electronics response to obtain the effective electronics response. This effective electronics response is the input to the fits to data on all three wire planes as described in section \ref{sec:signalshape_fits}. Figure \ref{fig:signalshape_tail} displays the data and the fit. The exponential describes the waveform shape well in the fit region. 

\subsubsection{Signal Shape Model}
\label{sec:signalshape_model}

We have developed a model of the signal shape measurement that takes the single electron field responses and the electronics response as input and produces the expected signal shape as a function of the track angle $\theta_{xw}$. The fit of the field and electronics responses are done by fitting this model to the measured signal shapes.

\begin{figure}[]
    \centering
    \includegraphics[width=0.32\textwidth]{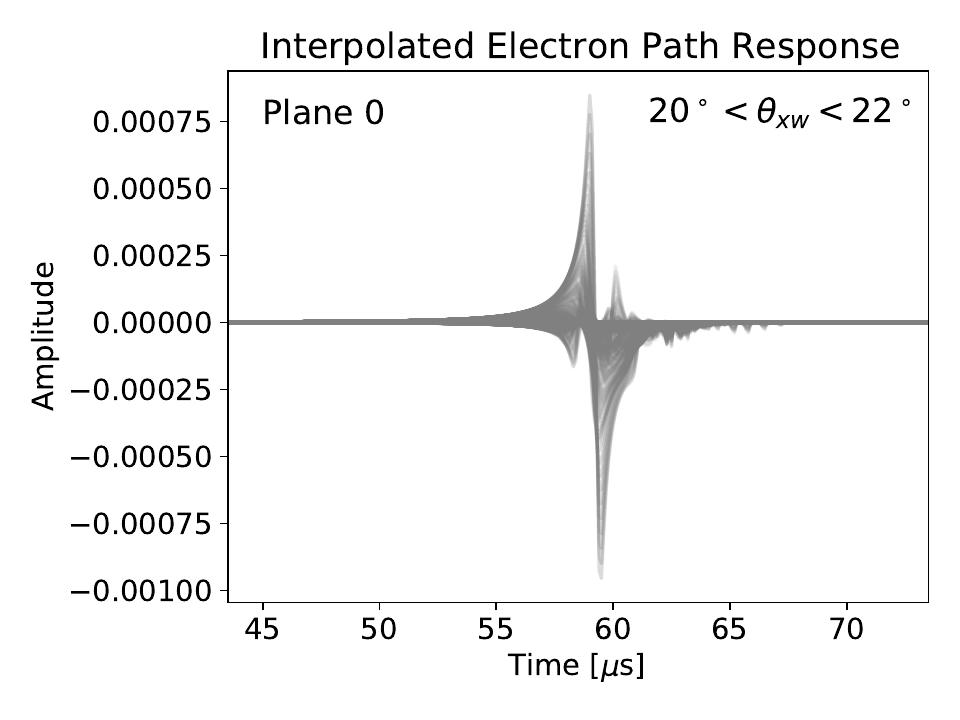}
    \includegraphics[width=0.32\textwidth]{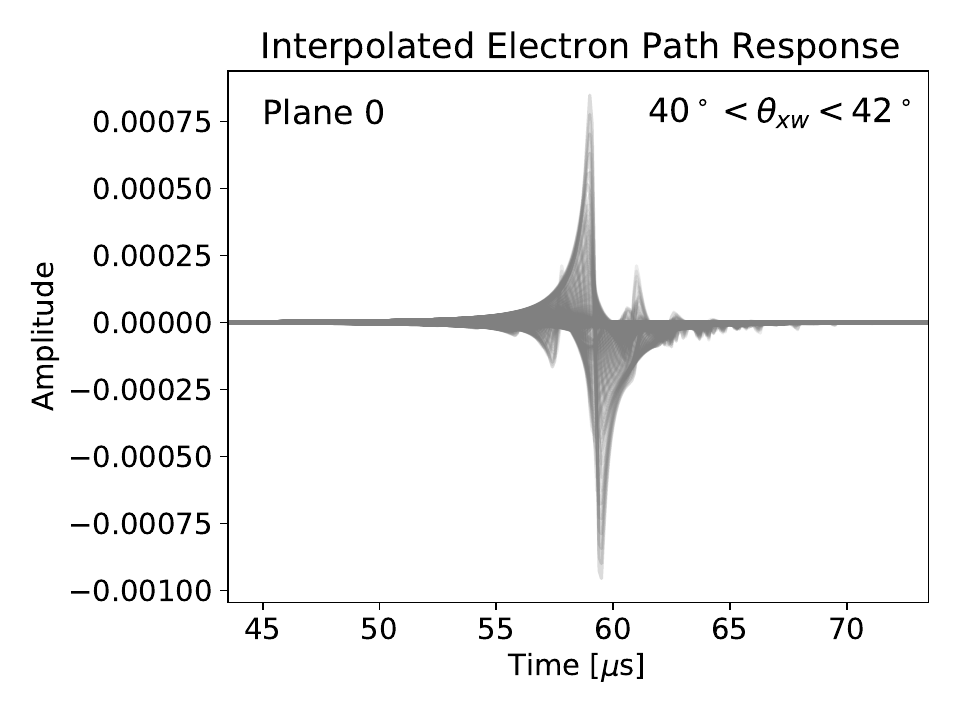}
    \includegraphics[width=0.32\textwidth]{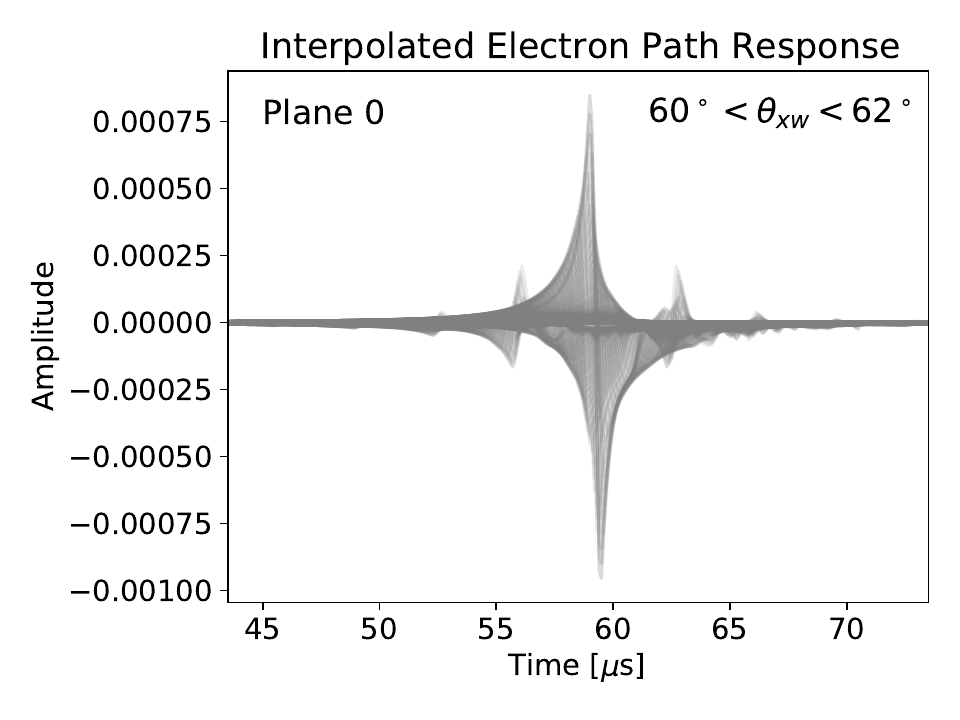}

    \includegraphics[width=0.32\textwidth]{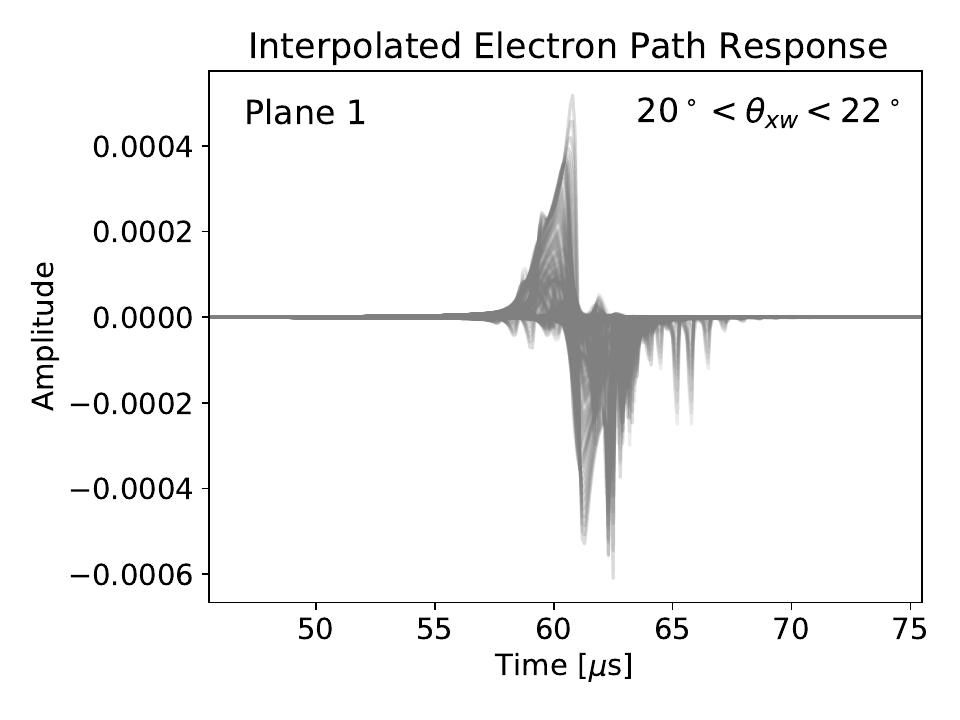}
    \includegraphics[width=0.32\textwidth]{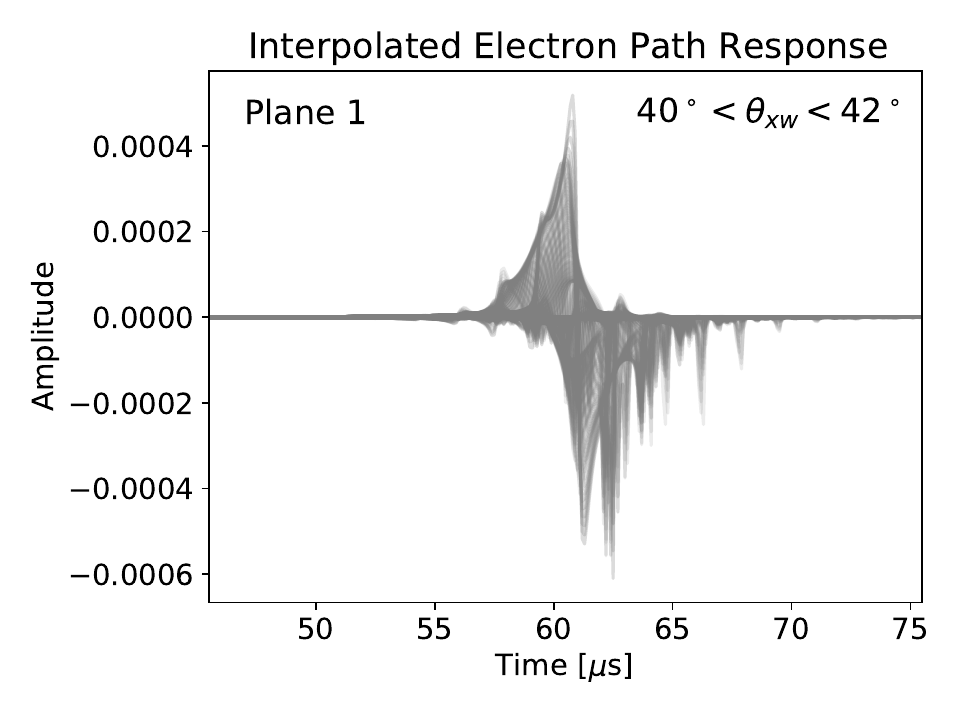}
    \includegraphics[width=0.32\textwidth]{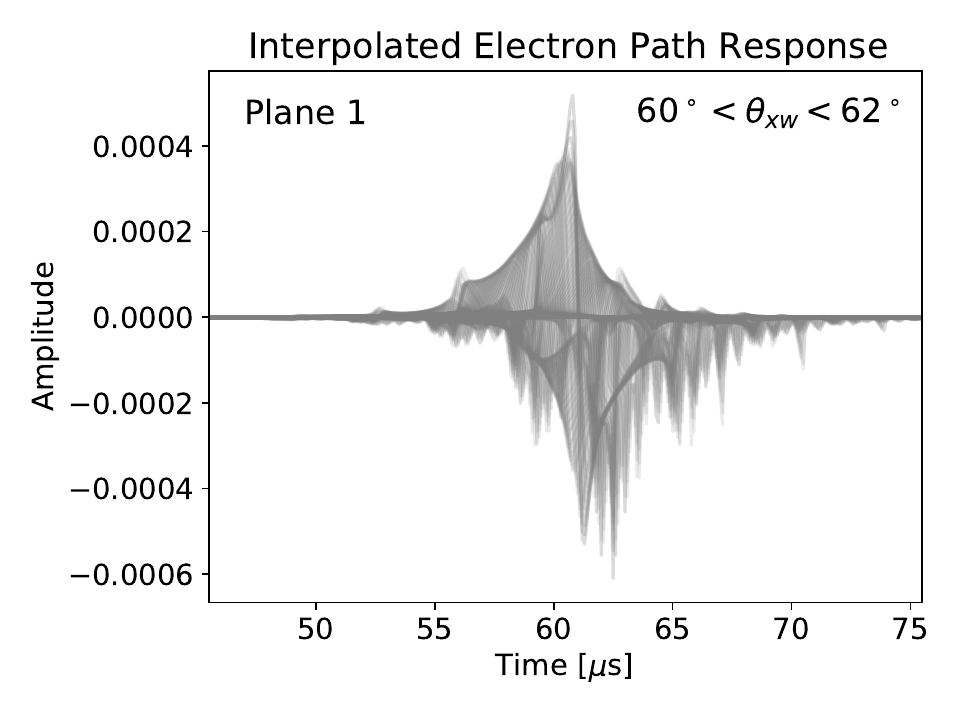}

    \includegraphics[width=0.32\textwidth]{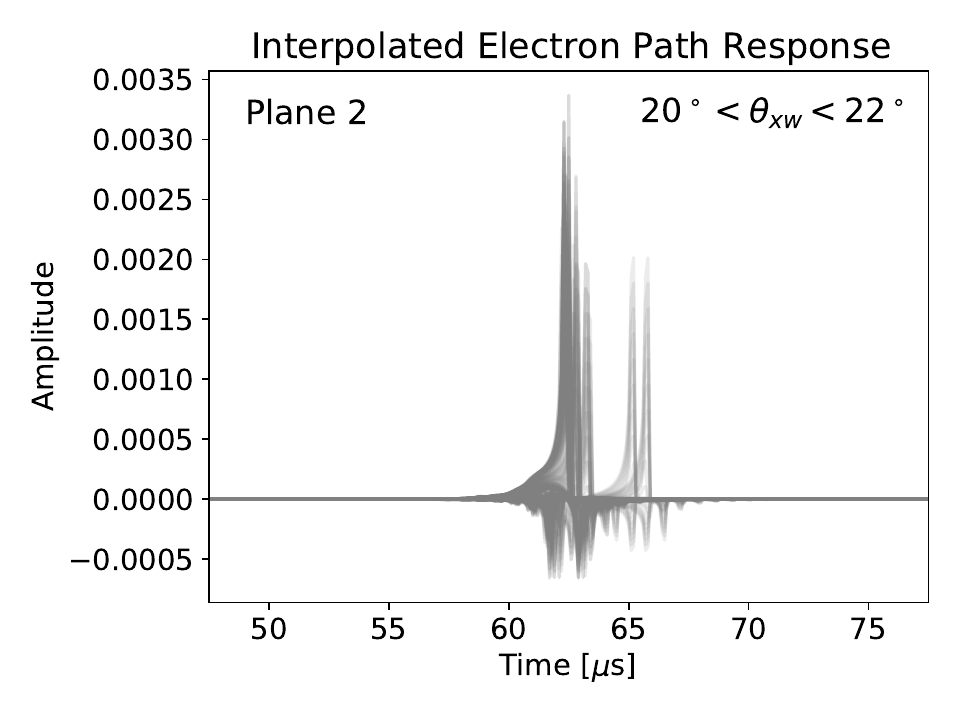}
    \includegraphics[width=0.32\textwidth]{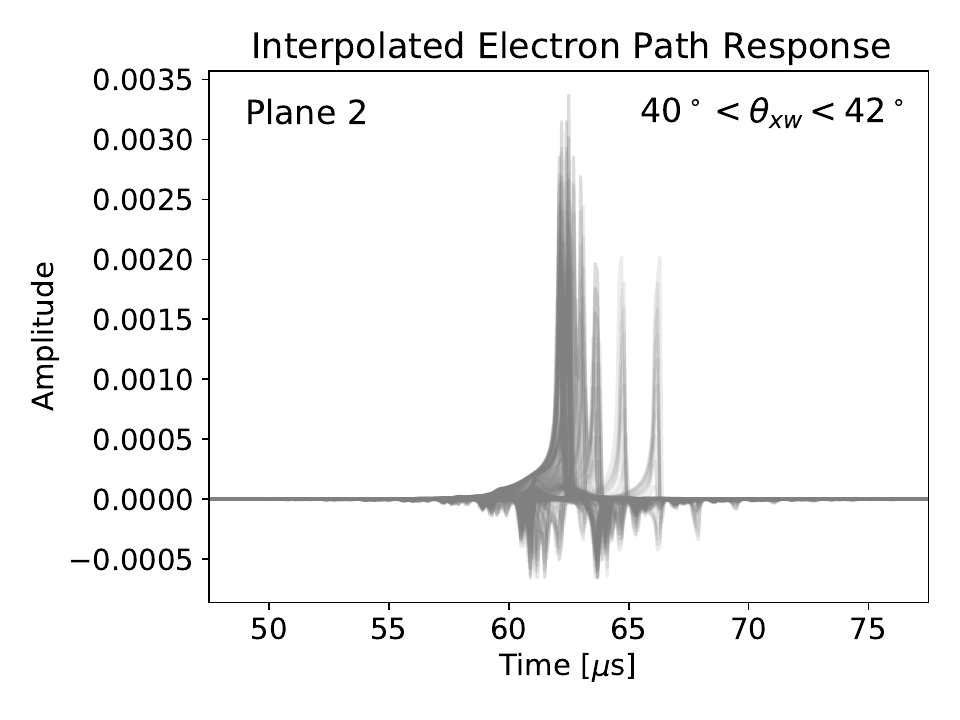}
    \includegraphics[width=0.32\textwidth]{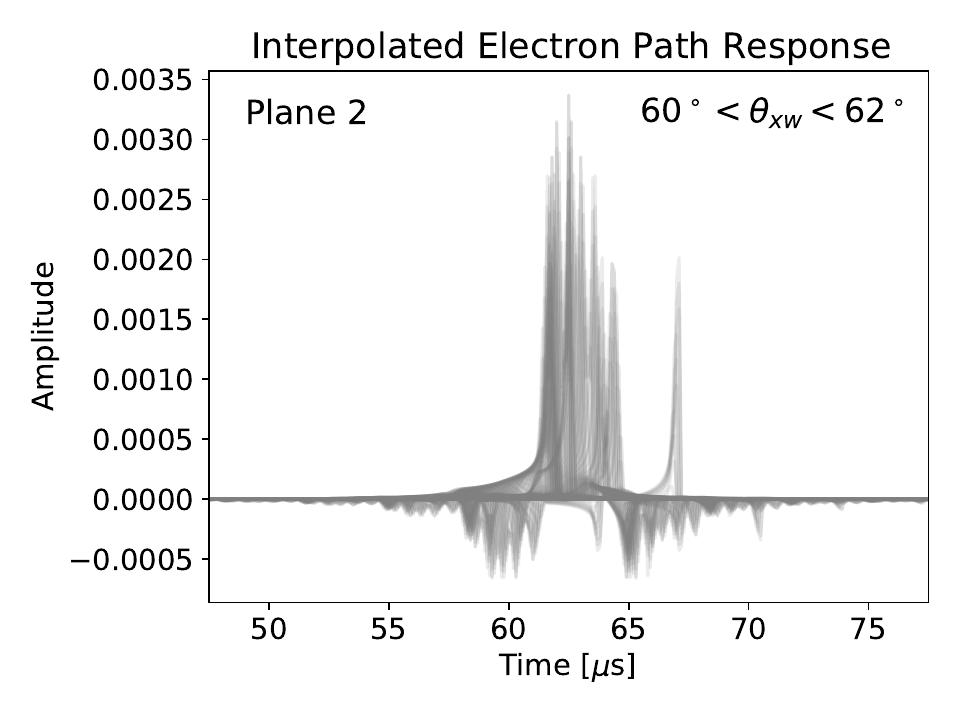}
    
    \caption{Sampled interpolated single electron field responses, shifted by track angle, as defined in equation \ref{eq:interpPaths}. Shown for the front induction plane (top row), middle induction plane (middle row), and the collection plane (bottom row).}
    \label{fig:interpPaths}
\end{figure}

\begin{figure}[]
    \centering
    \includegraphics[width=0.32\textwidth]{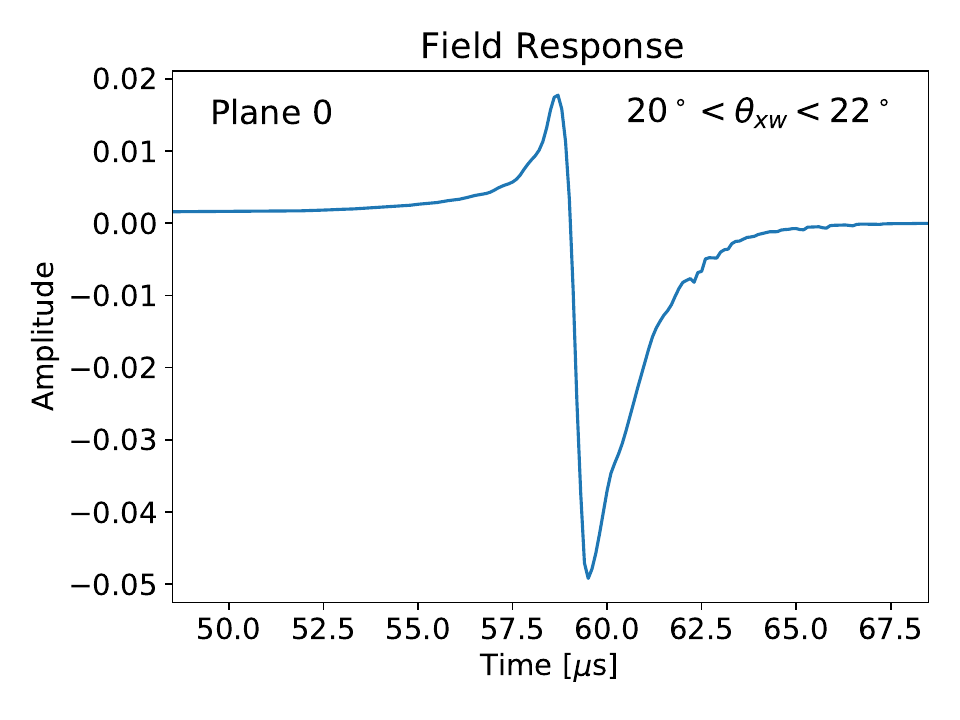}
    \includegraphics[width=0.32\textwidth]{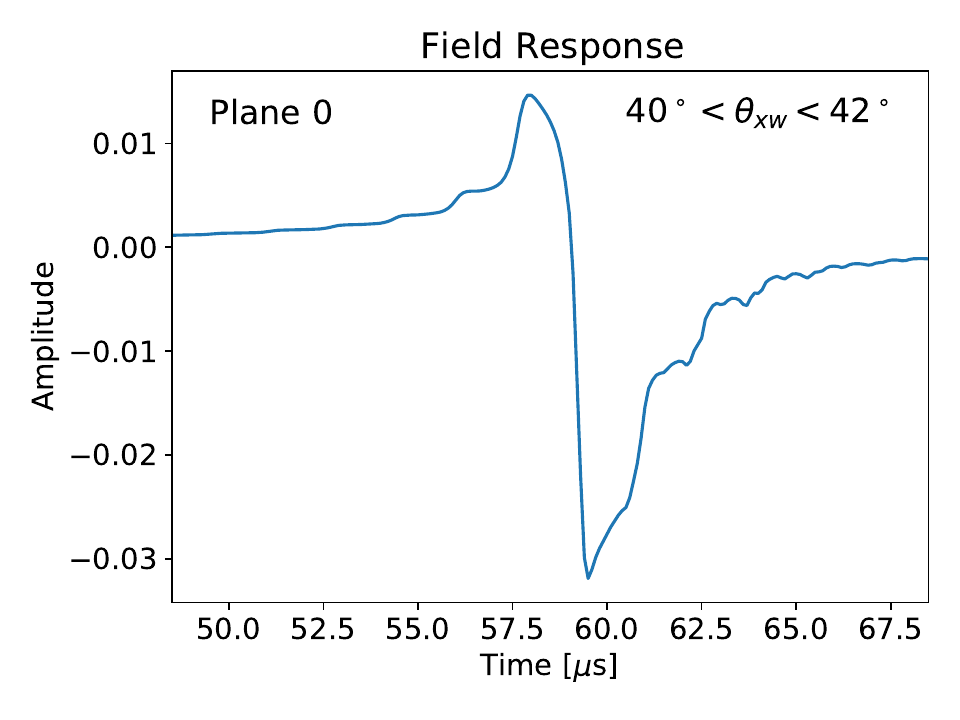}
    \includegraphics[width=0.32\textwidth]{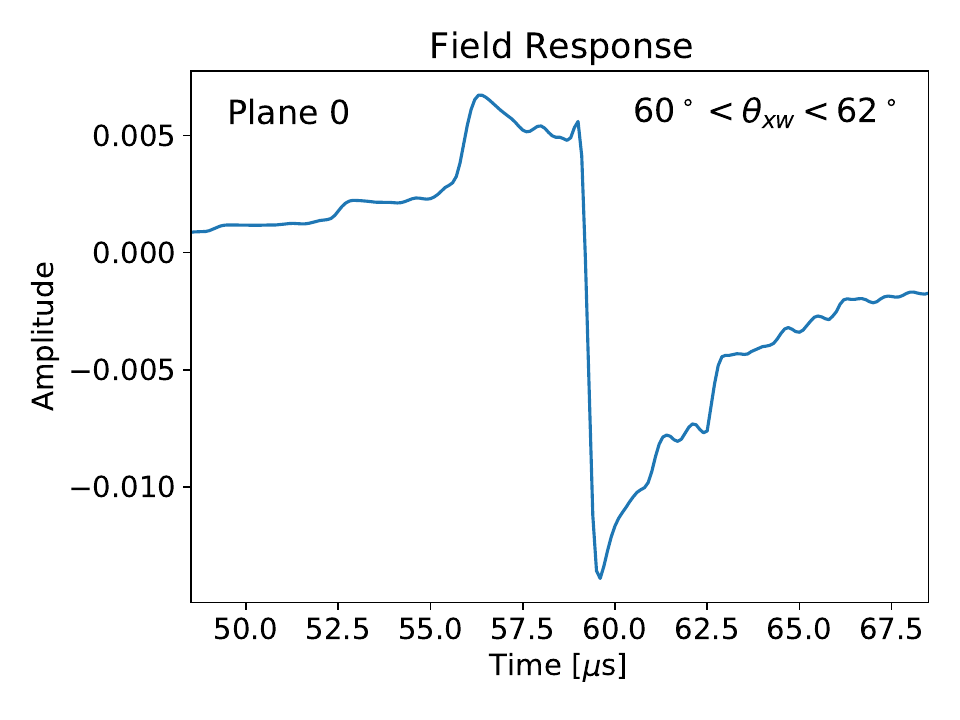}

    \includegraphics[width=0.32\textwidth]{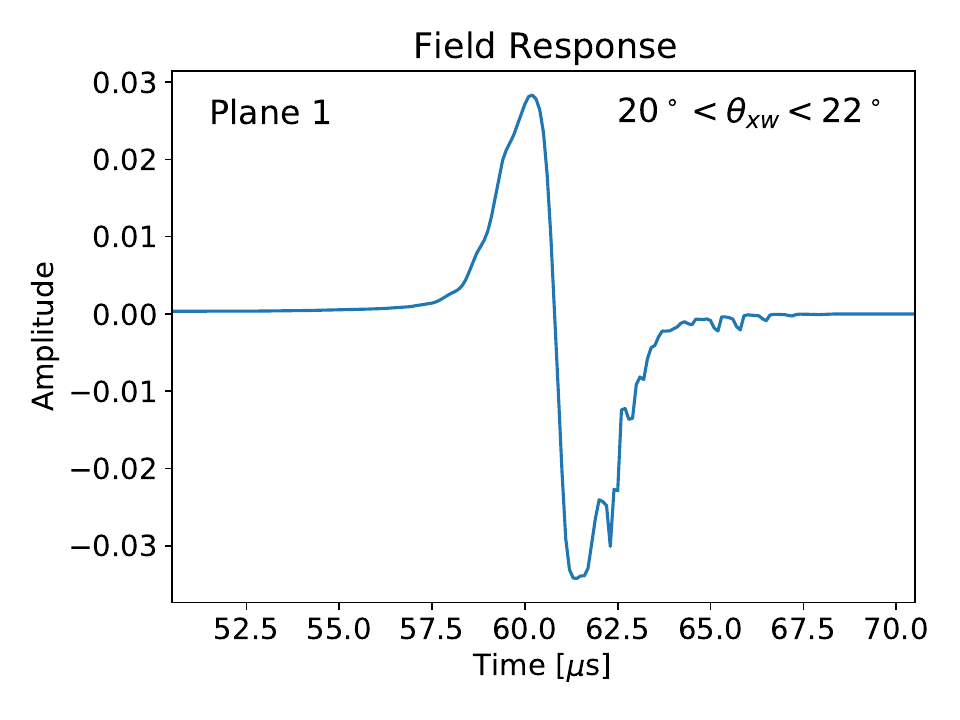}
    \includegraphics[width=0.32\textwidth]{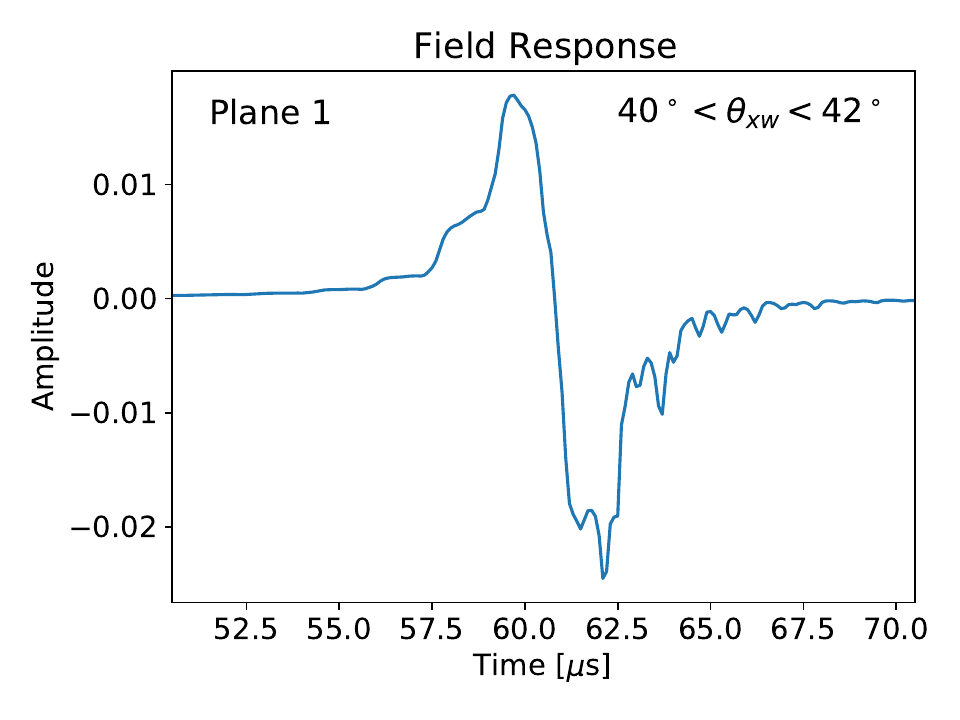}
    \includegraphics[width=0.32\textwidth]{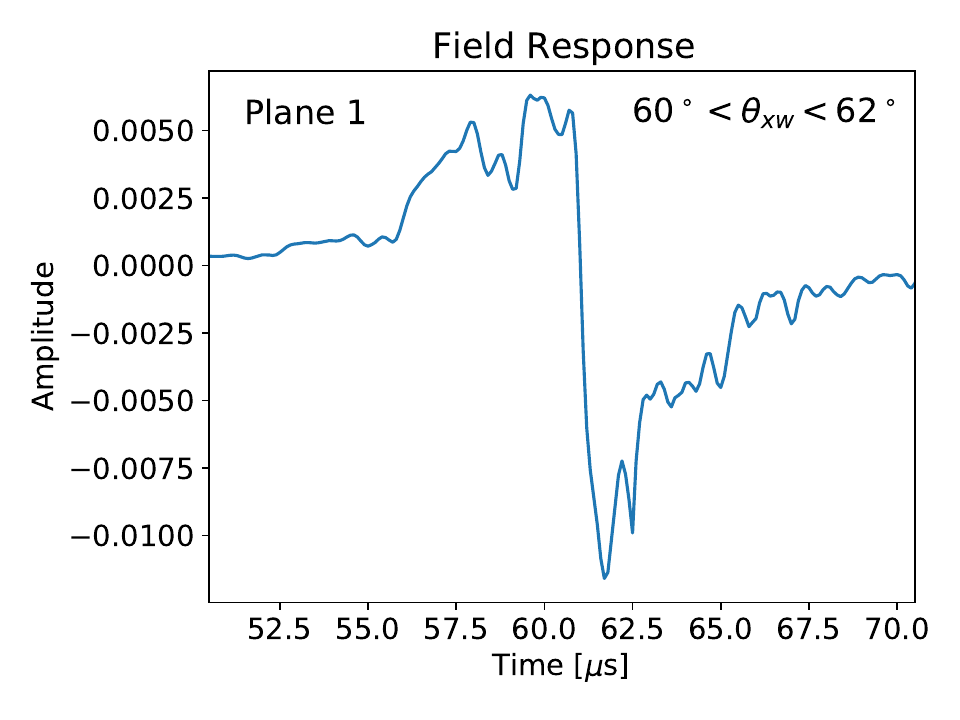}

    \includegraphics[width=0.32\textwidth]{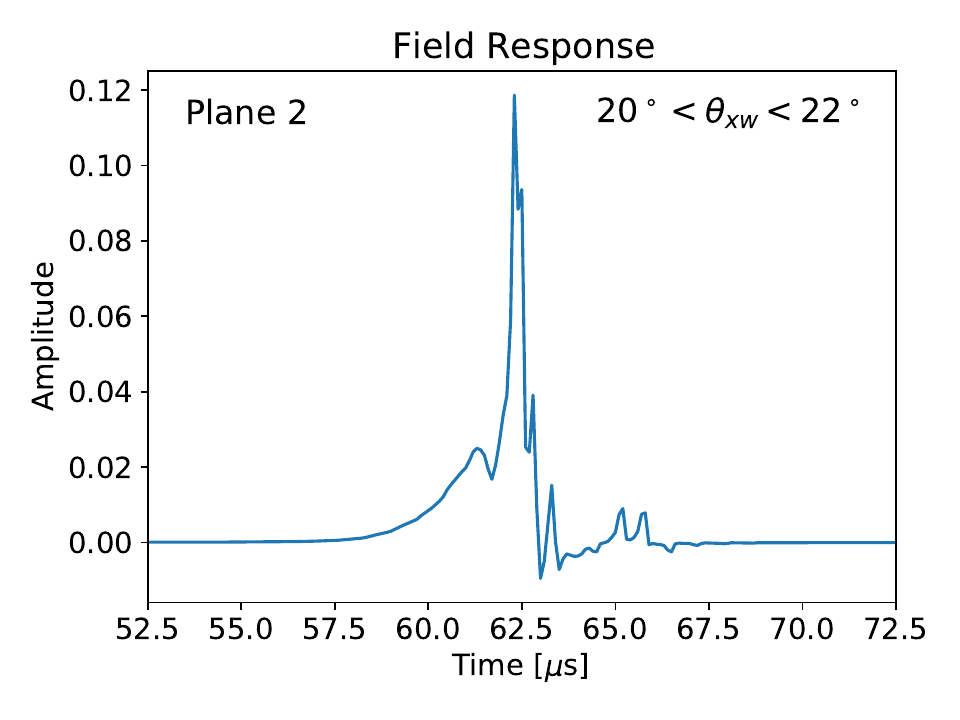}
    \includegraphics[width=0.32\textwidth]{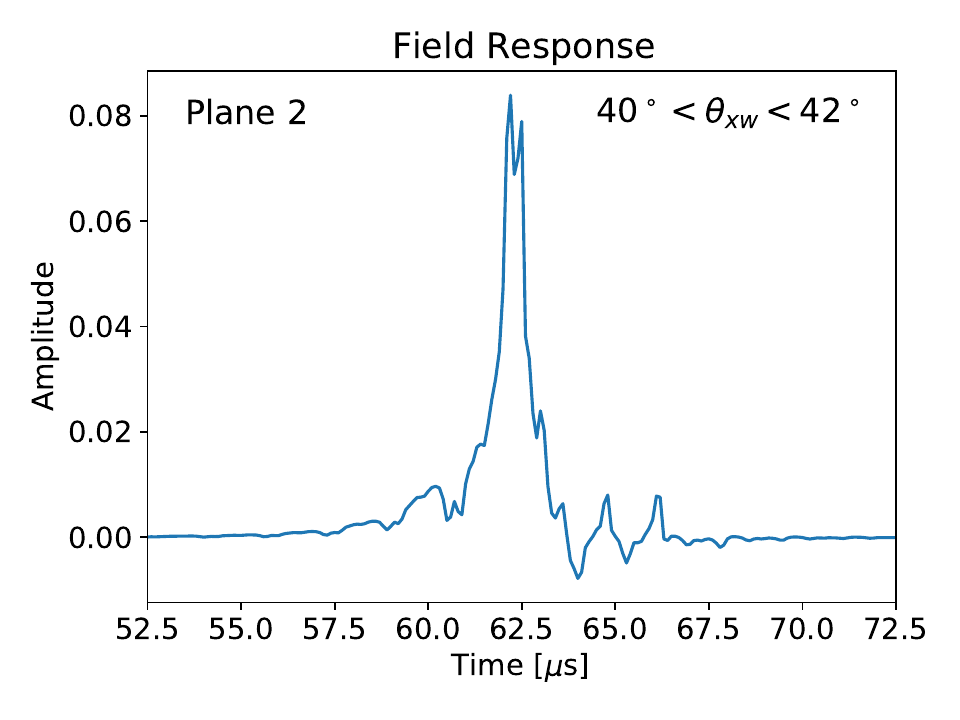}
    \includegraphics[width=0.32\textwidth]{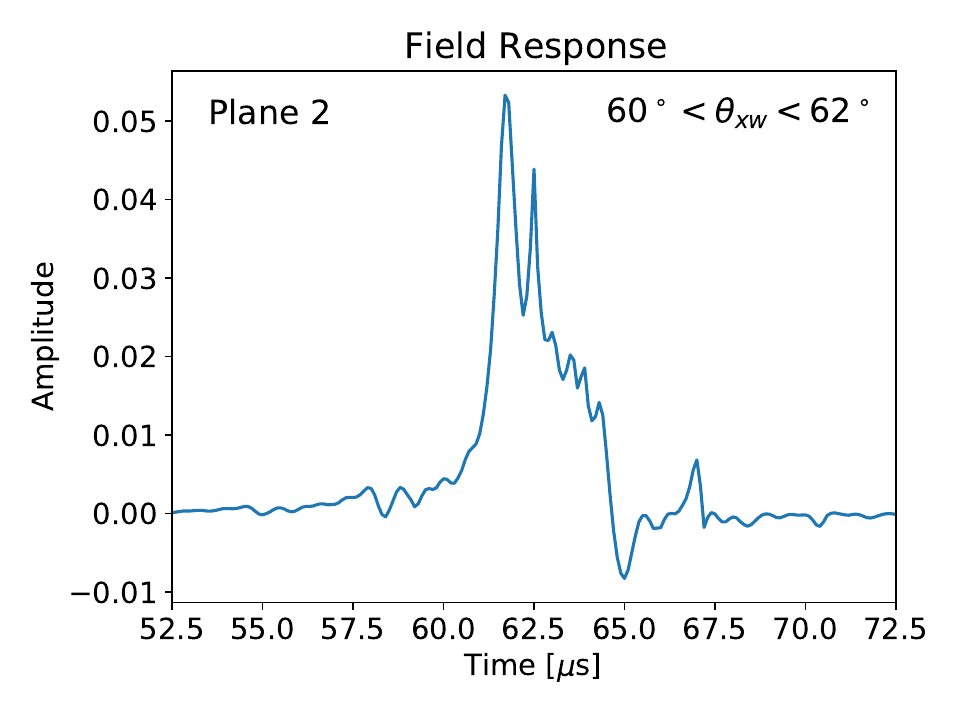}
    
    \caption{Track field response $f(\theta_{xw}; t)$, as defined in equation \ref{eq:interpPaths}. Shown for the front induction plane (top row), middle induction plane (middle row), and the collection plane (bottom row).}
    \label{fig:fieldResp}
\end{figure}

The model first turns the set of single electron field responses (201 total, each spaced \SI{0.3}{\milli\meter} apart) into an angle dependent ``track field response". This is done by sub-sampling the single electron field responses. Each sample linearly interpolates the responses on either side (as is done in Wire-Cell). The samples are shifted in time according to the chosen track angle and summed together. Given the single electron field responses $s_{-\SI{30}{\milli\meter}}(t), s_{-\SI{29.7}{\milli\meter}}(t), \ldots, s_{\SI{30}{\milli\meter}}(t)$ relative to the wire at time $t$, the track field response $f(\theta_{xw}; t)$ is equal to
\begin{equation}
    f(\theta_{xw}; t) = \sum\limits_{x_i} \left(1-\frac{x_i - \lfloor x_i \rfloor}{\SI{0.3}{\milli\meter}}\right)\cdot s_{\lfloor x_i \rfloor}\left(t - \frac{x_i \text{tan}\theta_{xw}}{v_D}\right) + \left(1- \frac{\lceil x_i \rceil - x_i}{\SI{0.3}{\milli\meter}}\right)\cdot s_{\lceil x_i \rceil}\left(t - \frac{x_i \text{tan}\theta_{xw}}{v_D}\right)\,,
    \label{eq:interpPaths}
\end{equation}
where $v_D$ is the drift velocity, $x_i$ are the sampled locations, $\lfloor x \rfloor$ is the position immediately below $x$ of a sampled single electron field response, and $\lceil x \rceil$ is the position immediately above $x$ of a sampled single electron field response. In our implementation, we sub-sampled the single electron field responses every \SI{0.03}{\milli\meter} for 2001 sub-samples. The sampled nominal field responses shifted by a few example track angles is shown in figure \ref{fig:interpPaths}. The sum of these samples (i.e., the track field response $f(\theta_{xw}; t)$) is shown for a few example track angles in figure \ref{fig:fieldResp}. The un-physical spikes in the field responses are caused by the finite sampling spacing of the single electron field responses and are smoothed out by the electronics response, as specified below.

The track field response is convolved with the electronics response $e(t)$ to obtain the track signal shape. In addition, the measurement of the signal shape will not perfectly reproduce the signal. The alignment of signals from different muons in the averaged waveform will not be perfect due to detector noise. The resolution from this misalignment smears the shape. (Other broadening effects, such as diffusion, are insignificant.) To account for this effect, the signal shape is convolved with a ``measurement kernel" $m(\theta_{xw}; t)$. The final measured track signal shape $S(\theta_{xw}; t)$ is thus equal to
\begin{equation}
    S(\theta_{xw}; t) = \left(f(\theta_{xw}) \circledast e \circledast m(\theta_{xw})\right)(t)\,,
    \label{eq:signalshape_model}
\end{equation}
where $\circledast$ denotes a convolution. 

\begin{figure}[t]
    \centering
    \includegraphics[width=0.49\textwidth]{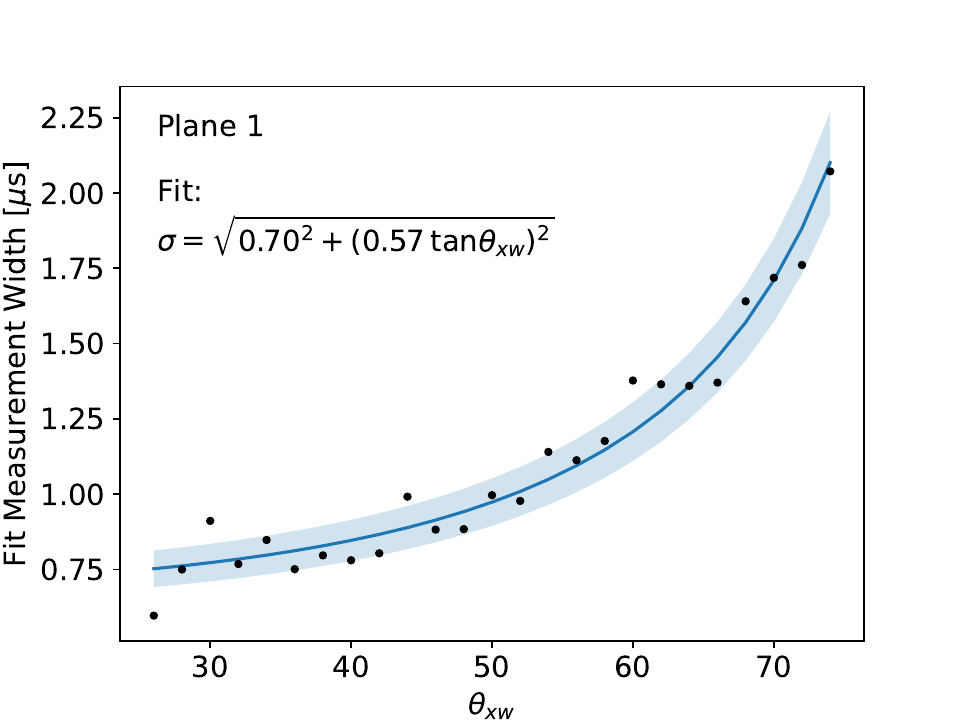}
    \includegraphics[width=0.49\textwidth]{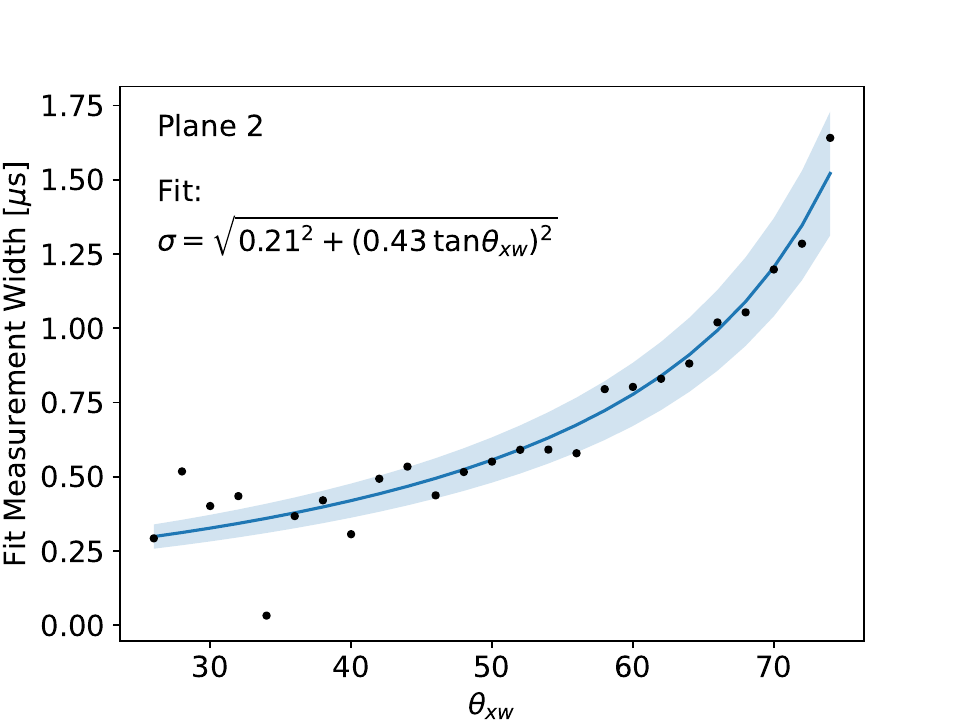}
    \caption{Broadening from resolution on track alignment in the signal shape measurement procedure as measured in ICARUS simulation. The width of the broadening is plotted as a function of the track angle which determines the signal shape, $\theta_{xw}$.}
    \label{fig:signalshape_mkernel}
\end{figure}

The measurement kernel is determined from a fit on ICARUS simulation where the underlying field and electronics response is known. It is found to be well approximated by a Gaussian with a width $\sigma$ depending on the track angle by a form $\sigma\left(\theta_{xw}\right) = \sqrt{a^2 + b^2 \text{tan}^2\theta_{xw}}$, where $a$ and $b$ are parameters individual to each wire plane. The measurement kernel width as determined in Monte Carlo simulation on each wire plane is shown in figure \ref{fig:signalshape_mkernel} for the middle induction and collection planes, on which the fit is performed.

The fit to data is done by matching the measured signal shape $S$ to the data by fitting the electronics response $e$ and the single electron field responses $s_i$ (implicit in the track field response $f$). In this fit, non-linear transformations parameterized by the fit are applied to the nominal field and electronic responses. The details of these transformations are in appendix \ref{sec:SignalShapeAppendix}. The fit is done on all measured track angles at once. 

\subsubsection{Fits}
\label{sec:signalshape_fits}

\begin{figure}[]
    \centering
    \includegraphics[width=0.24\textwidth]{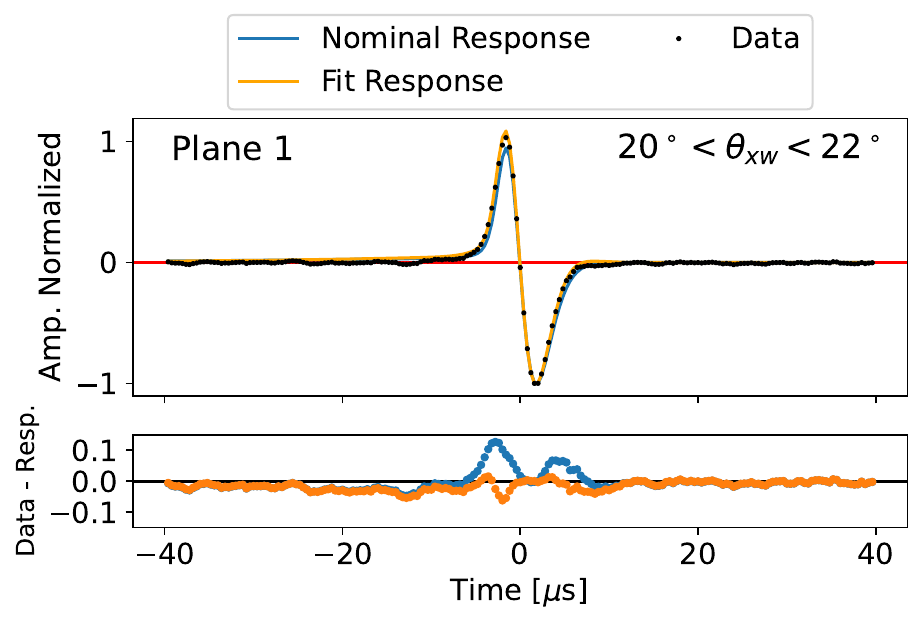}
    \includegraphics[width=0.24\textwidth]{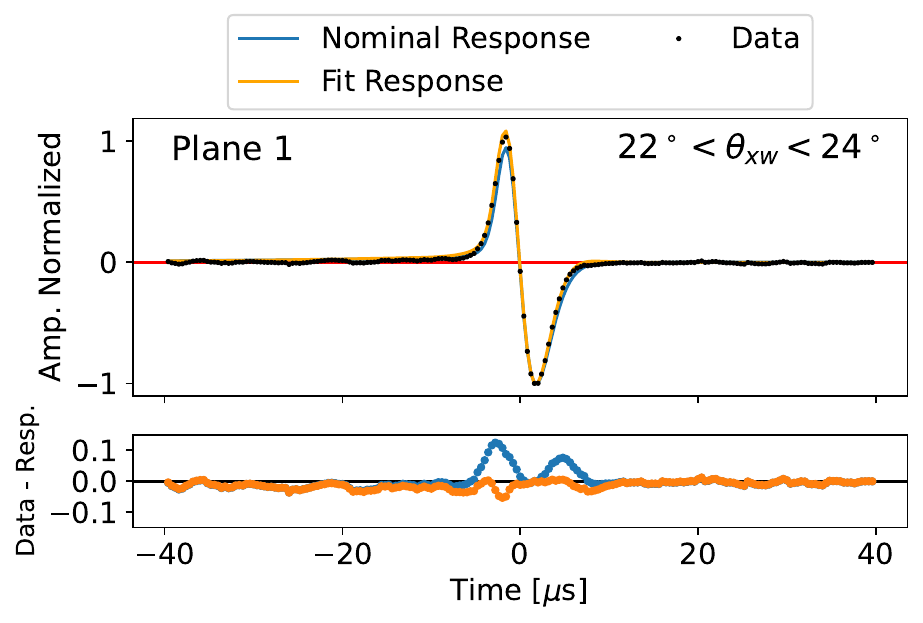}
    \includegraphics[width=0.24\textwidth]{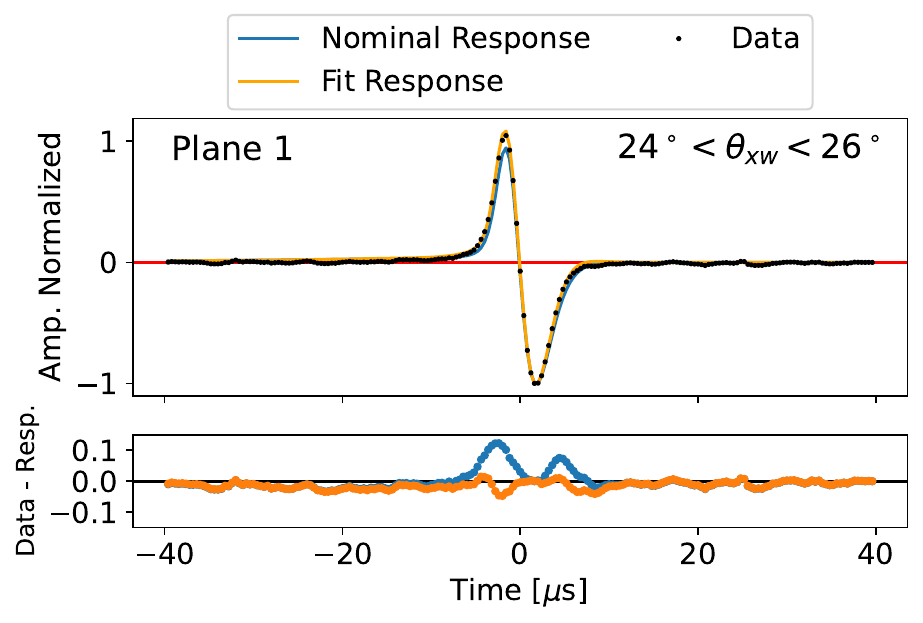}
    \includegraphics[width=0.24\textwidth]{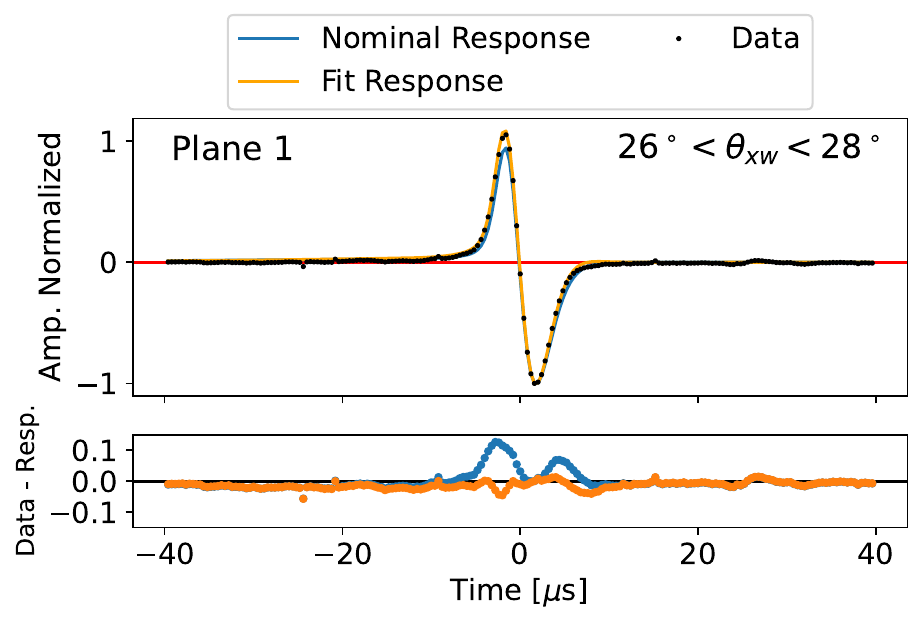}
    \includegraphics[width=0.24\textwidth]{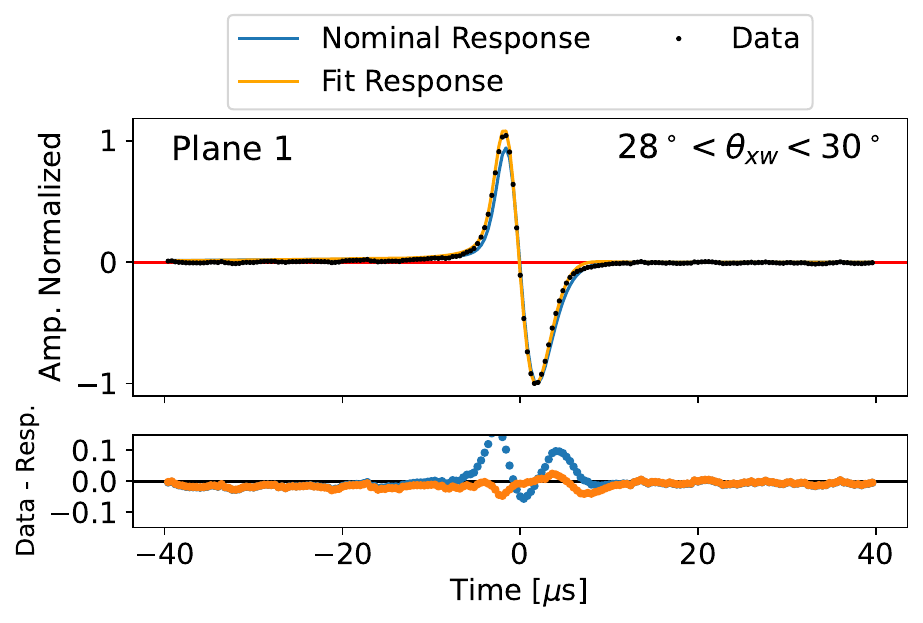}
    \includegraphics[width=0.24\textwidth]{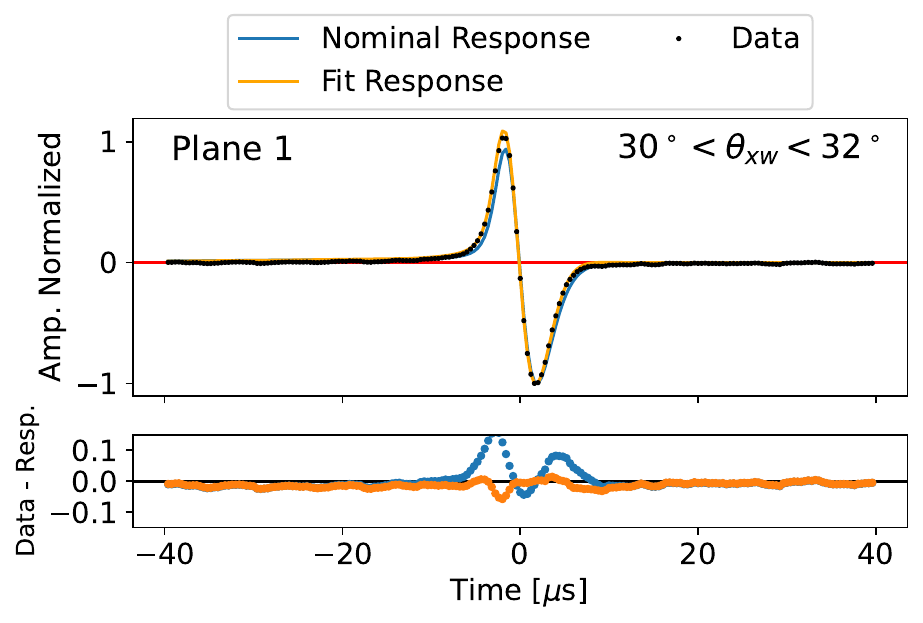}
    \includegraphics[width=0.24\textwidth]{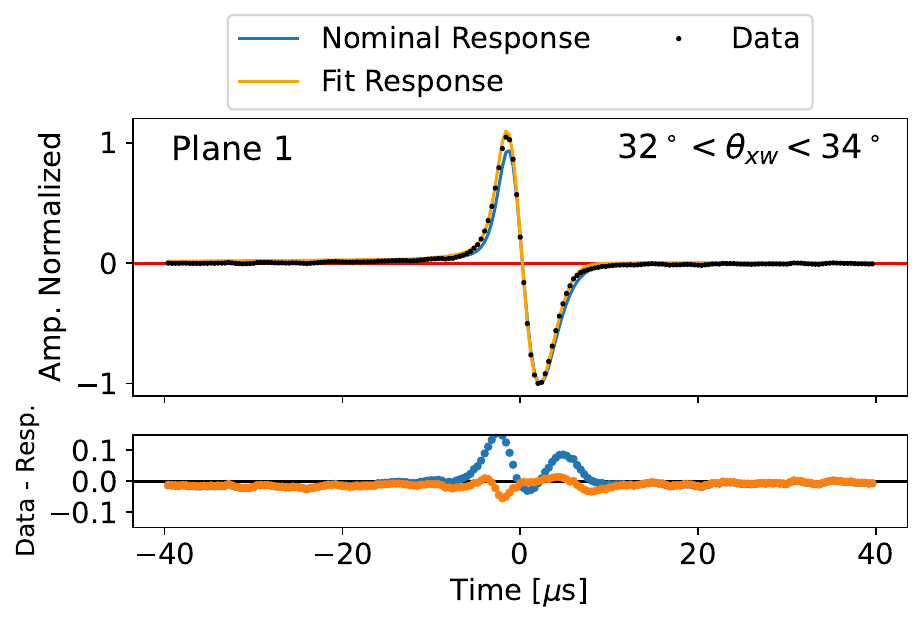}
    \includegraphics[width=0.24\textwidth]{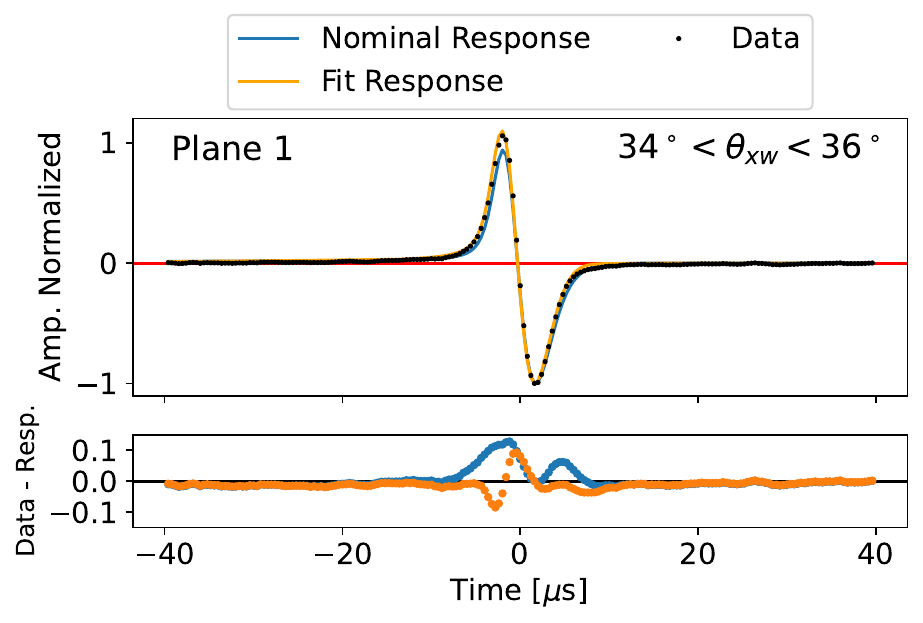}
    \includegraphics[width=0.24\textwidth]{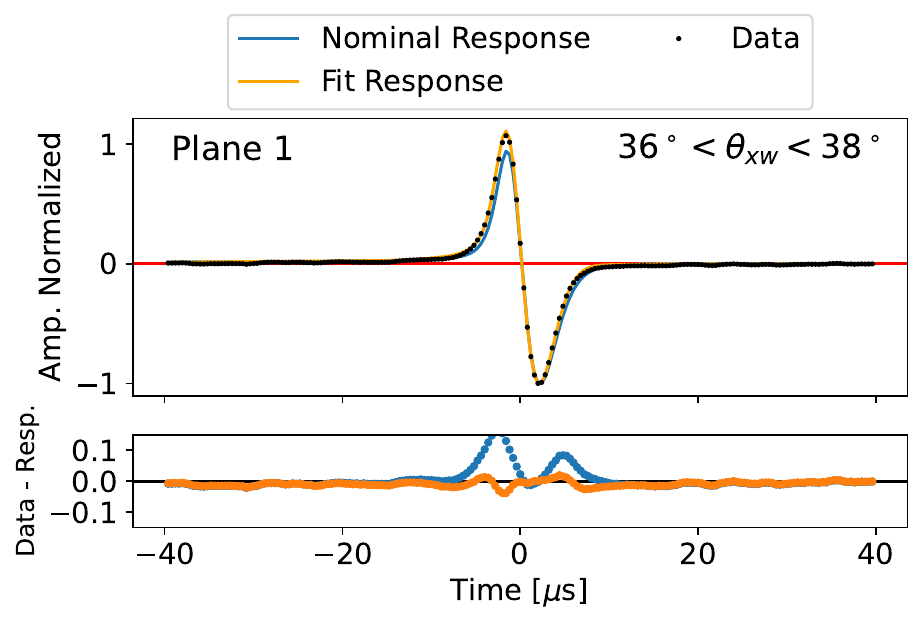}
    \includegraphics[width=0.24\textwidth]{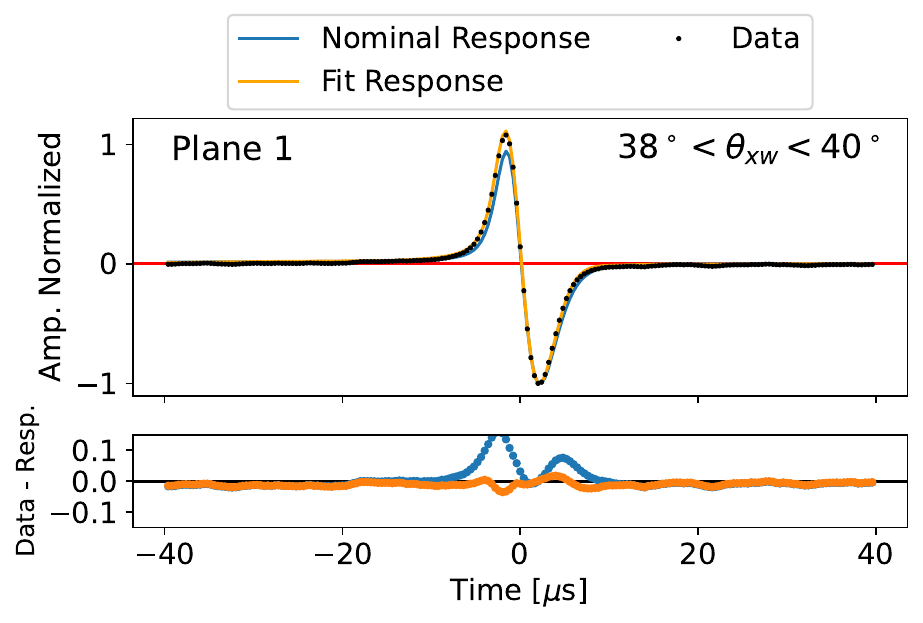}
    \includegraphics[width=0.24\textwidth]{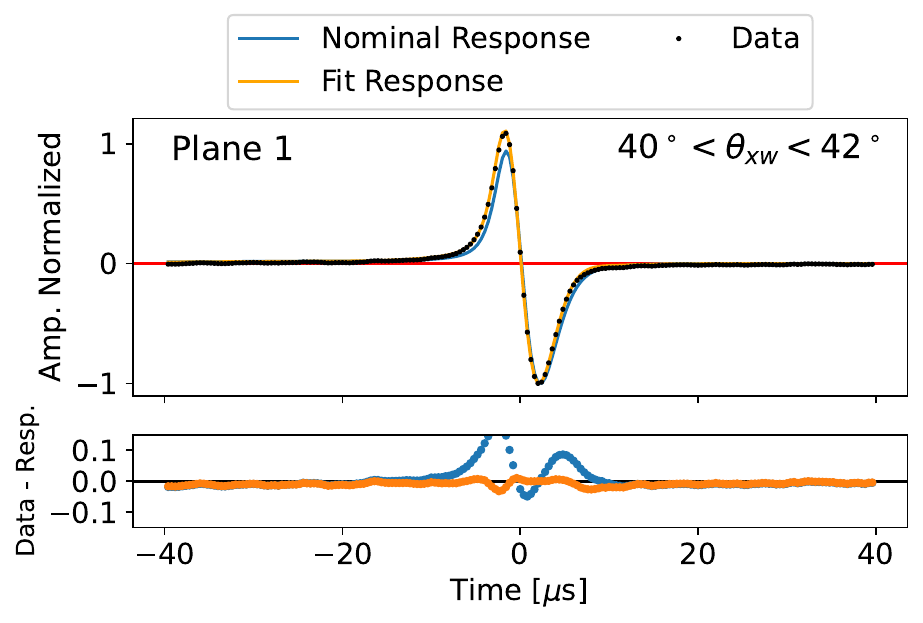}
    \includegraphics[width=0.24\textwidth]{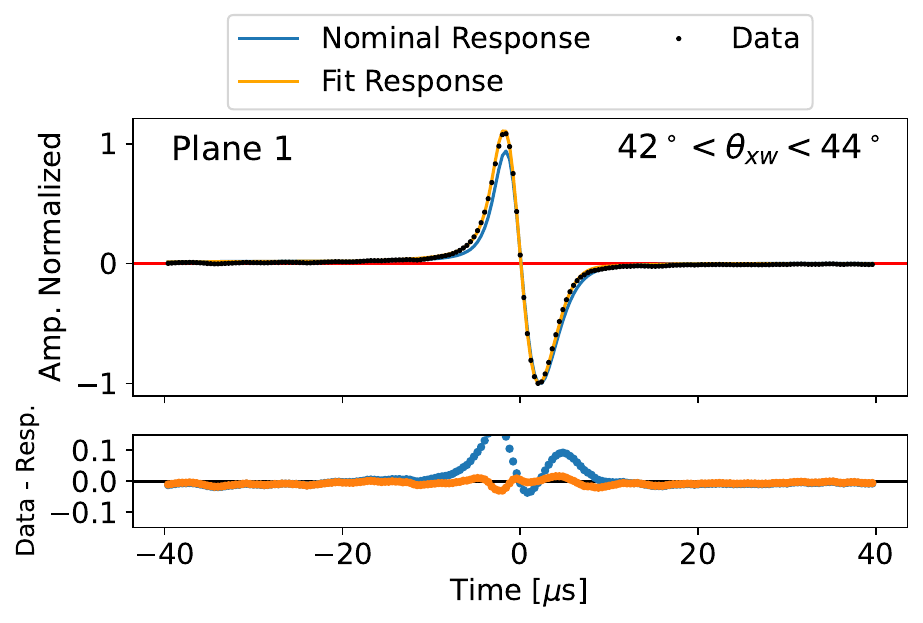}
    \includegraphics[width=0.24\textwidth]{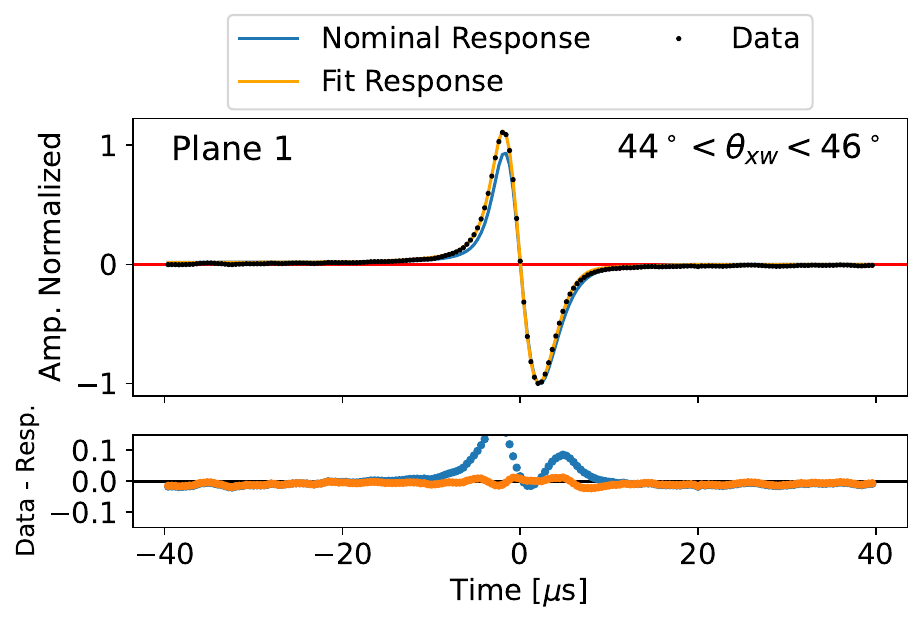}
    \includegraphics[width=0.24\textwidth]{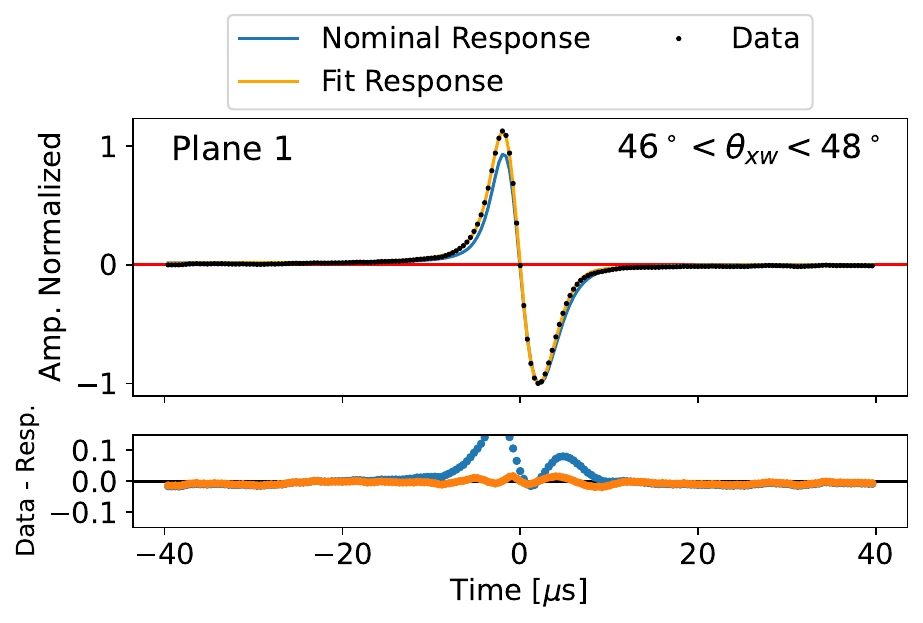}
    \includegraphics[width=0.24\textwidth]{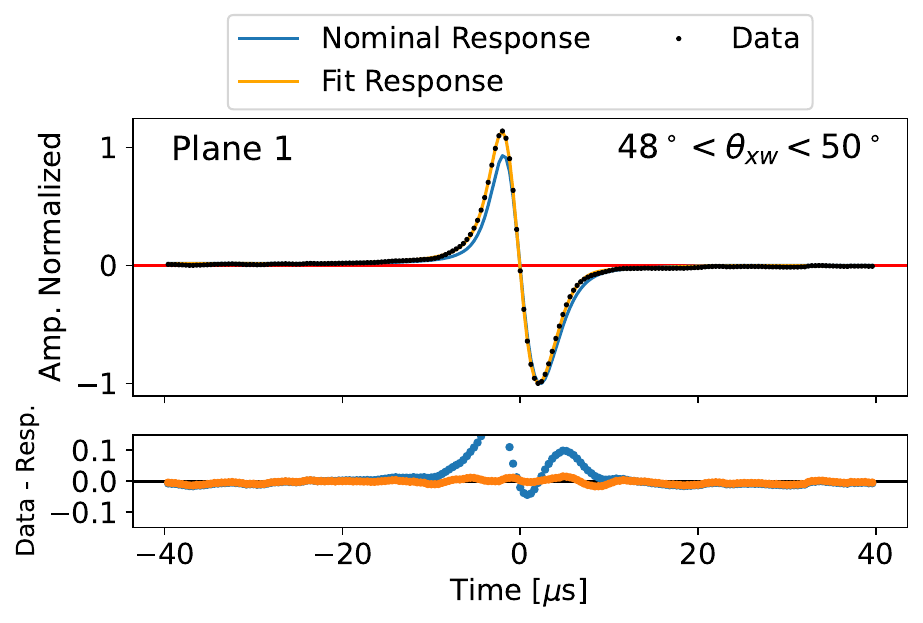}
    \includegraphics[width=0.24\textwidth]{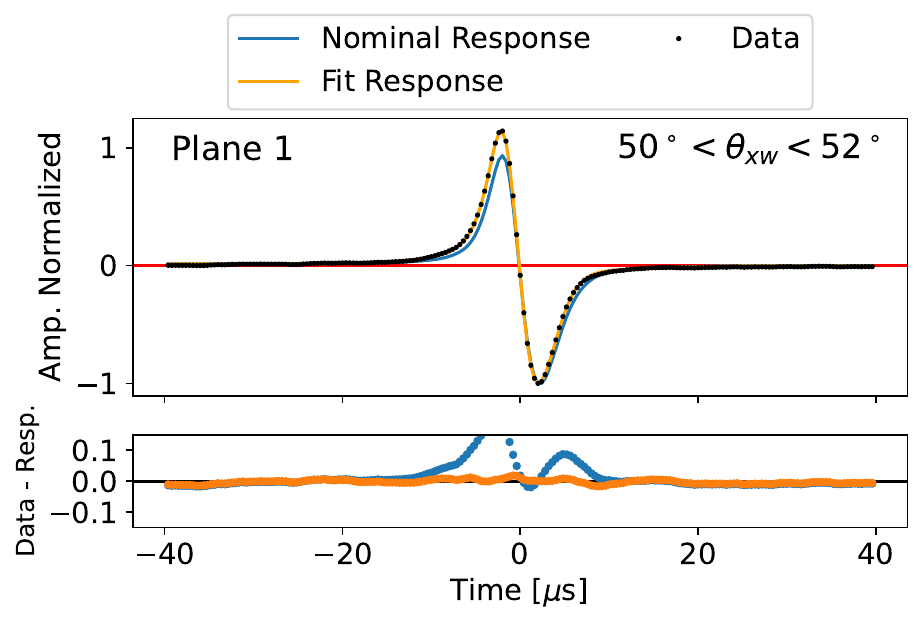}
    \includegraphics[width=0.24\textwidth]{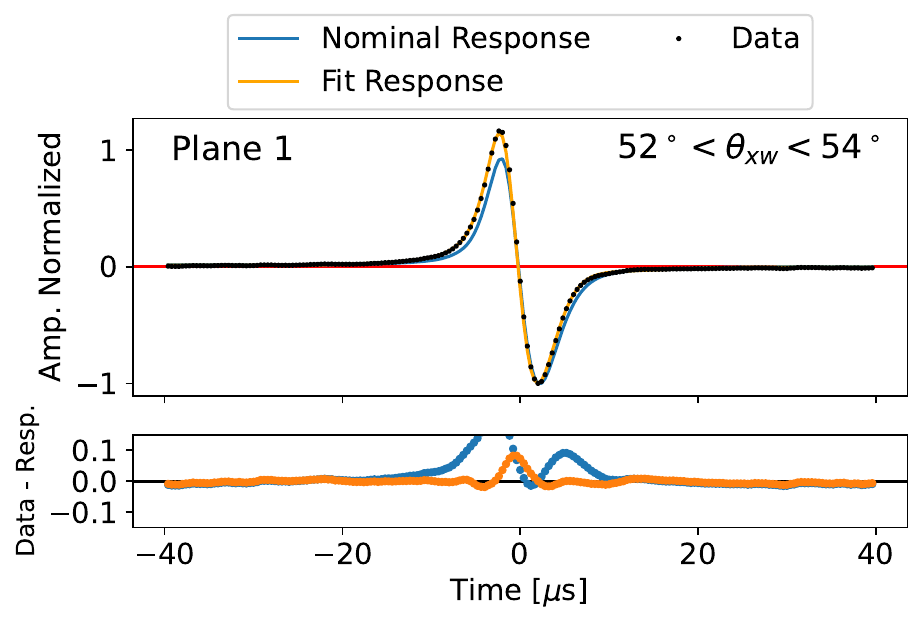}
    \includegraphics[width=0.24\textwidth]{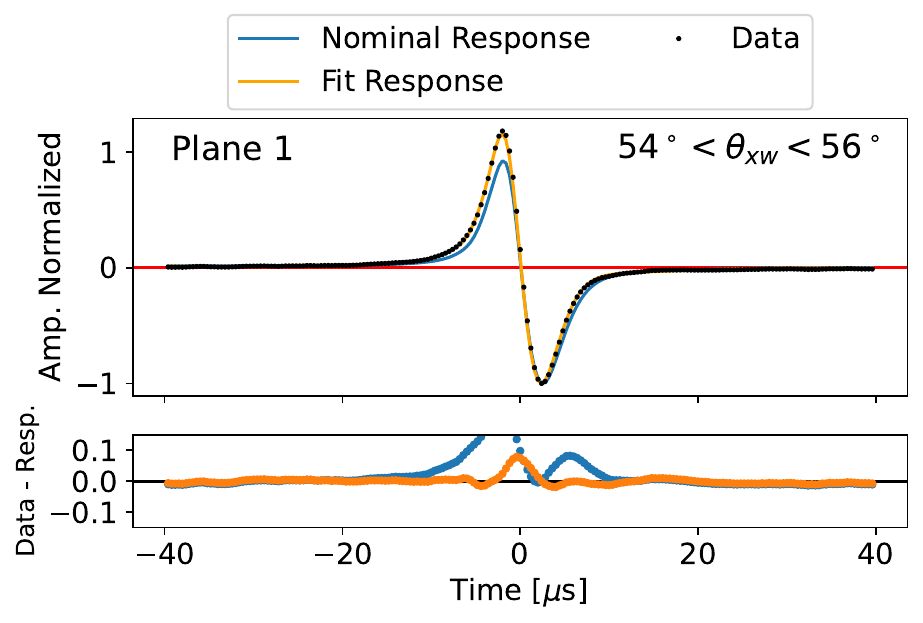}
    \includegraphics[width=0.24\textwidth]{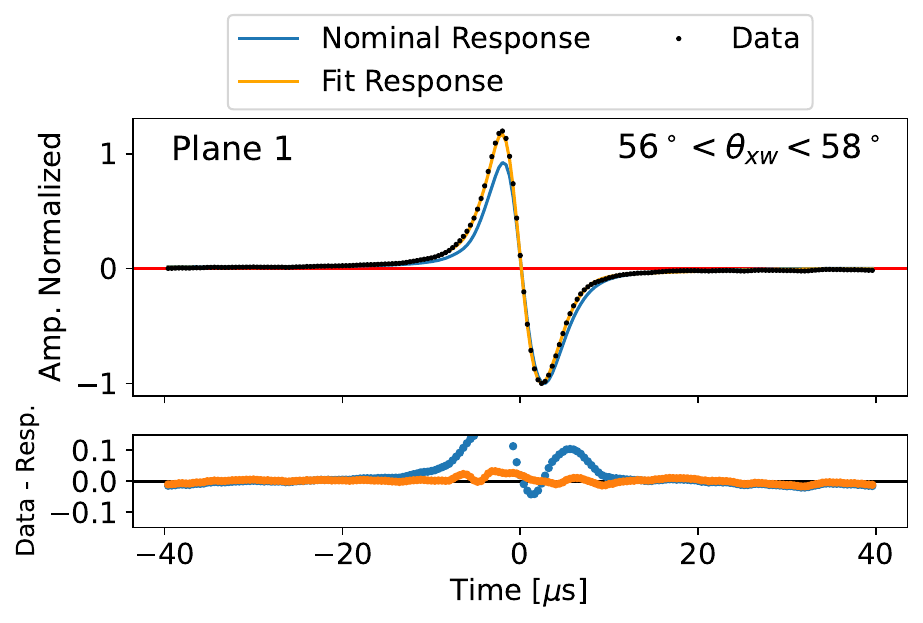}
    \includegraphics[width=0.24\textwidth]{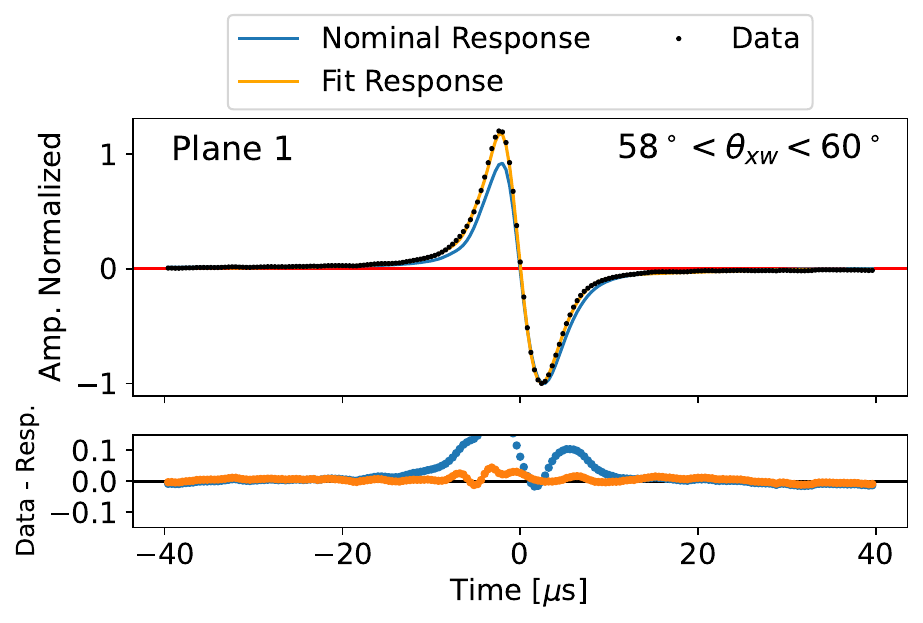}
    \includegraphics[width=0.24\textwidth]{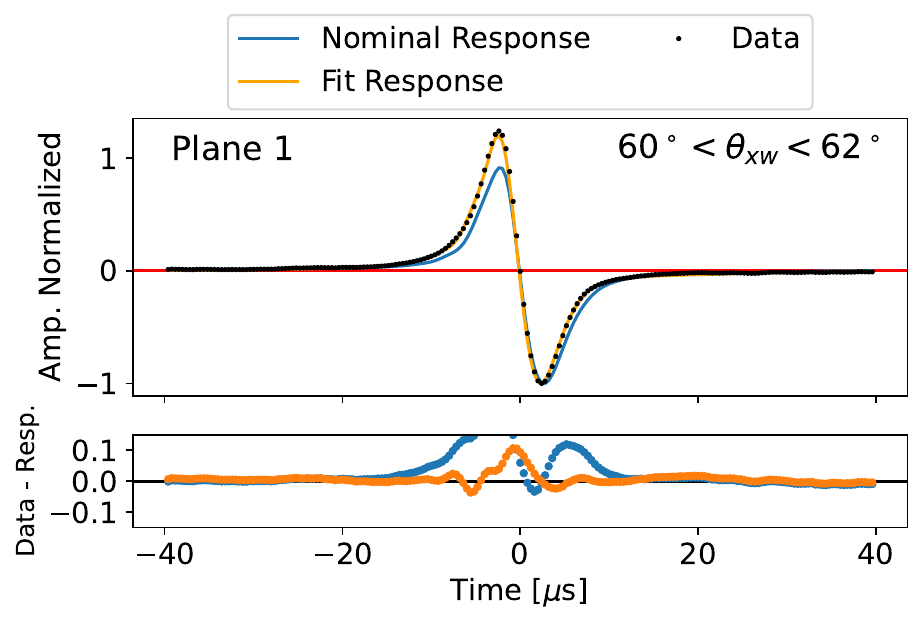}
    \includegraphics[width=0.24\textwidth]{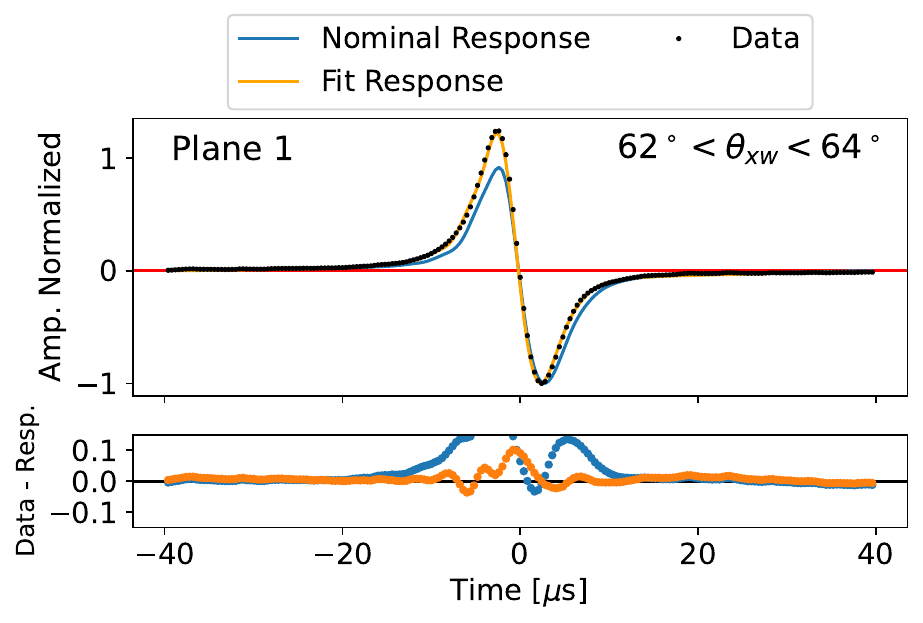}
    \includegraphics[width=0.24\textwidth]{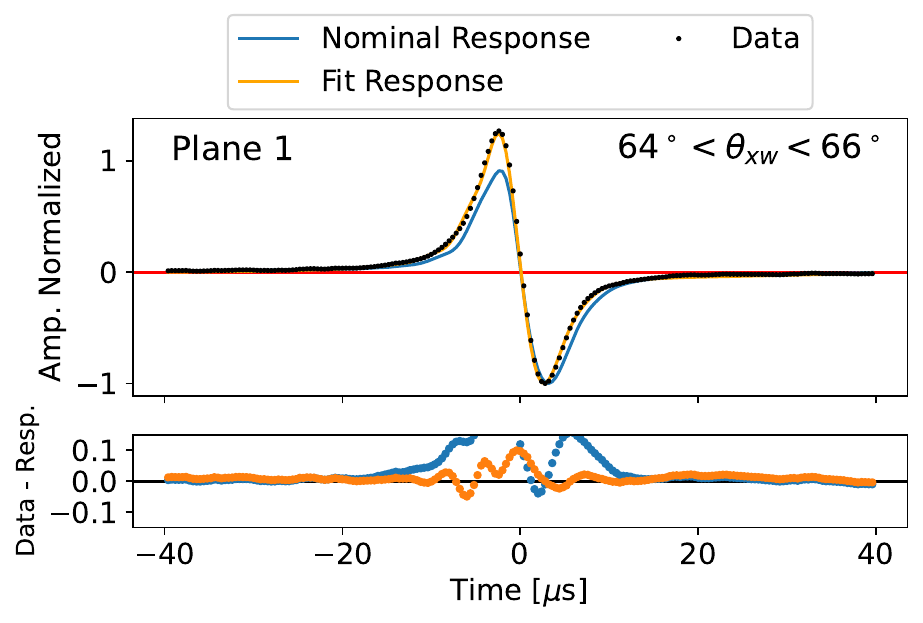}
    \includegraphics[width=0.24\textwidth]{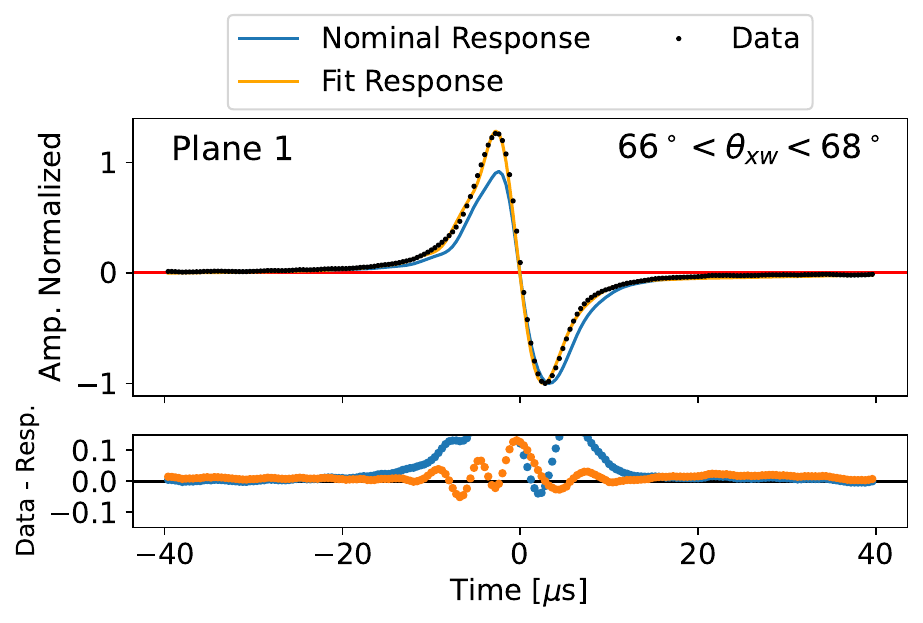}
    \includegraphics[width=0.24\textwidth]{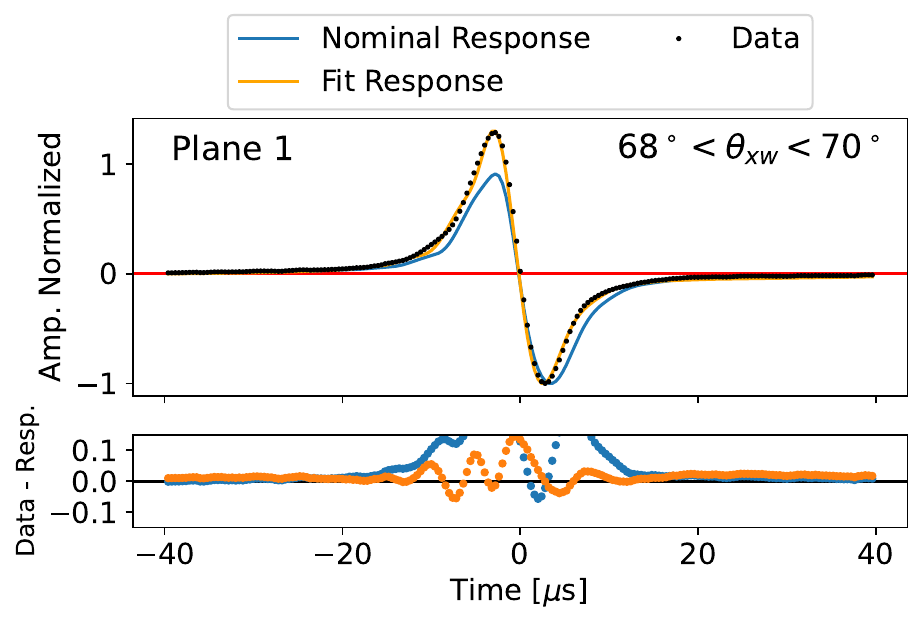}
    \includegraphics[width=0.24\textwidth]{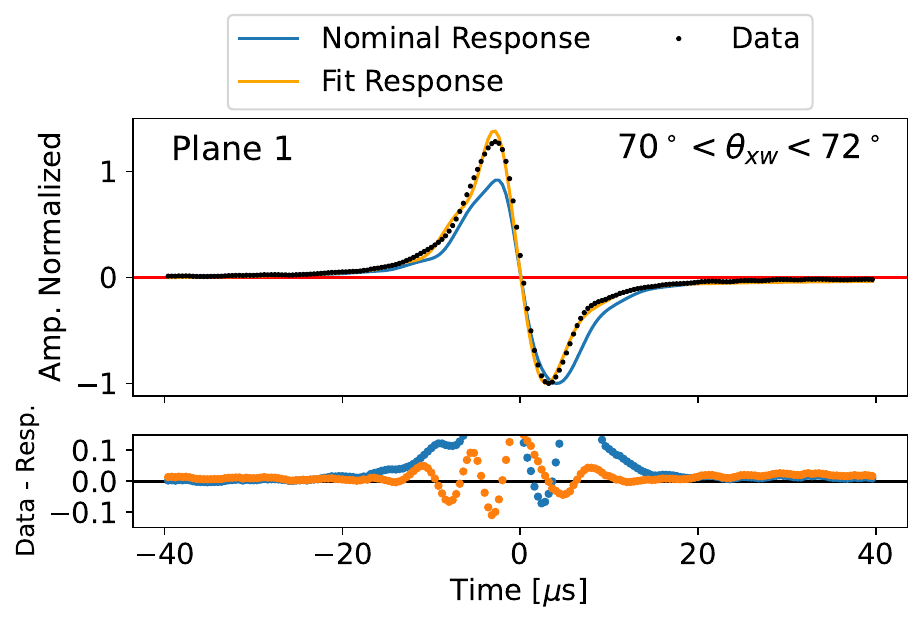}
    \includegraphics[width=0.24\textwidth]{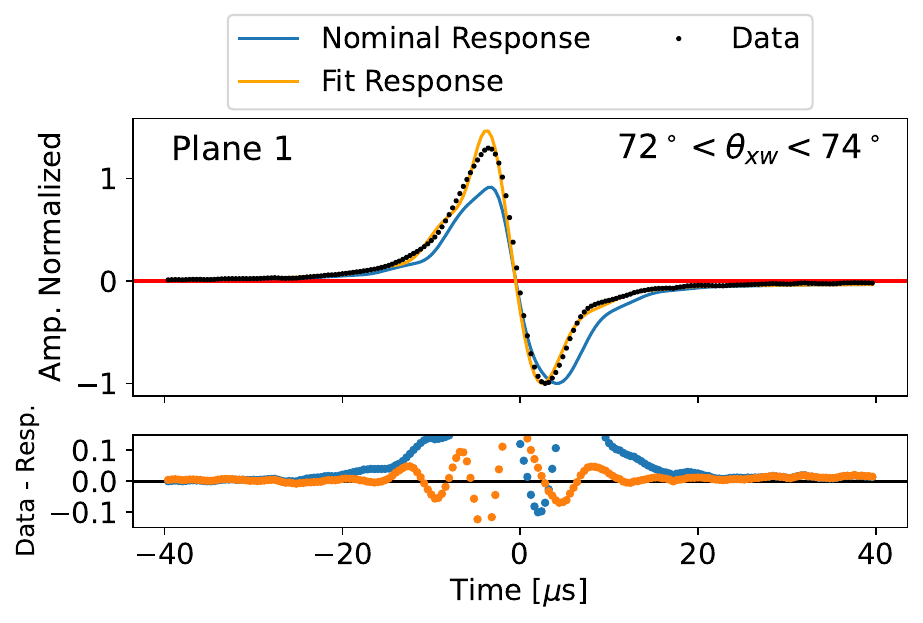}
    \includegraphics[width=0.24\textwidth]{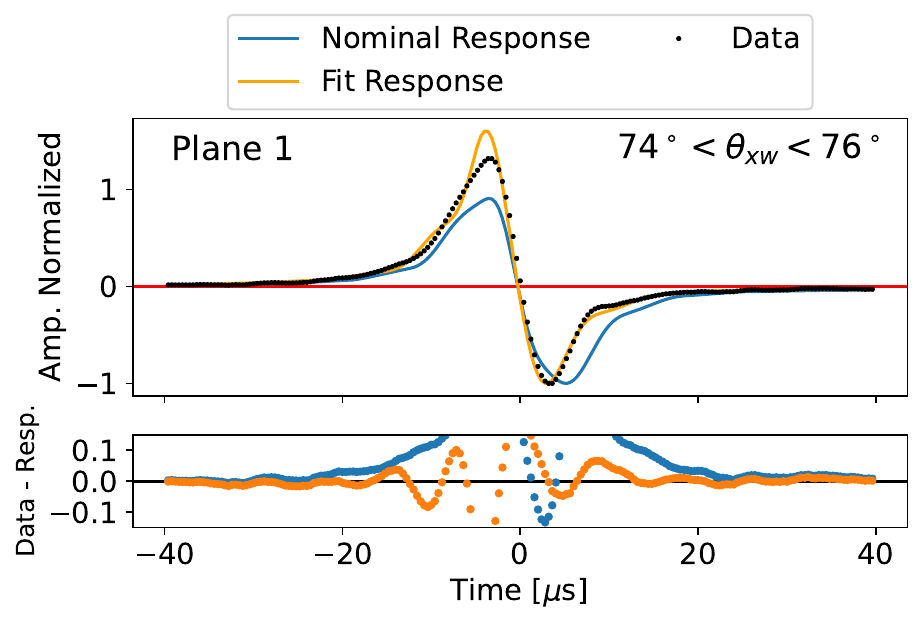}
    \caption{Signal shape fits on the middle induction plane. Each plot shows one angle bin between $20^\circ$ and $76^\circ$. The blue curve is the nominal ICARUS signal shape and the orange is the result of the fit.}
    \label{fig:signalshape_datafit_Plane1}
\end{figure}

\begin{figure}[]
    \centering
    \includegraphics[width=0.24\textwidth]{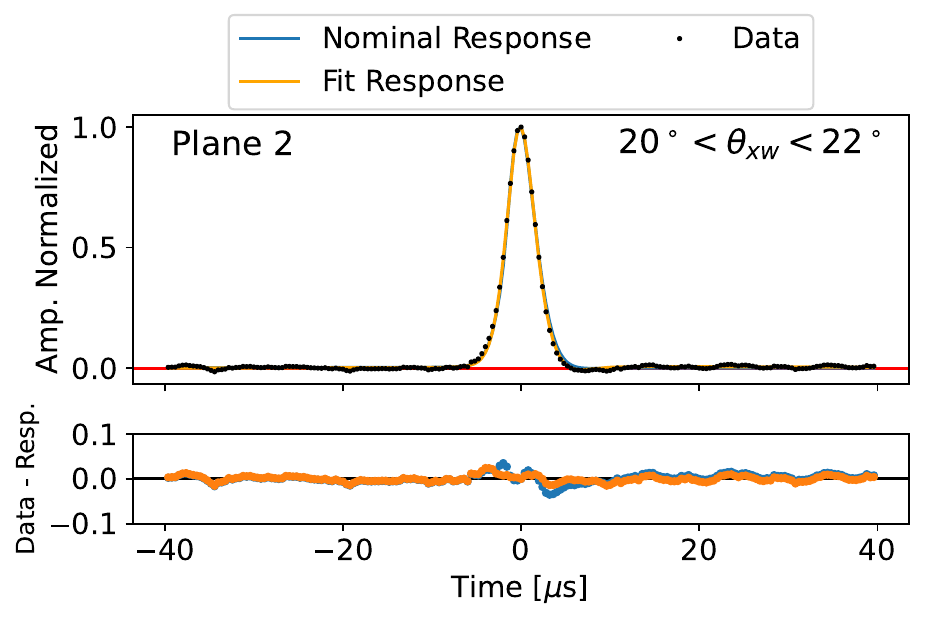}
    \includegraphics[width=0.24\textwidth]{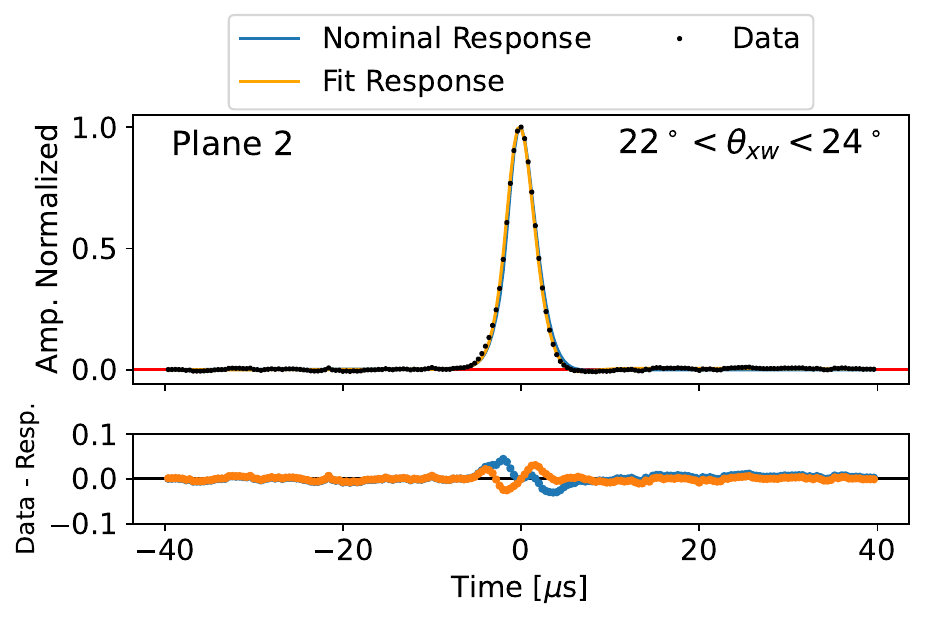}
    \includegraphics[width=0.24\textwidth]{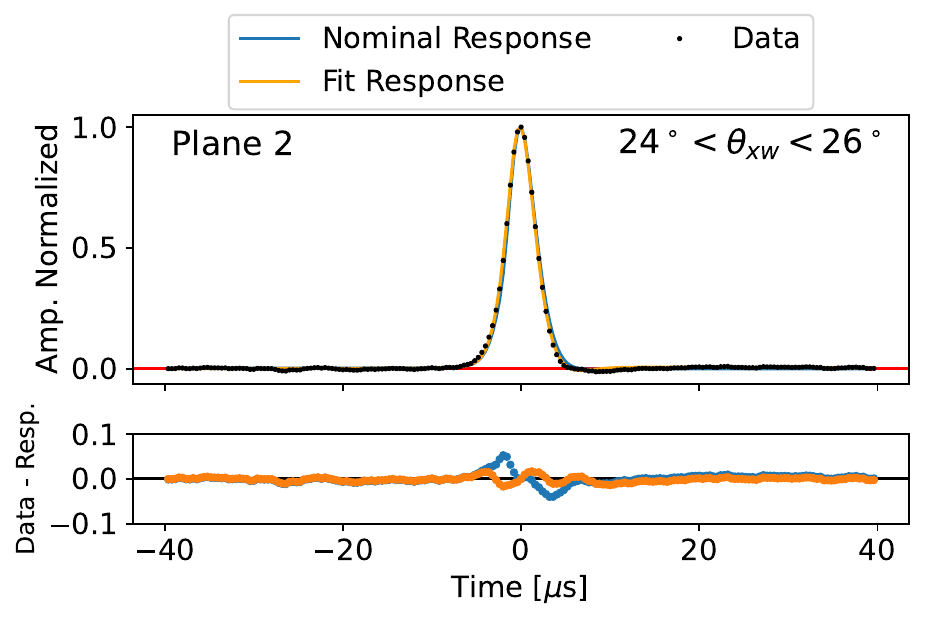}
    \includegraphics[width=0.24\textwidth]{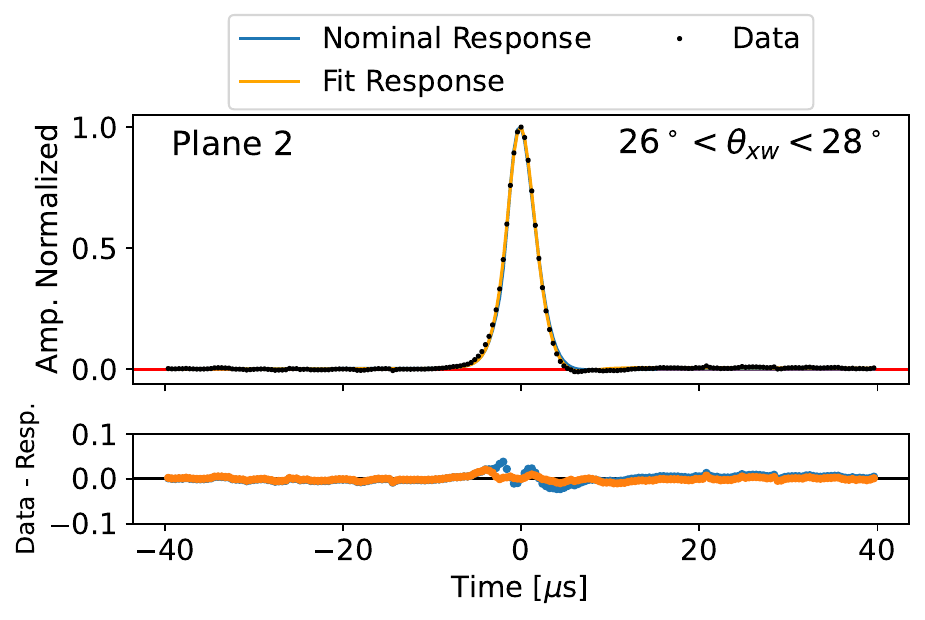}
    \includegraphics[width=0.24\textwidth]{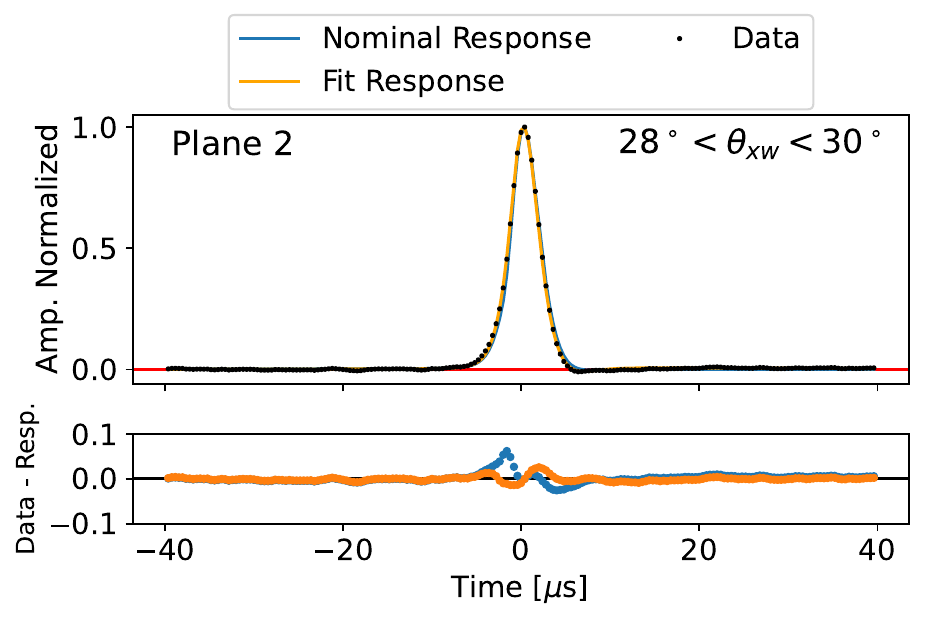}
    \includegraphics[width=0.24\textwidth]{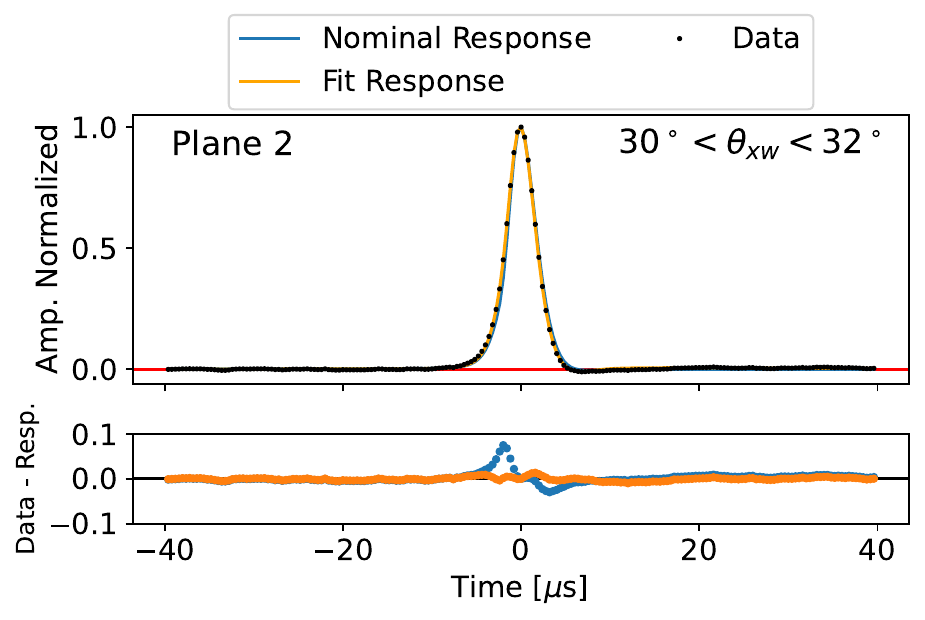}
    \includegraphics[width=0.24\textwidth]{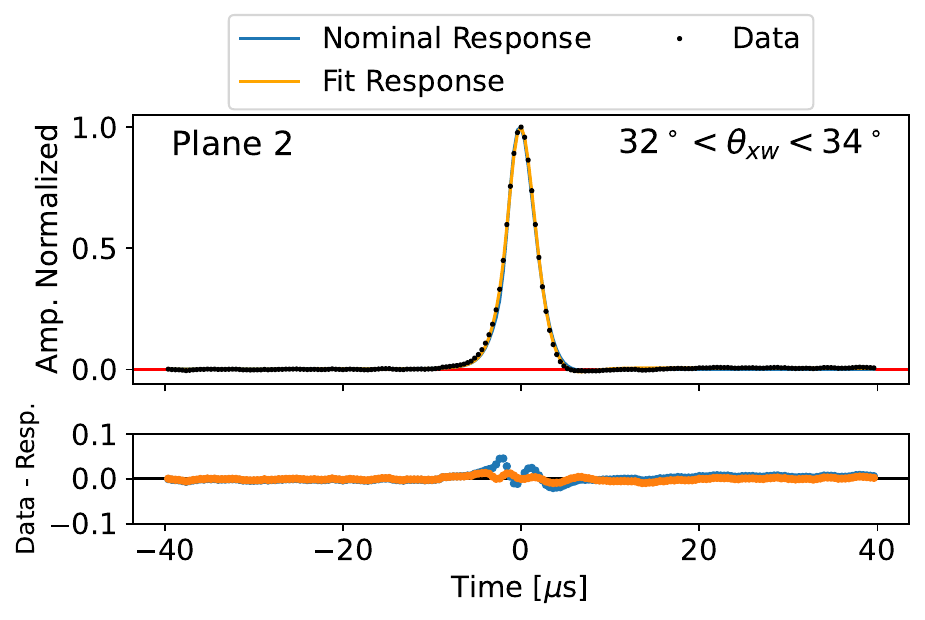}
    \includegraphics[width=0.24\textwidth]{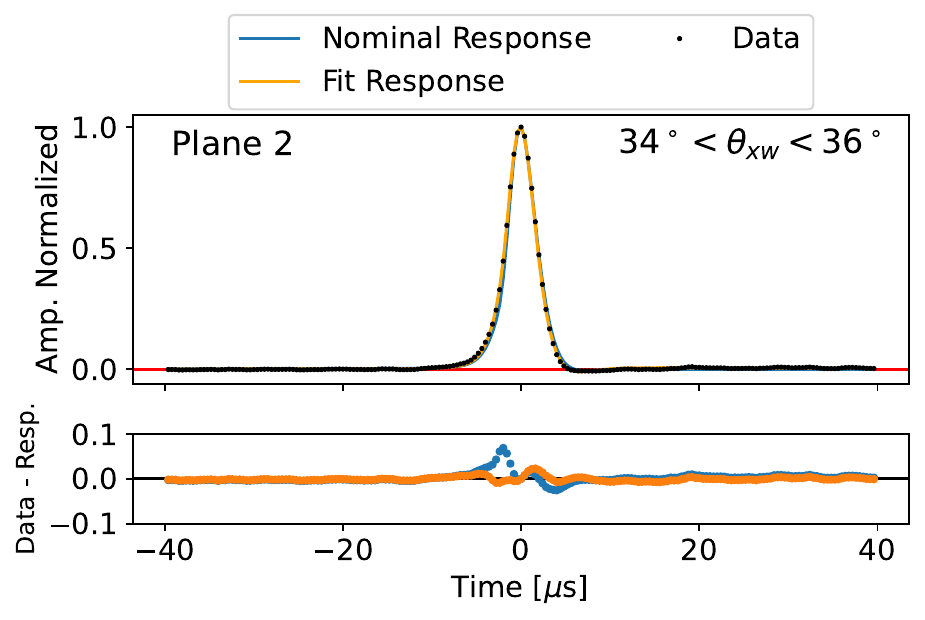}
    \includegraphics[width=0.24\textwidth]{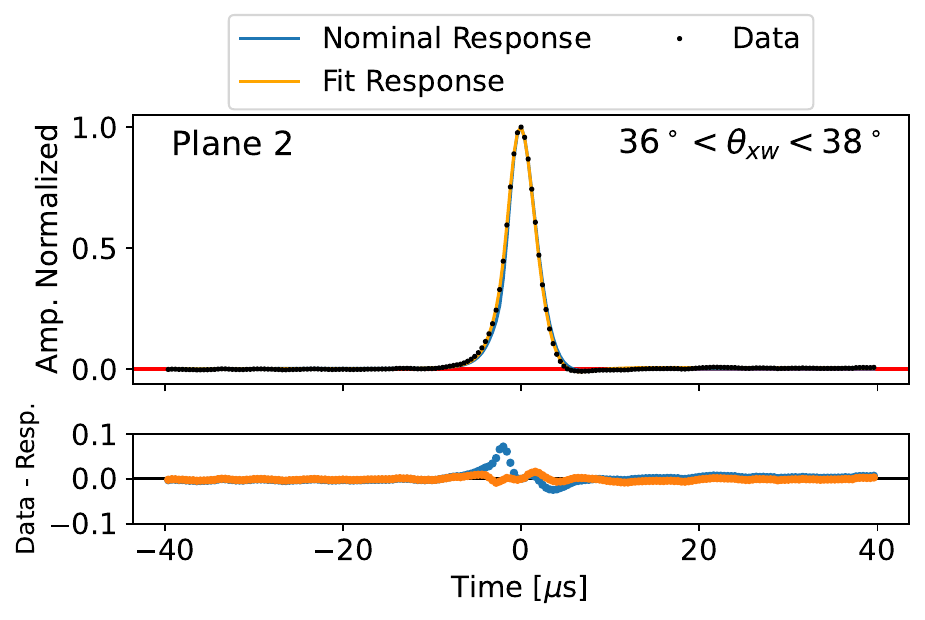}
    \includegraphics[width=0.24\textwidth]{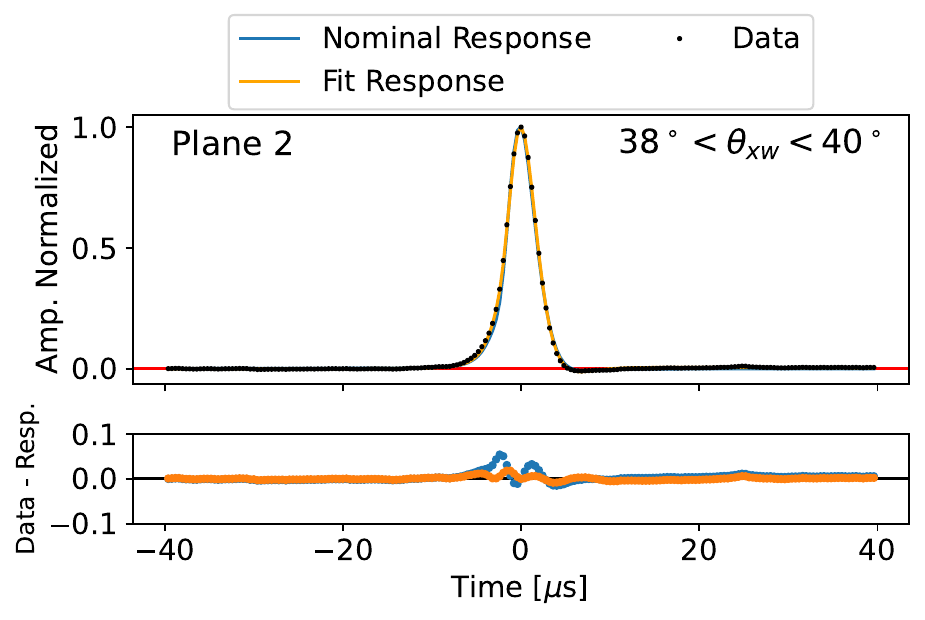}
    \includegraphics[width=0.24\textwidth]{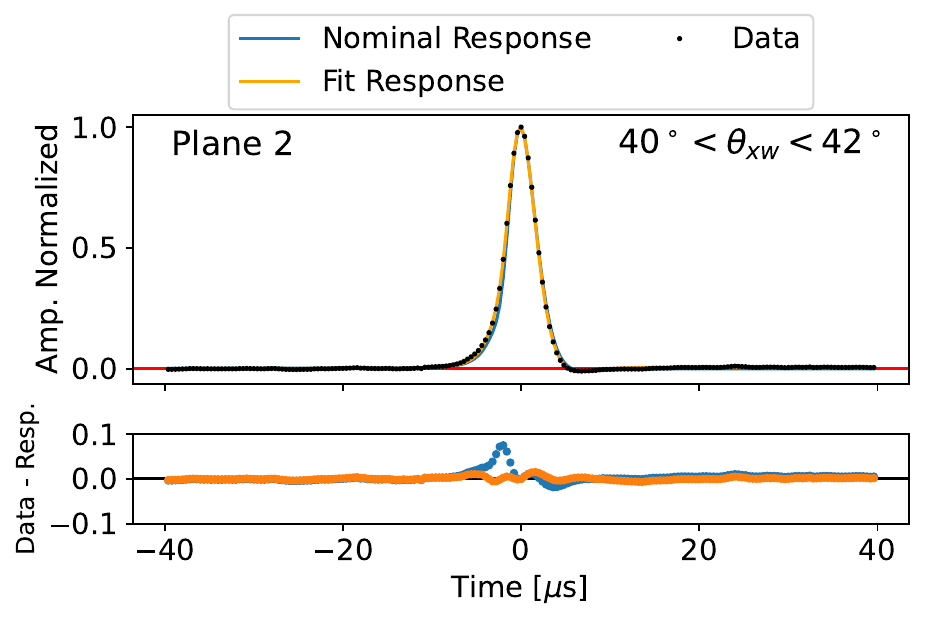}
    \includegraphics[width=0.24\textwidth]{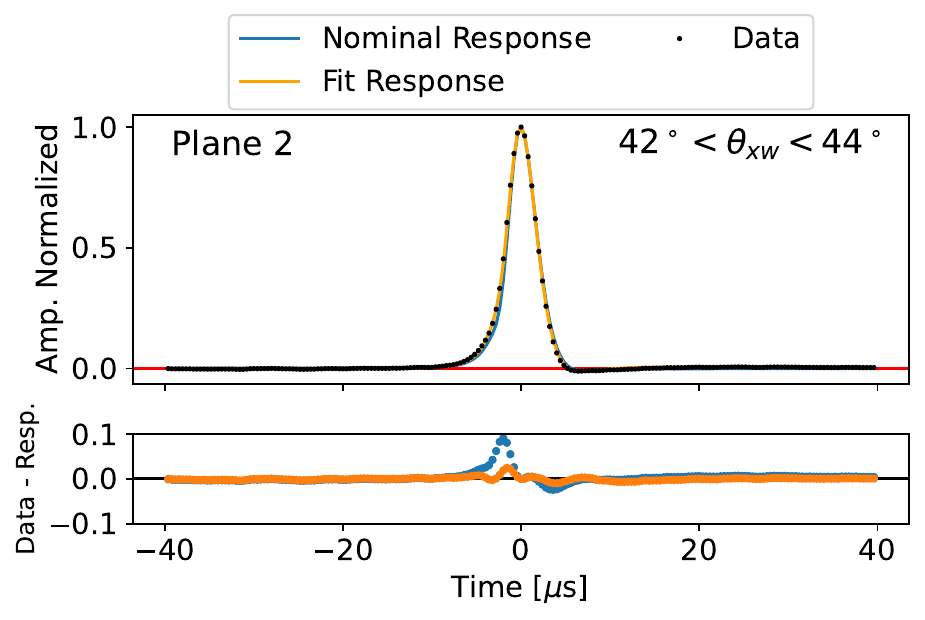}
    \includegraphics[width=0.24\textwidth]{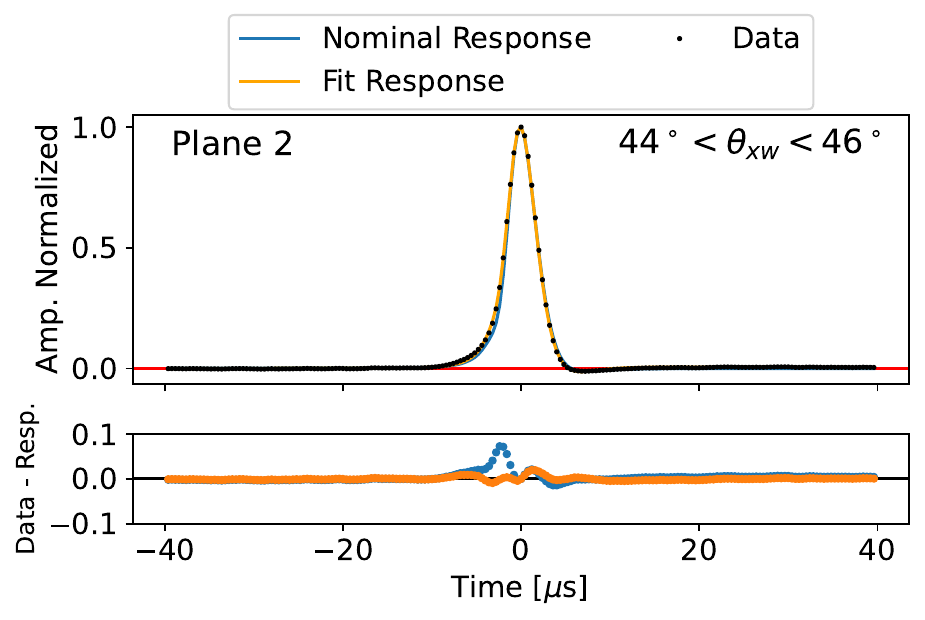}
    \includegraphics[width=0.24\textwidth]{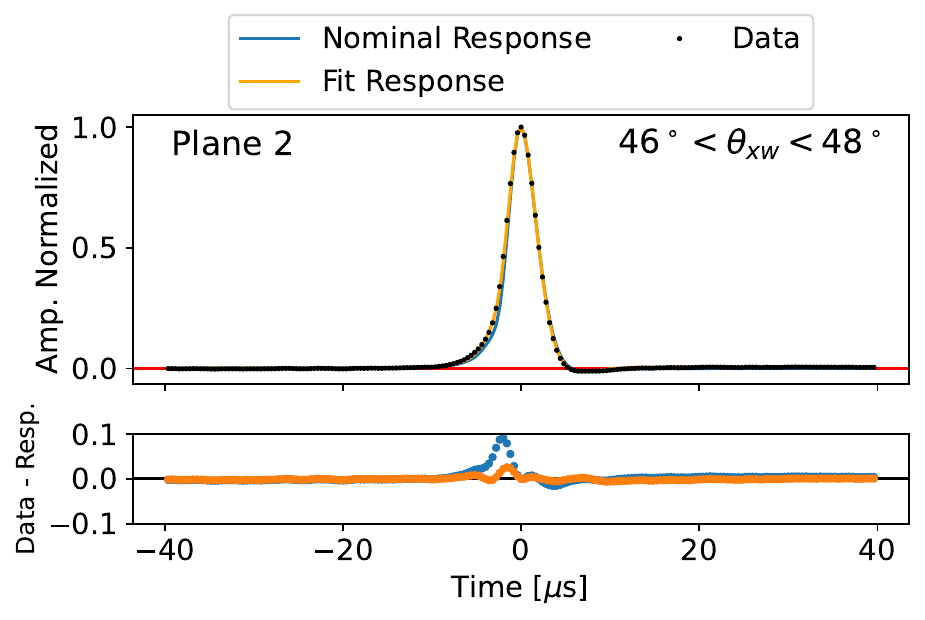}
    \includegraphics[width=0.24\textwidth]{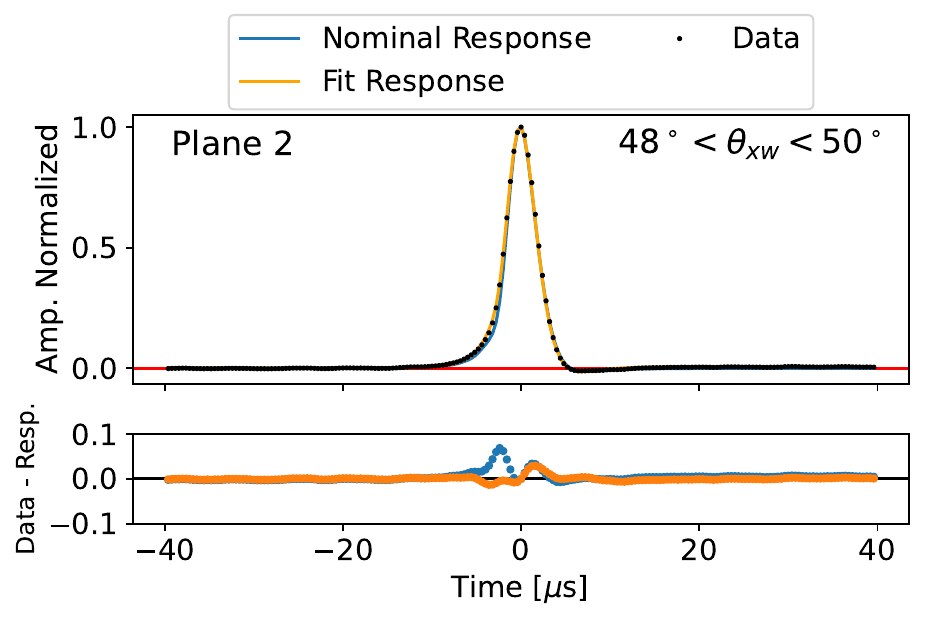}
    \includegraphics[width=0.24\textwidth]{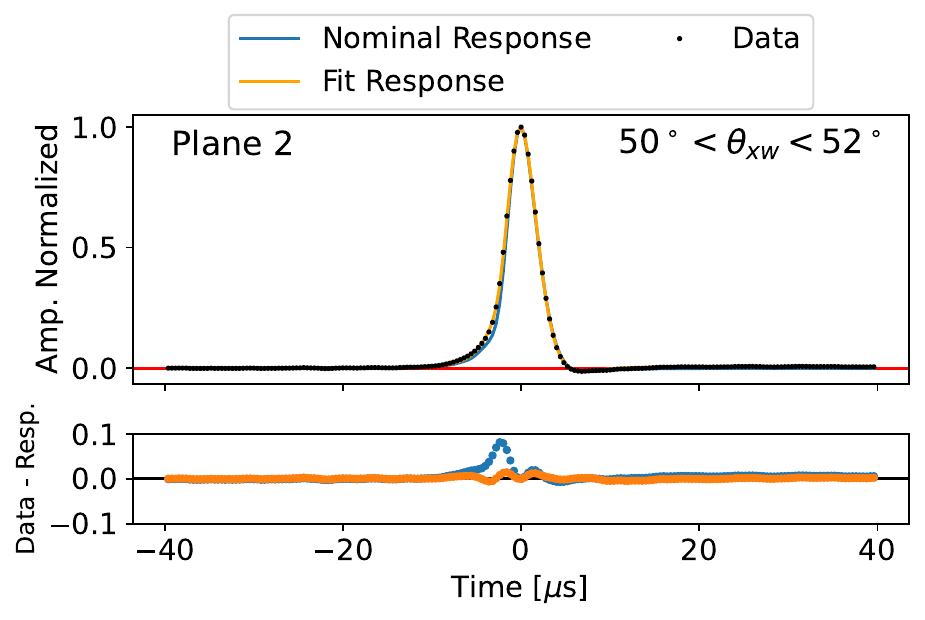}
    \includegraphics[width=0.24\textwidth]{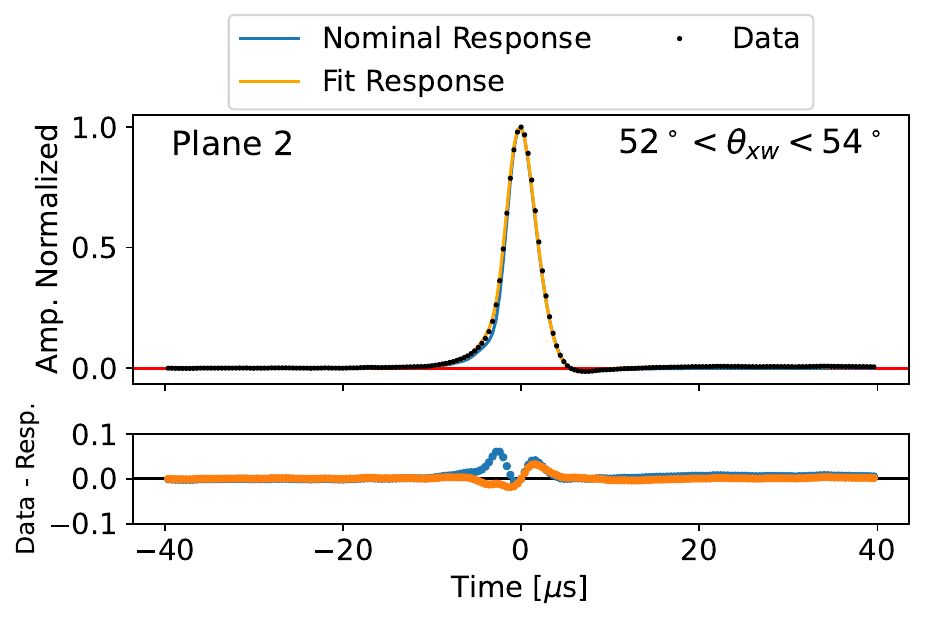}
    \includegraphics[width=0.24\textwidth]{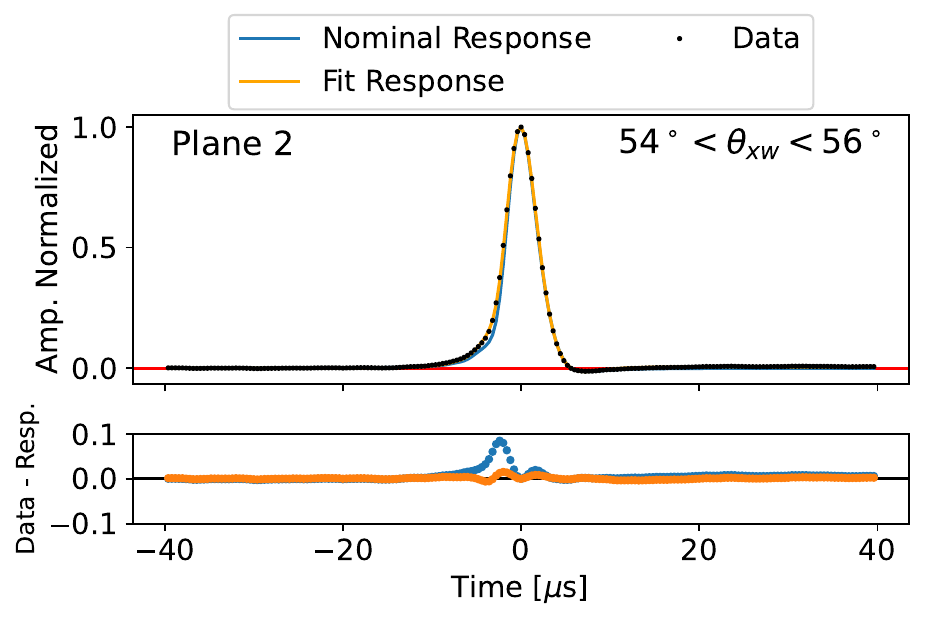}
    \includegraphics[width=0.24\textwidth]{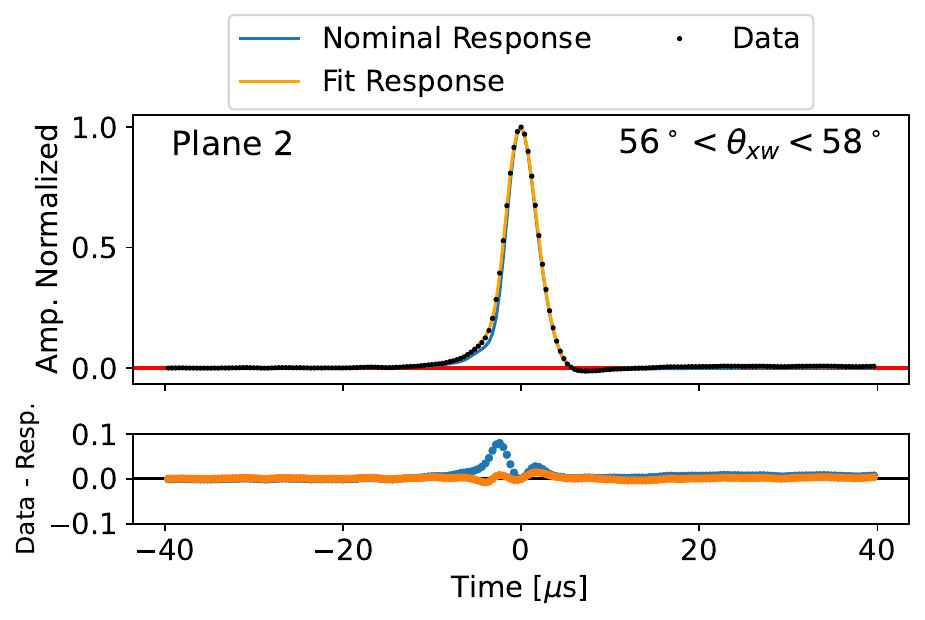}
    \includegraphics[width=0.24\textwidth]{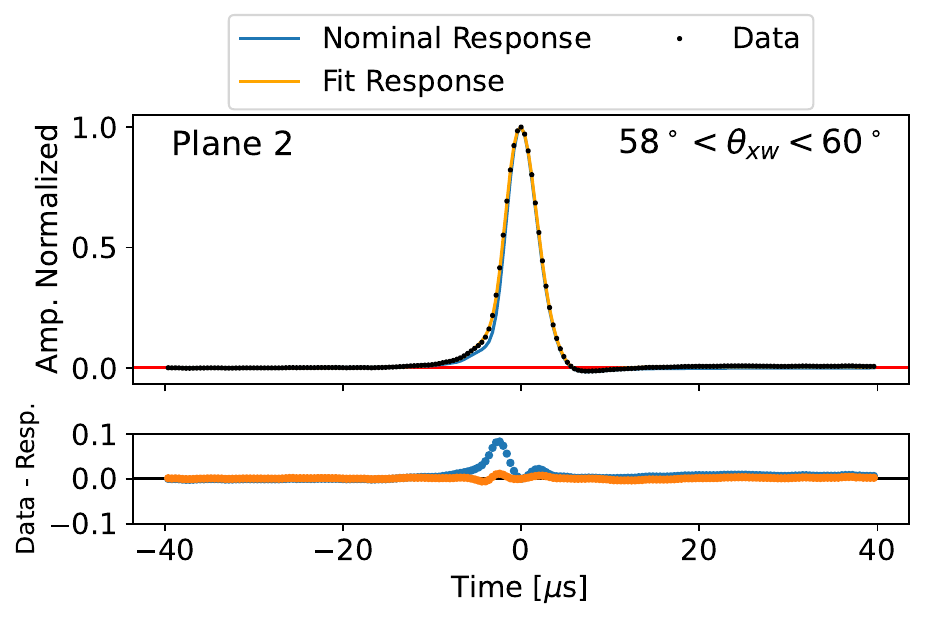}
    \includegraphics[width=0.24\textwidth]{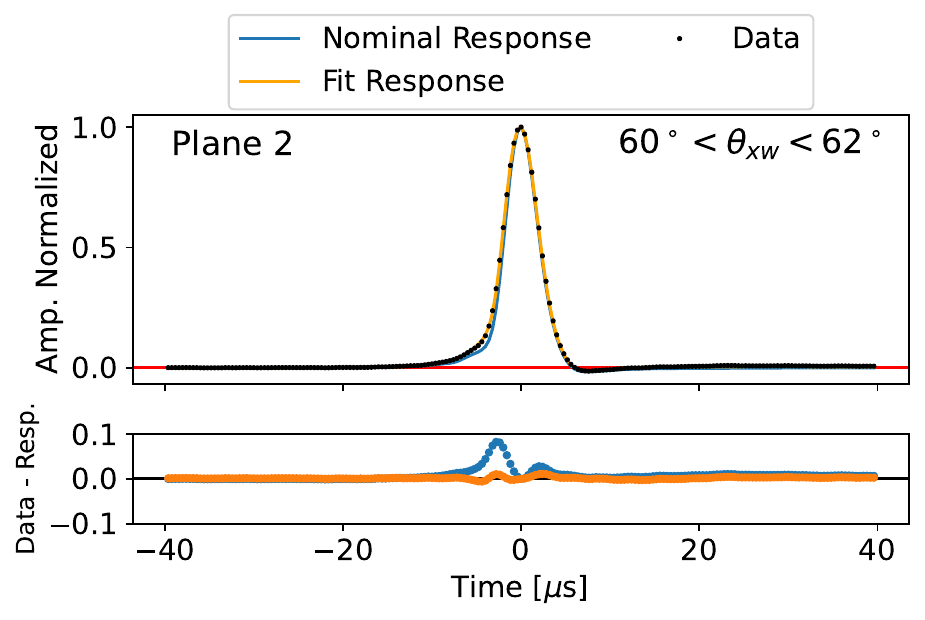}
    \includegraphics[width=0.24\textwidth]{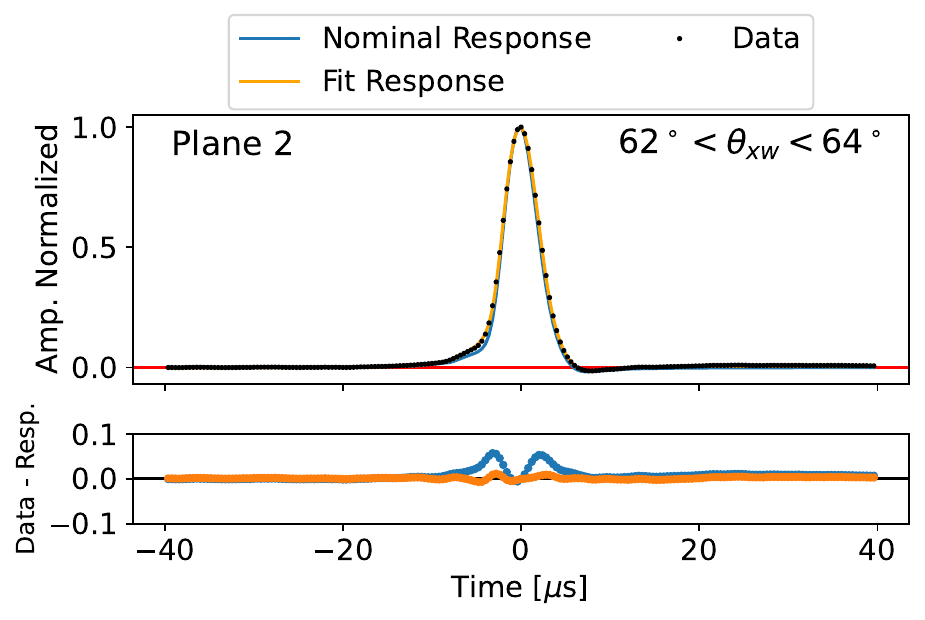}
    \includegraphics[width=0.24\textwidth]{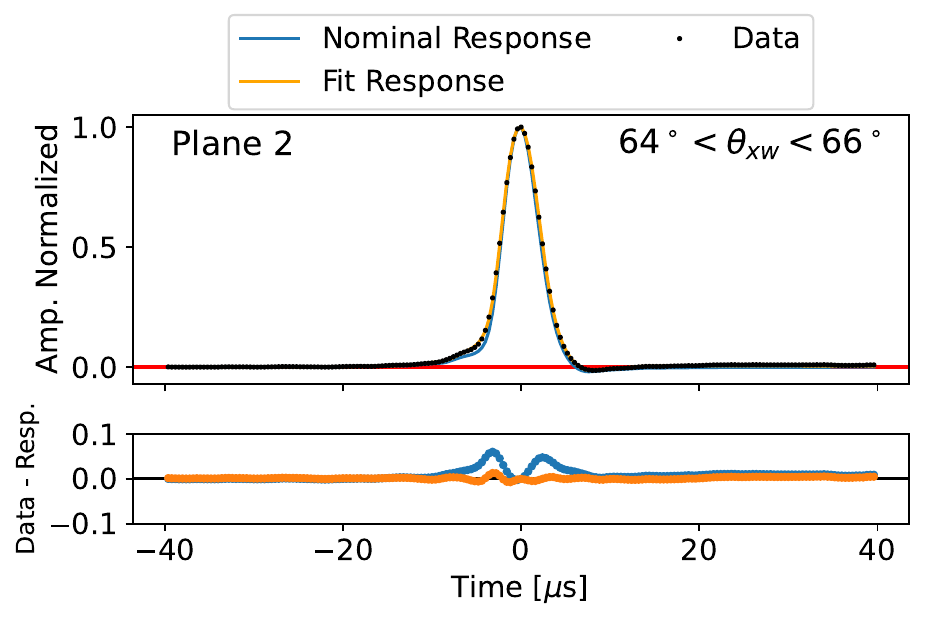}
    \includegraphics[width=0.24\textwidth]{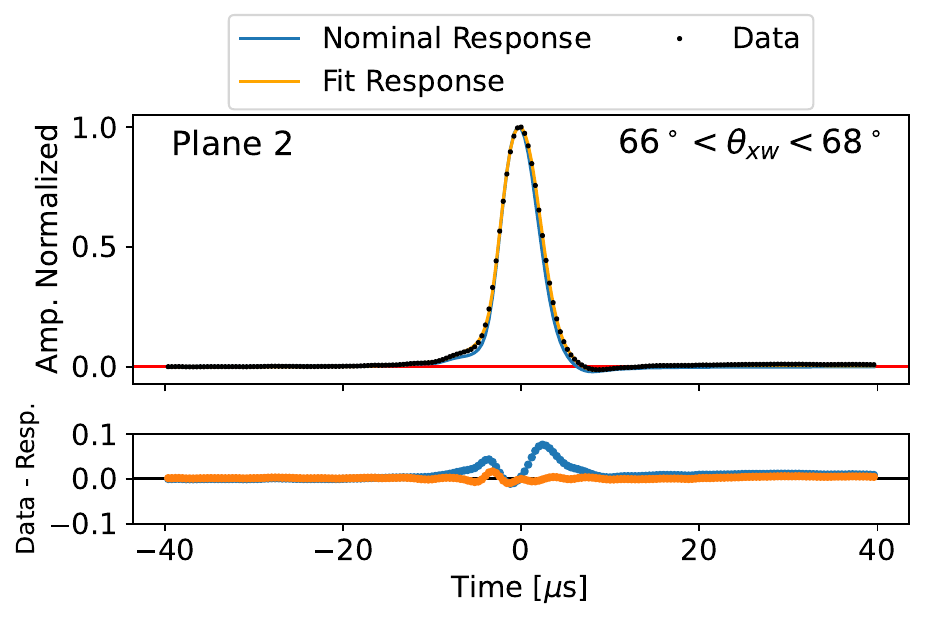}
    \includegraphics[width=0.24\textwidth]{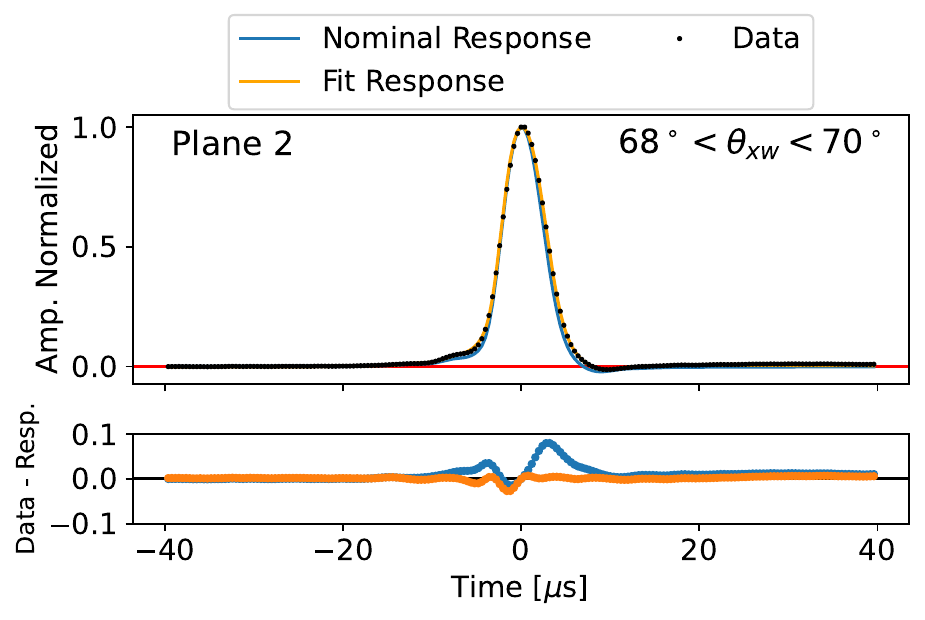}
    \includegraphics[width=0.24\textwidth]{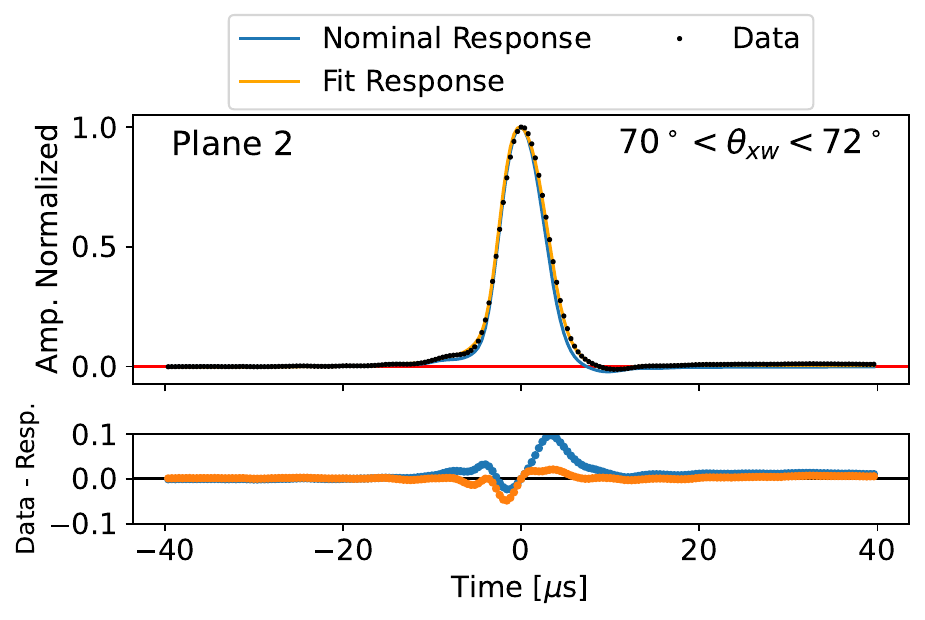}
    \includegraphics[width=0.24\textwidth]{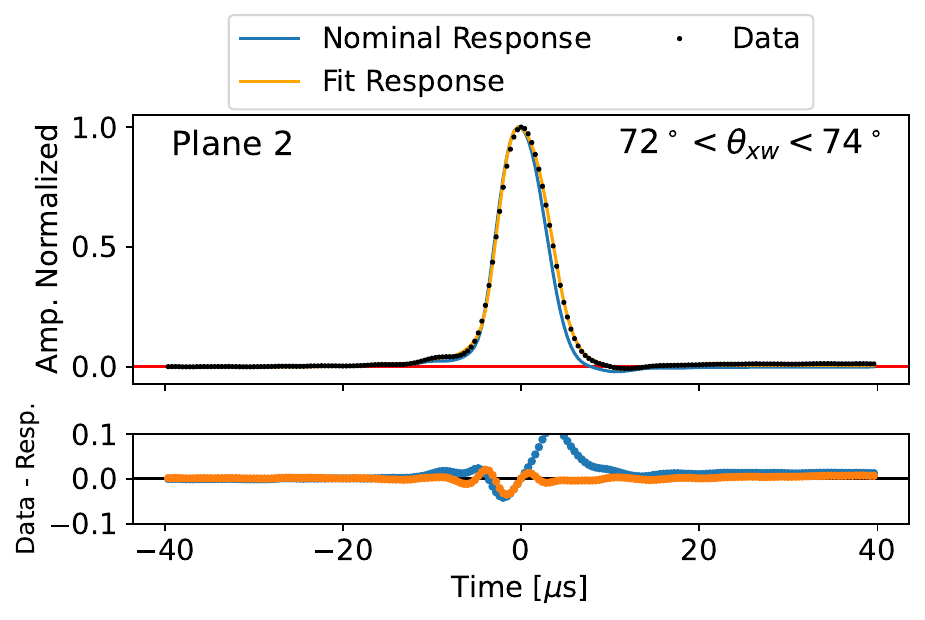}
    \includegraphics[width=0.24\textwidth]{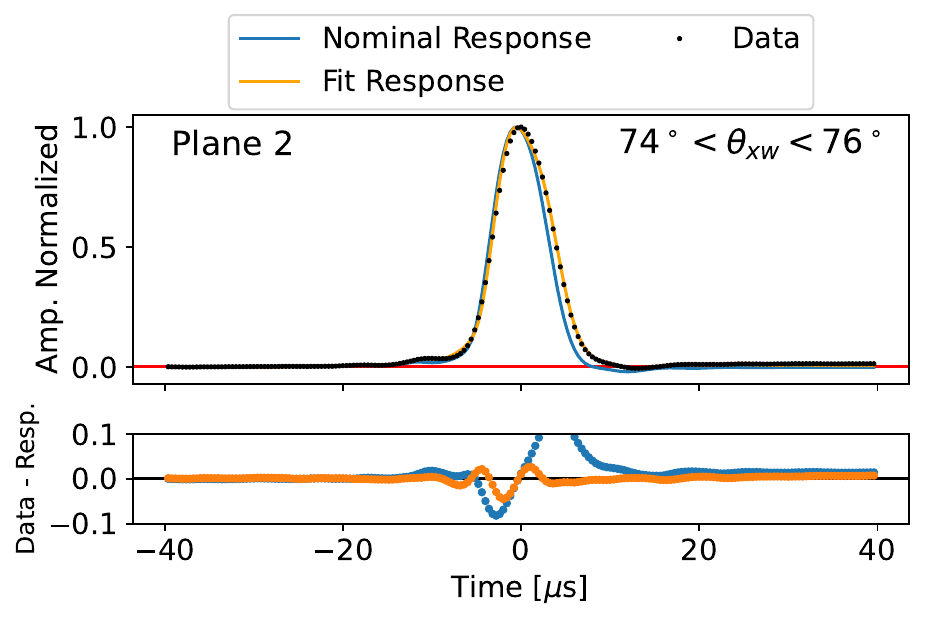}
    \caption{Signal shape fits on the collection plane. Each plot shows one angle bin between $20^\circ$ and $76^\circ$. The blue curve is the nominal ICARUS signal shape and the orange is the result of the fit.}
    \label{fig:signalshape_datafit_Plane2}
\end{figure}

The results of the fit are shown in figures \ref{fig:signalshape_datafit_Plane1} and \ref{fig:signalshape_datafit_Plane2} for the middle induction plane and collection plane respectively. The fit is done in angle bins $2^\circ$ in width from $\theta_{xw} = 20^\circ - 76^\circ$. The fit improves the signal shape model at all angles on each plane.

\subsection{Tuned Signal Shape Results}

\begin{figure}[]
    \centering
    \includegraphics[width=0.32\textwidth]{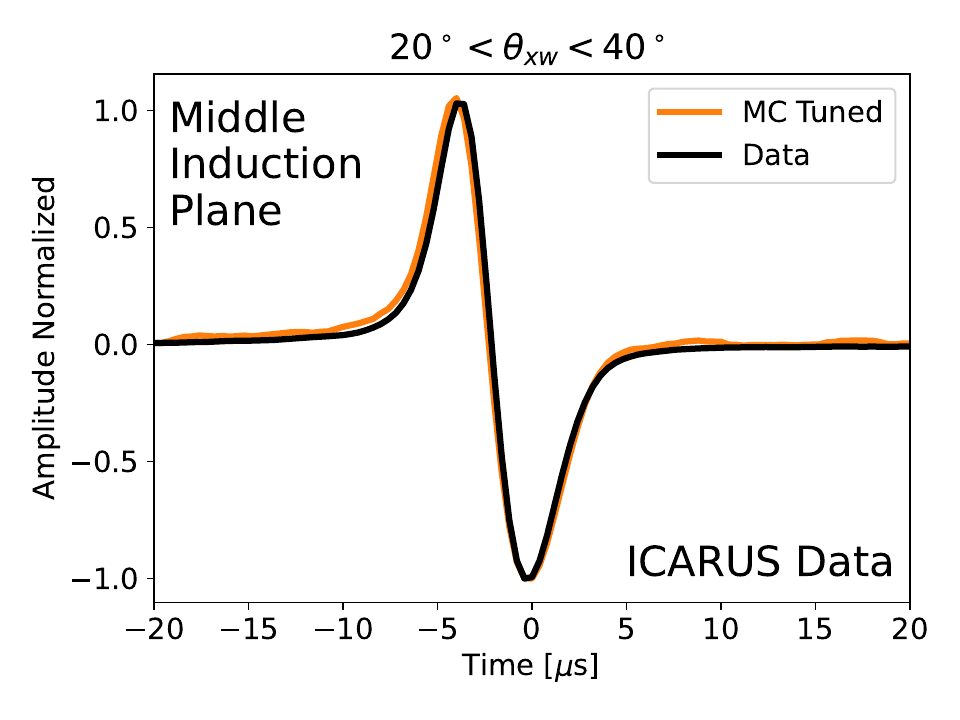}
    \includegraphics[width=0.32\textwidth]{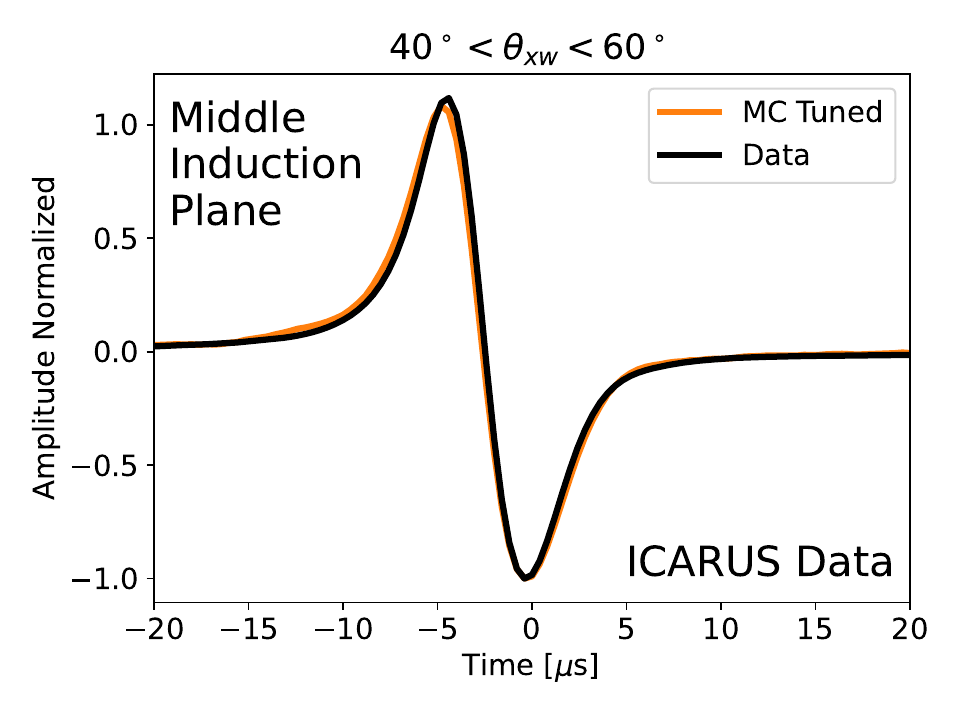}
    \includegraphics[width=0.32\textwidth]{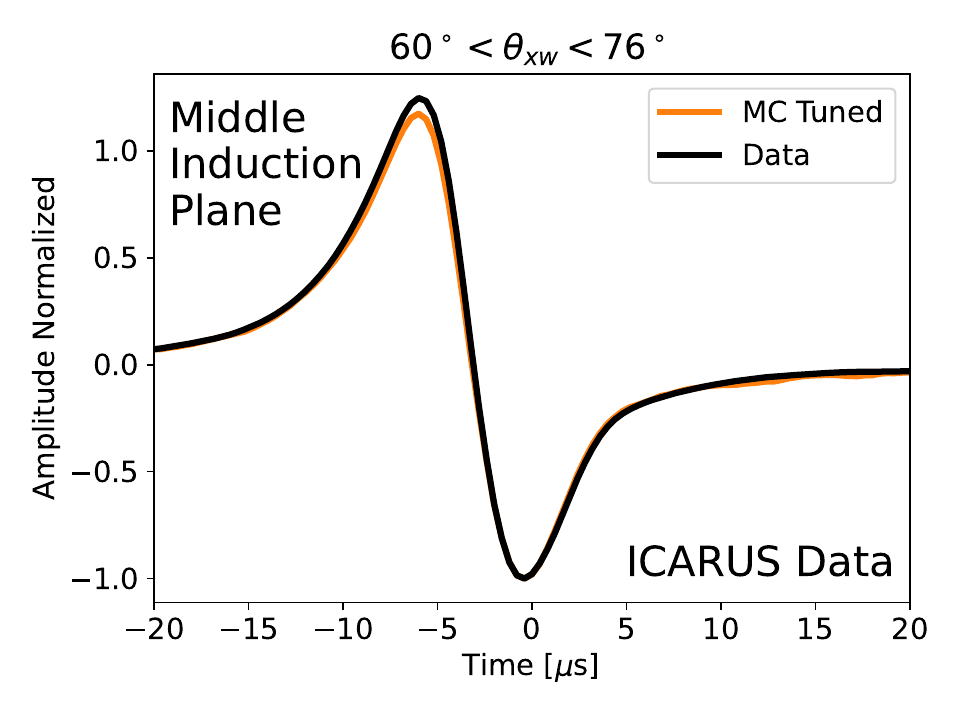}
    
    \includegraphics[width=0.32\textwidth]{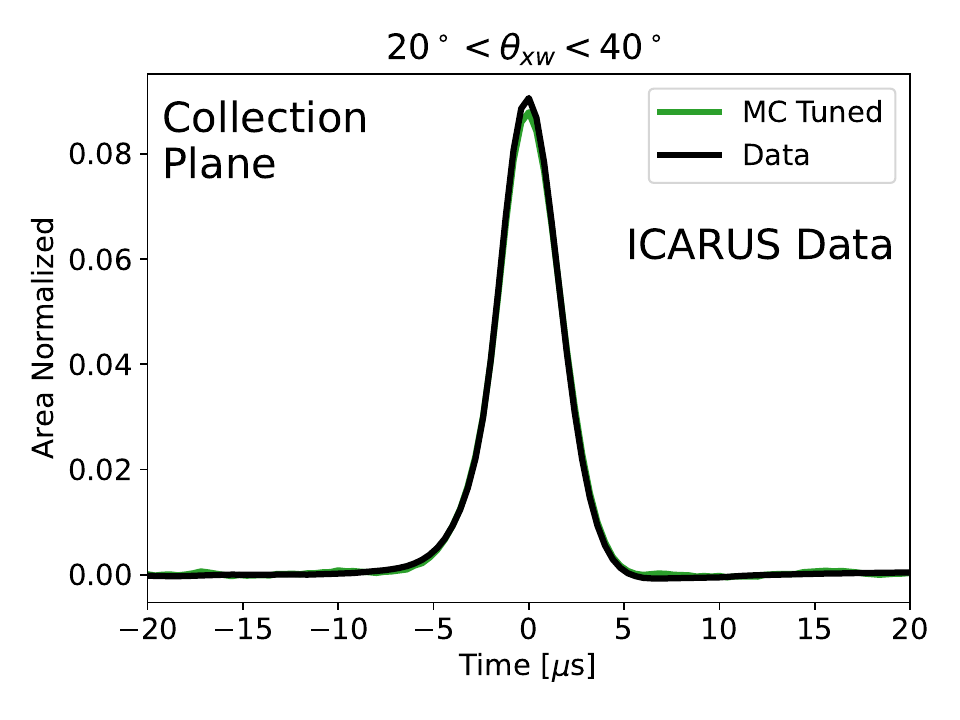}
    \includegraphics[width=0.32\textwidth]{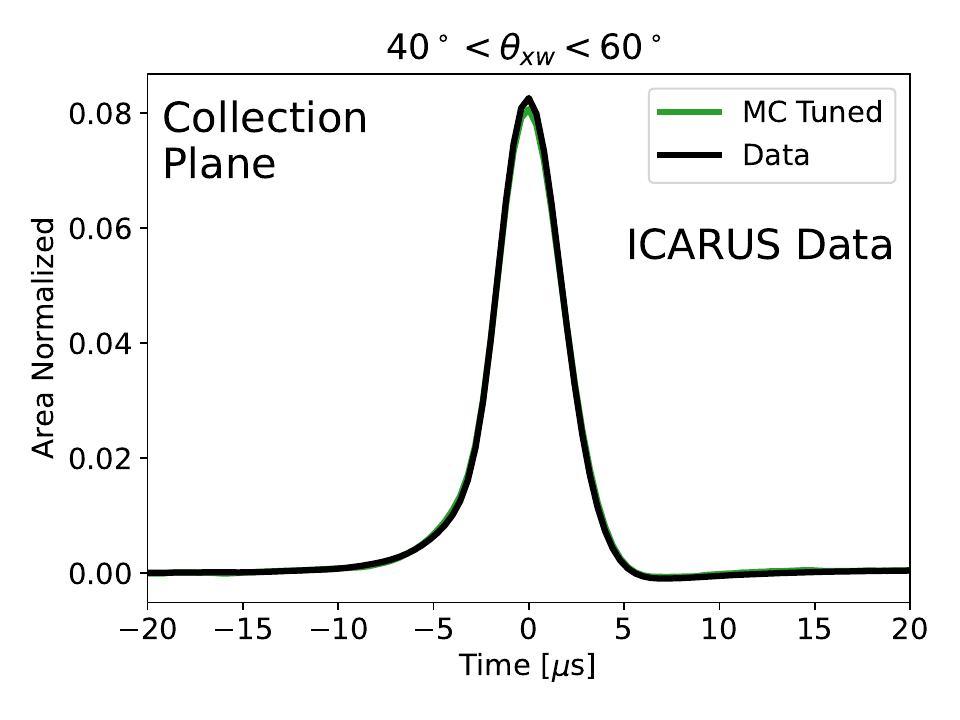}
    \includegraphics[width=0.32\textwidth]{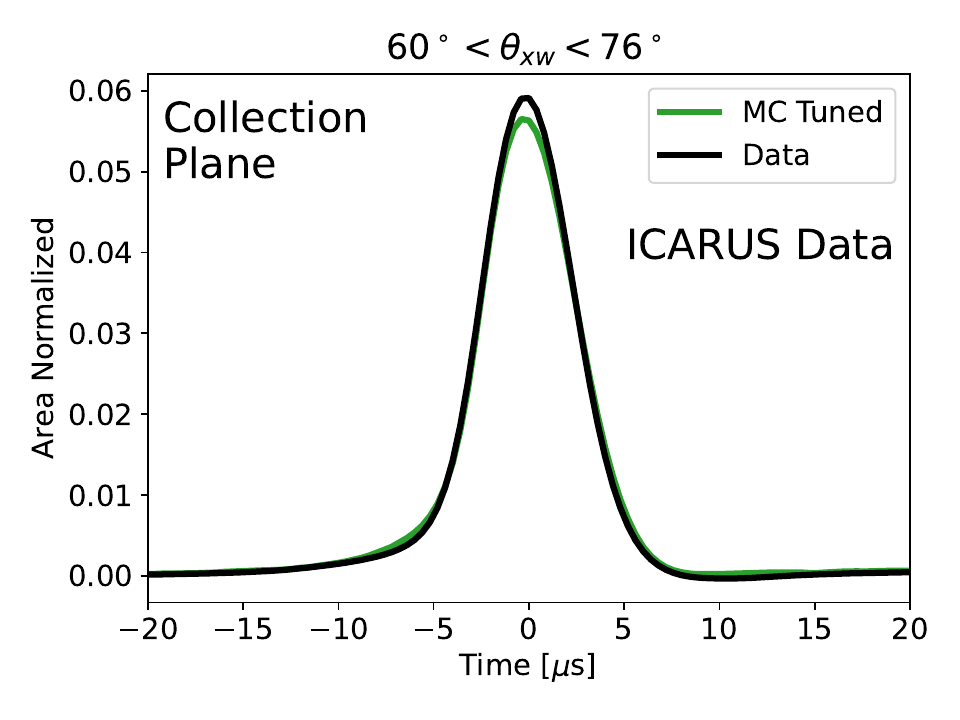}
    
    \caption{Comparison of the signal shape measurement between data and the tuned Monte Carlo simulation (``MC Modified") on the  middle induction plane (top row) and collection plane (bottom row). Shown are measured signal waveforms averaged across $\theta_{xw}$ ranges of [$20^\circ$, $40^\circ$] (left column), [$40^\circ$, $60^\circ$] (middle column), and [$60^\circ$, $76^\circ$] (right column).}
    \label{fig:signalshpae_validation_mod}
\end{figure}

As a validation of the tuned signal shapes, we compare the signal shape measurement from data against Monte Carlo simulation generated with the tuning applied.  The comparison is shown above for the nominal signal shapes in figure \ref{fig:signalshpae_validation_nom}. Figure \ref{fig:signalshpae_validation_mod} shows the comparison on the middle induction and collection planes with the tune applied. The modeling is improved in all angle bins on both planes.

\section{Charge Resolution Comparison}
\label{sec:ChargeResComp}
\begin{figure}[]
    \centering
    \includegraphics[width=0.32\textwidth]{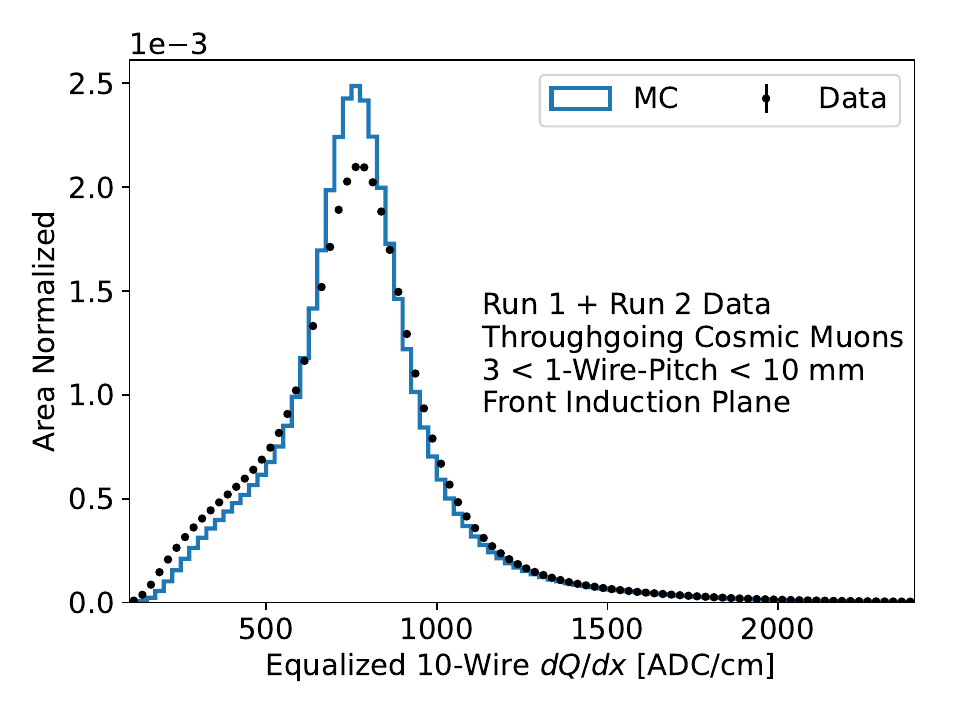}
    \includegraphics[width=0.32\textwidth]{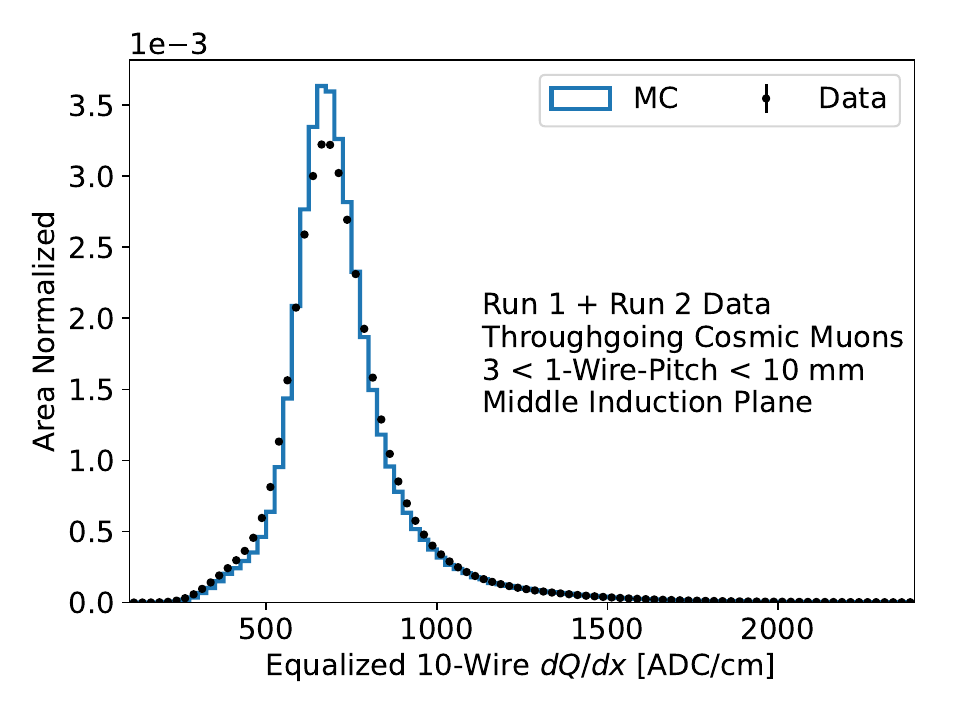}
    \includegraphics[width=0.32\textwidth]{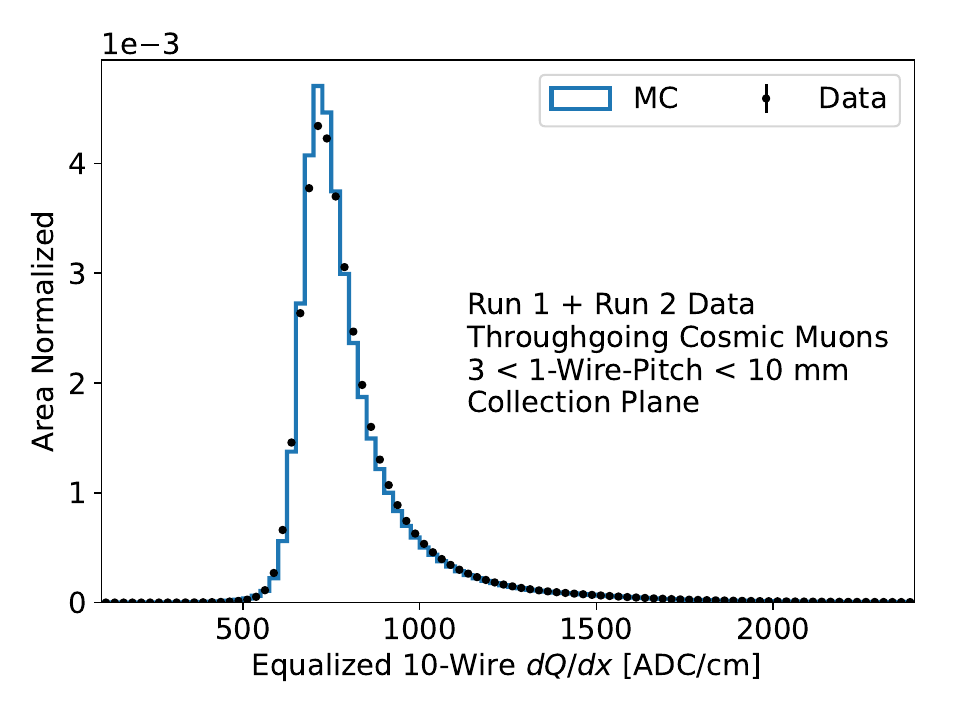}
    \caption{Distribution of equalized coarse-grained $dQ/dx$ values for throughgoing cosmic muons, compared between Monte Carlo simulation and data. The simulation uses the nominal GARFIELD signal shape on the front induction plane and the tuned signal shape on the other two planes. The data is taken from the Run 1 and Run 2 datasets. Shown for the front induction plane (left), the middle induction plane (middle), and the collection plane (right).}
    \label{fig:q=dist_finalMCdata}
\end{figure}

To validate the equalization and simulation results of this paper, we compare the distribution of equalized $dQ/dx$ for throughgoing cosmic muons between data and Monte Carlo simulation. This is shown in figure \ref{fig:q=dist_finalMCdata}. The data is shown after applying the corrections discussed in section \ref{sec:Equalization}. The simulation uses the noise simulation described in section \ref{sec:Noise}. The signal shape applies the nominal GARFIELD simulation on the front induction plane, and the tuned signal shape (as described in section \ref{sec:SignalShape}) on the middle and collection planes. The Monte Carlo simulation does not include any y-z detector response variations. It is simulated with a uniform \SI{3}{\milli\second} lifetime, which is corrected for using the same methodology as in the data. The simulated gain was tuned on each plane so that the peaks of the distributions matched.

Taken together, the final comparison shows very good agreement on all planes. There is a small residual underestimation of the charge resolution in simulation. This is observed on all three planes and is biggest on the front induction plane. There are a number of possibilities that could explain this effect: variations in the effective channel gain (from, e.g., the varying electron lifetime) not included in the simulation, deficiencies in the noise model, or differences in the inherent fluctuations from recombination, for example. The source of these residual disagreements are currently being investigated.

\section{Conclusion}
\label{sec:Conclusion}
This paper has described the procedure developed on ICARUS to equalize charge measurements in data and tune the simulated TPC noise and signal shapes to data. Charge measurements are equalized in the drift and wire plane directions. These corrections predominantly address the attenuation of charge signals due to argon impurities and a variable in-transparency to charge across the induction planes in ICARUS. The noise is simulated directly from measurements of signal-less wires. The signal shape is modeled by a GARFIELD simulation of the ICARUS wire planes, with tuning done on the middle induction and collection planes to match the simulated signal shapes to distortions observed in the data. This tuning is a novel procedure we have developed for ICARUS. 

At this stage of calibration, the modeling of charge resolution is satisfactory on all three wire planes of the ICARUS TPC. Residual uncertainties on the detector performance arise predominantly from variations in the detector response not included in the simulation. The impact of such variations on charge calorimetry are removed by the charge equalization procedure, but this calibration cannot remove the variation in detector performance that is baked-in by (e.g.) the varying signal-to-noise ratio. Future work to calibrate the ICARUS TPC will address the simulation of these variations in signal-to-noise across the runtime of the experiment. When these variations are included, it may also be possible to apply the signal shape tuning procedure on the front induction plane.

\section*{Acknowledgements}
\label{sec:acknowledgements}
This document was prepared by the ICARUS
Collaboration using the resources of the Fermi
National Accelerator Laboratory (Fermilab), a
U.S. Department of Energy, Office of Science,
HEP User Facility. Fermilab is managed by
Fermi Research Alliance, LLC (FRA), acting
under Contract No. DE-AC02-07CH11359.
This work was supported by the US Department of Energy, INFN,EU Horizon 2020 Research
and Innovation Program under the Marie Sklodowska-Curie Grant Agreement
No. 734303, 822185, 858199 and 101003460, and the Horizon Europe Research
and Innovation Program under the Marie Sklodowska-Curie Grant Agreement
No. 101081478
Part of the work resulted from the implementation of
the research Project No. 2019/33/N/ST2/02874
funded by the National Science Centre, Poland.
We also acknowledge the contribution of
many SBND colleagues, in particular for the development of a number of simulation, reconstruction and analysis tools which are shared
within the SBN Program.

\bibliographystyle{JHEP}
\bibliography{cite}

\appendix

\section{Field and Electronic Response Transformations in Signal Shape Fit}
\label{sec:SignalShapeAppendix}
The single electron field response fit applies a set of non-linear transformations to the nominal Wire-Cell responses. The transformations depend on the time $t$ and the location $x$ along the direction perpendicular to the wire orientation ($\hat{w}$). The fit is done by splitting each field response into a left (denoted with a subscript $\ell$) and right (denoted with a subscript $r$) side of a central time tick. The time tick is defined as the peak of the field response on the collection plane and the zero-cross point on the induction planes. Both the shape of the field response $s$ and the time input to the field response $t$ is transformed. All position dependence is encoded in an ``offset parameter" $o$. The fit single electronics field response $s(x,t)$, in terms of the nominal WireCell single electronics field response $s^0(x,t)$, is defined below.
\begin{equation}
        \begin{split}
        s(t, x) =& s_\ell(t'(t, x), x)\cdot (t'(t, x) < 0) + s_r(t'(t, x), x) \cdot (t'(t, x) \geq
        0)\\
        s_{\ell,r}(t,x) =& a_{\ell,r}^0(x) s^s_{\ell,r}(t,x) +
        a_{\ell,r}^1(x)
        (s^s_{\ell,r}(t,x))^2\cdot\text{sign}(s^s_{\ell,r}(t,x))\\
        s^s_{\ell,r}(t,x) =& (s^0(t, x) + ds_{\ell, r}(t, x))
        \cdot\text{exp}\left[(t >
        \bm{t^\text{start}_{\ell,r}})\cdot \bm{e_{\ell,r}}|t|\right]\\
        a_{\ell,r}^{0,1}(x) =& \bm{a0_{\ell,r}^{0,1}} + \bm{a1_{\ell,r}^{0,1}}\cdot
        o(x)\\
        ds_{\ell}(t, x) =& 0\,,\quad ds_{r}(t, x)
        \frac{\bm{c^p_{r}} \cdot\text{exp}[-|x|/\bm{\ell^p_{r}}]\cdot(|x| \leq
        1.5 \text{mm})}{1 +
        (|t|/\bm{\tau^p_{r}})^{\bm{a^p_{r}}}}\cdot(|t| >
        \bm{t^{p-\text{start}}_{r}})\\
        t'(t, x) =& t'_{\ell}(t, x)\cdot(t < 0) + t'_{r}(t, x)\cdot(t
        \geq 0) + \bm{c}\cdot o(x)\\
        t'_{\ell,r}(t,x) =& t\cdot\left(s_{\ell,r}^0(x) +
        \frac{s_{\ell,r}^2(x)}{1 + \left(t/\tau_{\ell,r}^2(x)\right)^2} +
        \frac{s_{\ell,r}^4(x)}{1 +
        \left(t/\tau_{\ell,r}^4(x)\right)^4}\right)\\
        s^{0,2,4}_{\ell, r}(x) =& \bm{s1^{0,2,4}_{\ell, r}}\cdot o(x) +
        \bm{s2^{0,2,4}_{\ell, r}} \cdot o(x)^2\\
        \tau^{2,4}_{\ell, r}(x) =& \bm{\tau1^{2,4}_{\ell, r}} +
        \bm{\tau2^{2,4}_{\ell, r}} \cdot o(x)\\
        o(x) =& 1 - e^{-|x|/1.5\text{mm}}\,.
    \end{split}
\end{equation}
The fit parameters in these equations are in bold. In these equation, there are 16 fit parameters on both sides of $t = 0$ $(\ell, r)$: 
$t^\text{start}_{\ell, r}$, $e_{\ell,r}$,
$a0^0_{\ell, r}$, $a0^1_{\ell, r}$, 
$a1^0_{\ell, r}$, $a1^1_{\ell, r}$, 
$s1^0_{\ell,r}$, $s1^2_{\ell,r}$, $s1^4_{\ell,r}$, $s2^0_{\ell,r}$, $s2^2_{\ell,r}$, $s2^4_{\ell,r}$, 
$\tau 1^2_{\ell,r}$, $\tau 1^4_{\ell,r}$, $\tau 2^2_{\ell,r}$ and $\tau 2^4_{\ell,r}$. There are 5 parameters only on the right side of $t=0$: $c^p_{r}$, $\ell^p_{r}$, $\tau^p_{r}$, $a^p_{r}$, and $t^{p-start}_{r}$. Finally, there is the time shift parameter $c$. In total, there are 38 parameters in the field response fit.

The electronics response is also fit for. The nominal electronics response $e^0(\tau; t)$ is a Bessel shaping function with a nominal shaping time $\tau = $\SI{1.3}{\micro\second}. This nominal shape is convolved with the long RC-tail measured externally to the signal shape fit as described in section \ref{sec:signalshape_ertail}. To allow for further distortions, the response is convolved with an RC-RC tail function with a time constant $\tau_\text{RCRC}$. The full fit electronics response $e(t)$ is
\begin{equation}
    \begin{split}
        e(t) &= (e^0(\bm{\tau}) \circledast \text{RC}(\bm{A}, \bm{\tau_\text{RC}}) \circledast\text{RC-RC}(\bm{\tau_\text{RCRC}}))(t)\\
        \text{RC}(A, \tau; t) &= \delta(t) + A e^{-t/\tau}\\
        \text{RC-RC}(\tau; t) &= \left( \frac{t}{\tau} - 2 \right) \frac{e^{-t/\tau}}{\tau}\,,
    \end{split}
\end{equation}
where $\circledast$ denotes a convolution and $\delta$ is the dirac-delta function. The four fit parameters, $\tau$, $\tau_\text{RC}$, $\tau_\text{RCRC}$, and $A$, are all in bold.

\end{document}